\pdfoutput=1
\RequirePackage{ifpdf}
\ifpdf 
\documentclass[pdftex]{sigma}
\else
\documentclass{sigma}
\fi

\numberwithin{equation}{section}

\newtheorem{Theorem}{Theorem}[section]
\newtheorem{Corollary}[Theorem]{Corollary}
\newtheorem{Lemma}[Theorem]{Lemma}
\newtheorem{Proposition}[Theorem]{Proposition}
{\theoremstyle{definition}
\newtheorem{Definition}[Theorem]{Definition}
\newtheorem{Remark}[Theorem]{Remark}
}

\DeclareMathOperator{\tr}{tr}
\DeclareMathOperator{\diag}{diag}

\begin{document}

\allowdisplaybreaks

\renewcommand{\thefootnote}{$\star$}

\renewcommand{\PaperNumber}{079}

\FirstPageHeading

\ShortArticleName{Master Symmetries of the Ablowitz--Ladik Hierarchy}

\ArticleName{A Centerless Virasoro Algebra of Master Symmetries\\
for the Ablowitz--Ladik Hierarchy\footnote{This paper is a~contribution to the Special Issue
in honor of Anatol Kirillov and Tetsuji Miwa.
The full collection is available at \href{http://www.emis.de/journals/SIGMA/InfiniteAnalysis2013.html}{http://www.emis.de/journals/SIGMA/InfiniteAnalysis2013.html}}}

\Author{Luc HAINE and Didier VANDERSTICHELEN}

\AuthorNameForHeading{L.~Haine and D.~Vanderstichelen}

\Address{Institut de Recherche en Math\'ematique et
Physique, Universit\'e catholique de Louvain,\\
Chemin du Cyclotron 2, 1348 Louvain-la-Neuve, Belgium}

\Email{\href{mailto:luc.haine@uclouvain.be}{luc.haine@uclouvain.be},
\href{mailto:didier.vanderstichelen@uclouvain.be}{didier.vanderstichelen@uclouvain.be}}

\ArticleDates{Received July 31, 2013, in f\/inal form November 30, 2013; Published online December 12, 2013}

\Abstract{We show that the (semi-inf\/inite) Ablowitz--Ladik (AL) hierarchy admits a~centerless Virasoro
algebra of master symmetries in the sense of Fuchssteiner [\textit{Progr.\ Theoret.\ Phys.} \textbf{70} (1983), 1508--1522].
An explicit expression for these symmetries is given in terms of a~slight generalization of the Cantero,
Moral and Vel\'azquez (CMV) matrices [\textit{Linear Algebra Appl.} \textbf{362} (2003), 29--56] and their
action on the tau-functions of the hierarchy is described.
The use of the CMV matrices turns out to be crucial for obtaining a~Lax pair representation of the master
symmetries.
The AL hierarchy seems to be the f\/irst example of an integrable hierarchy which admits a~\textit{full
centerless} Virasoro algebra of master sym\-metries, in contrast with the Toda lattice and Korteweg--de~Vries
hierarchies which possess only ``half of'' a~Virasoro algebra of master sym\-metries, as explained in Adler
and van Moerbeke [\textit{Duke Math.~J.} \textbf{80} (1995), 863--911], Damianou [\textit{Lett.\
Math.\ Phys.} \textbf{20} (1990), 101--112] and Magri and Zubelli [\textit{Comm.\ Math.\ Phys.} \textbf{141} (1991), 329--351].}

\Keywords{Ablowitz--Ladik hierarchy; master symmetries; Virasoro algebra}

\Classification{37K10; 17B68}

\renewcommand{\thefootnote}{\arabic{footnote}}
\setcounter{footnote}{0}

\tableofcontents

\section{Introduction}

The group $U(n)$ of $n\times n$ unitary matrices, with Haar measure $\text{d}U$ normalized as a~probability
measure, has eigenvalue probability distribution given by the Weyl formula
\begin{gather*}
\frac{1}{n!}\vert\Delta_n(z)\vert^2\prod_{k=1}^{n}\frac{\text{d}z_k}{2\pi iz_k},
\qquad
z_k=e^{i\varphi_k}\in S^1,
\qquad
\varphi_k\in{}]{-}\pi,\pi],
\end{gather*}
with $S^1=\{z\in \mathbb{C}: \vert z\vert=1\}$ the unit circle, and $\Delta_n(z)$ the Vandermonde
determinant
\begin{gather}
\label{Vandermonde}
\Delta_n(z)=\det\big(z_l^{k-1}\big)_{1\leq k,l\leq n}=\prod_{1\leq k<l\leq n}(z_l-z_k).
\end{gather}
Thus, for $\eta, \theta \in {}]{-}\pi,\pi]$, with $\eta\leq \theta$, the probability that a~randomly chosen
matrix from $U(n)$ has no eigenvalues within the arc of circle $\{z\in S^1: \eta<\text{arg}(z)< \theta\}$
is given by
\begin{gather*}
\tau_n(\eta,\theta)=\frac{1}{(2\pi)^n n!}\int_{\theta}^{2\pi+\eta}\cdots\int_{\theta}^{2\pi+\eta}
\prod_{1\leq k<l\leq n}\vert e^{i\varphi_k}-e^{i\varphi_l}\vert^2\text{d}\varphi_1\cdots\text{d}
\varphi_n.
\end{gather*}
Obviously, this probability depends only on the length $\theta-\eta$.

The starting motivation for the present work was our attempt in~\cite{HV} to understand a~dif\/fe\-ren\-ti\-al
equation satisf\/ied by the function $-\frac{1}{2}\frac{\text{d}}{\text{d}\theta}\log\tau_n(-\theta,\theta)$, obtained by
Tracy and Widom in~\cite{TW}, from the point of view of the Adler--Shiota--van~Moerbeke
approach~\cite{ASVM}, in terms of Virasoro constraints.
Introducing the 2-Toda time-dependent tau-functions
\begin{gather}
\label{tauint}
\tau_n(t,s;\eta,\theta)=\frac{1}{n!}\int_{[\theta,2\pi+\eta]^n}\text{d}I_{n}(t,s,z)
\end{gather}
with $(t,s)=(t_1,t_2,\ldots,s_1,s_2,\ldots)$ and
\begin{gather*}
\text{d}I_{n}(t,s,z)=\vert\Delta_n(z)\vert^2\prod_{k=1}^{n}\left(e^{\sum\limits_{j=1}^\infty(t_jz_k^j+s_jz_k^{-j})}
\frac{\text{d}z_k}{2\pi iz_k}\right),
\end{gather*}
deforming the probabilities $\tau_n(\eta,\theta)=\tau_n(0,0;\eta,\theta)$, we discovered that they satisfy
a~set of Virasoro constraints indexed by \emph{all} integers, decoupling into a~boundary-part and
a~time-part
\begin{gather}
\label{virtauint}
\frac{1}{i}\left(e^{ik\theta}\frac{\partial}{\partial\theta}+e^{ik\eta}\frac{\partial}{\partial\eta}\right)
\tau_n(t,s;\eta,\theta)=L_k^{(n)}\tau_n(t,s;\eta,\theta),
\qquad
k\in\mathbb{Z},
\qquad
i=\sqrt{-1},
\end{gather}
with the time-dependent operators $L_k^{(n)}$ providing a~\emph{centerless} representation of the
\emph{full} Virasoro algebra, that is
\begin{gather}
\label{Virasoro L}
\big[L_k^{(n)},L_l^{(n)}\big]=(k-l)L_{k+l}^{(n)},
\qquad
\forall\, k,l\in\mathbb{Z}.
\end{gather}
The basic trick for this result was to use the Lagrangian approach~\cite{MM} for obtaining Virasoro
constraints in matrix models, showing that the following variational formulas hold $\forall\, k\geq 0$
\begin{gather*}
\frac{\text{d}}{\text{d}\varepsilon}\text{d}
I_n\big(z_\alpha\mapsto z_\alpha e^{\varepsilon(z_\alpha^k-z_\alpha^{-k})}\big)\Big|_{\varepsilon=0}
=\big(L_k^{(n)}-L_{-k}^{(n)}\big)\text{d}I_n,
\\
\frac{\text{d}}{\text{d}\varepsilon}\text{d}
I_n\big(z_\alpha\mapsto z_\alpha e^{i\varepsilon(z_\alpha^k+z_\alpha^{-k})}\big)\Big|_{\varepsilon=0}
=i\big(L_k^{(n)}+L_{-k}^{(n)}\big)\text{d}I_n,
\end{gather*}
with $L_k^{(n)}$ given by
\begin{gather}
L_k^{(n)}=\sum_{j=1}^{k-1}\frac{\partial^2}{\partial t_{j}\partial t_{k-j}}+n\frac{\partial}{\partial t_k}
+\sum_{j=1}^{\infty}jt_j\frac{\partial}{\partial t_{j+k}}\nonumber
\\
\phantom{L_k^{(n)}=}
{}-\sum_{j=k+1}^{\infty}js_j\frac{\partial}{\partial s_{j-k}}-\sum_{j=1}^{k-1}js_j\frac{\partial}
{\partial t_{k-j}}-nks_k,
\qquad
k\geq1,
\label{virplus}
\\
L_0^{(n)}=\sum_{j=1}^{\infty}jt_j\frac{\partial}{\partial t_{j}}-\sum_{j=1}^{\infty}js_j\frac{\partial}
{\partial s_{j}},
\label{virzero}
\\
L_{-k}^{(n)}=-\sum_{j=1}^{k-1}\frac{\partial^2}{\partial s_j\partial s_{k-j}}-n\frac{\partial}
{\partial s_k}-\sum_{j=1}^{\infty}js_j\frac{\partial}{\partial s_{j+k}}\nonumber
\\
\phantom{L_{-k}^{(n)}=}
{}+\sum_{j=k+1}^{\infty}jt_j\frac{\partial}{\partial t_{j-k}}+\sum_{j=1}^{k-1}jt_j\frac{\partial}
{\partial s_{k-j}}+nkt_k,
\qquad
k\geq1
\label{virminus}
.
\end{gather}

When $\eta=\theta$, the integral~\eqref{tauint} is obviously independent of $\theta$, and the left-hand
side of~\eqref{virtauint} is equal to zero.
By using Weyl's integration formula, one can recognize it as the partition function of the unitary matrix
model, introduced in~\cite{M}.
After~\cite{HV} was completed, we found out that our result in this case had already been obtained by
Bowick, Morozov and Shevitz~\cite{BMS}, though these authors didn't notice the commutation
relations~\eqref{Virasoro L} of the centerless Virasoro algebra (see Corollary~\ref{unitary matrix model}
and Remark~\ref{Virasoro unitary matrix model}).
Kharchev and Mironov~\cite{KM} f\/irst recognized that the partition function of the unitary matrix model
is a~special tau function of the two-dimensional Toda lattice (in short 2DTL) hierarchy of Ueno and
Takasaki~\cite{UT}, by using bi-orthogonal polynomials on the circle.
Then, Kharchev, Mironov and Zhedanov~\cite{KMZ1, KMZ2} showed that the coef\/f\/icients entering the
Szeg\"o type recursion relations satisf\/ied by these bi-orthogonal polynomials solve the semi-inf\/inite
Ablowitz--Ladik (AL in short) hierarchy, a~result which is already implicitly contained in~\cite{KM}.
We remind the reader that the f\/irst vector f\/ield of the AL hierarchy is the system of
dif\/ferential-dif\/ference equations introduced by Ablowitz and Ladik~\cite{AL1, AL2} in the form
\begin{gather}
\dot{x}_n=x_{n+1}-2x_n+x_{n-1}-x_n y_n(x_{n+1}+x_{n-1}),
\nonumber
\\
\dot{y}_n=-y_{n+1}+2y_n-y_{n-1}+x_n y_n(y_{n+1}+y_{n-1}).
\label{ALequations}
\end{gather}
Upon making the change of variable $t\to it$, when $y_n=\mp\overline{x_n}$ the system reduces to the
equation¨
\begin{gather*}
-i\dot{x}_n=x_{n+1}-2x_n+x_{n-1}\pm\vert x_n\vert^2(x_{n+1}+x_{n-1}),
\end{gather*}
which is a~discrete version of the focusing/defocusing nonlinear Schr\"odinger equation.

The functions $\tau_n(t,s;\eta,\theta)$ are thus \emph{special} instances of tau-functions of the
semi-inf\/inite AL hierarchy.
The Virasoro constraints they satisfy suggest that the semi-inf\/inite AL hierarchy admits a~\emph{full
centerless} Virasoro algebra of additional symmetries (so-called master symmetries), a~notion which will be
explained below.
The goal of this paper is to identify the Virasoro algebra of master symmetries both on the variables
$x_n$, $y_n$, $n\geq 0$, as well as on the \emph{general} tau-functions of the AL hierarchy.
Since the pioneering works~\cite{KM, KMZ1, KMZ2} the fact that the semi-inf\/inite AL hierarchy is related
to (bi)-orthogonal polynomials on the circle in the same way as the semi-inf\/inite Toda lattice hierarchy
is related to orthogonal polynomials on the line, has been rediscovered several times, see for
instance~\cite{AVM2, AVM3, C, N}.
We now introduce the necessary tools to explain this connection.

We denote by $\mathbb{C}\big[z,z^{-1}\big]$ the ring of Laurent polynomials over $\mathbb{C}$.
A bilinear form
\begin{gather}
\label{bi-functional}
\mathcal{L}: \ \mathbb{C}\big[z,z^{-1}\big]\times\mathbb{C}\big[z,z^{-1}\big]\to\mathbb{C},
\qquad
(f,g)\mapsto\mathcal{L}[f,g],
\end{gather}
will be called a~bi-moment functional.
The bi-moments associated to $\mathcal{L}$ are
\begin{gather}
\label{bimoments first form}
\mu_{mn}=\mathcal{L}\big[z^m,z^n\big],
\qquad
\forall\, m,n\in\mathbb{Z}.
\end{gather}
We assume that $\mathcal{L}$ satisf\/ies the \emph{Toeplitz condition}
\begin{gather}
\label{Toeplitz condition}
\mathcal{L}\big[z^m,z^n\big]=\mathcal{L}\big[z^{m-n},1\big],
\qquad
\forall\, m,n\in\mathbb{Z}.
\end{gather}
Because of the Toeplitz condition~\eqref{Toeplitz condition}, the bi-moments depend only on the
dif\/ference $m-n$ and we shall often write
\begin{gather}
\label{bimoments second form}
\mu_{mn}:=\mu_{m-n}.
\end{gather}
In the rest of the paper, we shall freely use both notations for the bi-moments.
An important example of a~Toeplitz bi-moment functional is provided by
\begin{gather}
\label{bimoment integral}
\mathcal{L}[f,g]=\oint_{S^1}f(z)g\big(z^{-1}\big)w(z)\frac{\text{d}z}{2\pi iz},
\end{gather}
with $w(z)$ some weight function on the unit circle $S^1$ which is not necessarily positive or even real
valued.
We shall also assume $\mathcal{L}$ to be \emph{quasi-definite}, that is
\begin{gather}
\label{quasi-definite}
\det\big(\mu_{kl}\big)_{0\leq k,l\leq n-1}\neq0,
\qquad
\forall\, n\geq1.
\end{gather}
This is a~necessary and suf\/f\/icient condition for the existence of a~sequence of bi-orthogonal
polynomials $\big\{p_n^{(1)}(z),p_n^{(2)}(z)\big\}_{n\geq 0}$ with respect to $\mathcal{L}$, that is $p_n^{(1)}(z)$
and $p_n^{(2)}(z)$ are polynomials of degree $n$, satisfying the orthogonality conditions
\begin{gather*}
\mathcal{L}\big[p_m^{(1)}(z),p_n^{(2)}(z)\big]=h_n\delta_{m,n},
\qquad
h_n\neq0,
\qquad
\forall\, m,n\in\mathbb{N}.
\end{gather*}

Introducing the variables
\begin{gather}
\label{Szegovariables}
x_n=p_n^{(1)}(0),
\qquad
y_n=p_n^{(2)}(0),
\qquad
n\geq0,
\end{gather}
the \emph{monic} bi-orthogonal polynomials $\big\{p_n^{(1)}(z), p_{n}^{(2)}(z)\big\}_{n\geq 0}$ satisfy the Szeg\"o
type recurrence relations
\begin{gather}
p_{n+1}^{(1)}(z)-zp_n^{(1)}(z)=x_{n+1}z^np_n^{(2)}\big(z^{-1}\big),
\qquad
p_{n+1}^{(2)}(z)-zp_n^{(2)}(z)=y_{n+1}z^np_n^{(1)}\big(z^{-1}\big),
\label{Szegorecurrence}
\end{gather}
from which it easily follows that
\begin{gather}
\label{condition h}
\frac{h_{n+1}}{h_n}=1-x_{n+1}y_{n+1},
\qquad
n\geq0.
\end{gather}
In~\cite{AVM2, AVM3, KM, KMZ1, KMZ2} the AL hierarchy\footnote{In~\cite{AVM2, AVM3} the terminology
``Toeplitz hierarchy'' instead of ``AL hierarchy'' is used.} is embedded in the 2DTL hierarchy by using
a~pair $(L_1, L_2)$ of Hessenberg matrices representing respectively the operator of multiplication
$\mathbb{C}[z]\to \mathbb{C}[z]: f(z)\to zf(z)$ in the bases $p^{(1)}(z)=\big(p_n^{(1)}(z)\big)_{n\geq 0}$
and $p^{(2)}(z)=\big(p_n^{(2)}(z)\big)_{n\geq 0}$ of bi-orthogonal polynomials
\begin{gather*}
zp^{(1)}(z)=L_1p^{(1)}(z),
\qquad
zp^{(2)}(z)=L_2p^{(2)}(z).
\end{gather*}
However, to represent the Virasoro algebra of master symmetries, what we shall need is a~basis of the ring
$\mathbb{C}\big[z,z^{-1}\big]$ of \emph{Laurent polynomials} in which both the operators of multiplication by $z$
and $z^{-1}$ admit nice matrix representations.
Thus, we shall adopt the more recent point of view of Nenciu~\cite{N} who used the celebrated Cantero,
Moral and Vel\'azquez matrices (CMV matrices in short) to obtain a~Lax pair representation for the AL
hierarchy in the special defocusing case, that is when $y_n=\overline{x_n}$.
We can now describe the content of our paper.

To deal with the general AL hierarchy, in Section~\ref{Section2}, we f\/irst develop a~slight generalization of
the CMV matrices as introduced in~\cite{CMV}.
The generalized CMV matrices are \emph{pentadiagonal} (semi-inf\/inite) matrices $A_1$, $A_2$ which will
represent multiplication by $z$ in bases of \emph{bi-orthogonal Laurent polynomials}\footnote{The
paper~\cite{CMV} considers the case of a~sesquilinear \emph{hermitian} quasi-def\/inite form on
$\mathbb{C}[z,z^{-1}]$ satisfying the Toeplitz condition, dealing thus with orthogonal instead of
bi-orthogonal Laurent polynomials.}, which will be denoted by $f(z)=(f_n(z))_{n\geq 0}$ and
$g(z)=(g_n(z))_{n\geq 0}$, satisfying $\mathcal{L}[f_m,g_n]=\delta_{m,n}h_n$ and the f\/ive-term recurrence
relations
\begin{gather}
\label{Laurent bi-ortho}
zf(z)=A_1f(z),
\qquad
zg(z)=A_2g(z).
\end{gather}
In these bases, we shall have that
\begin{gather*}
z^{-1}f(z)=A_1^{*}f(z),
\qquad
z^{-1}g(z)=A_2^{*}g(z),
\end{gather*}
with $A_1^{*}=hA_2^Th^{-1}$, $A_2^{*}=hA_1^Th^{-1}$ and $h$ the diagonal matrix $\diag(h_n)_{n\geq
0}$, so that $A_1^*=A_1^{-1}$ and $A_2^{*}=A_2^{-1}$.
Putting $z_n=1-x_n y_n$, with $x_n$ and $y_n$ def\/ined as in~\eqref{Szegovariables} (note that
$x_0=y_0=1$), the matrix $A_1$ reads
\begin{gather*}
A_1=
\begin{pmatrix}
-x_1y_0&y_0&0
\\
-x_2z_1&-x_2y_1&-x_3&1
\\
z_1z_2&y_1z_2&-x_3y_2&y_2&0&&&O
\\
&0&-x_4z_3&-x_4y_3&-x_5&1
\\
&&z_3z_4&y_3z_4&-x_5y_4&y_4&0
\\
&&&0&*&*&*&1
\\
&O&&&*&*&*&*&0
\\
&&&&&\ddots&\ddots&\ddots&\ddots&\ddots
\end{pmatrix},
\end{gather*}
and $A_2$ is obtained from $A_1$ by exchanging the roles of the variables $x_n$ and $y_n$.
This will be proven at the end of Section~\ref{Section2}.
To make contact with the work of Nenciu~\cite{N} as well as with the authoritative treatises on OPUC by
Simon~\cite{S1,S2}, it suf\/f\/ices to specialize to the case $x_{n+1}=-\overline{\alpha_n}$,
$y_{n+1}=-\alpha_{n}$, $n\geq 0$, where $\alpha_n$ are the so-called Verblunsky coef\/f\/icients, remembering
that $x_0=y_0=1$.\footnote{With these notations, the transpose $\mathcal{C}^T$
of the CMV matrix in~\cite{N,S1,S2} is given by $\mathcal{C}^T=(\sqrt{h})^{-1}A_1\sqrt{h}$, with
$\sqrt{h}=\diag(\sqrt{h_n})_{n\geq 0}$ and $h_{n+1}/h_{n}$ as in~\eqref{condition h}.}
We notice that Gesztesy, Holden, Michor and Teschl~\cite{GHMT} have obtained a~Lax pair representation for
the doubly inf\/inite AL hierarchy, involving a~matrix similar to $A_1$ above (up to some conjugation).
According to them, the proof is based on ``fairly tedious computations''.
Our approach via bi-orthogonal Laurent polynomials and the ``dressing method'' explained below, is more
conceptual.

In Section~\ref{Section3}, we put this theory to use to obtain Lax pair representations both for the AL hierarchy
and its Virasoro algebra of master symmetries.
Our approach is based on a~Favard like theorem which states that there is a~one-to-one correspondence
between pairs of CMV matrices $(A_1,A_2)$, with entries built in terms of $x_n$ and $y_n$ satisfying
$x_0=y_0=1$ and $x_ny_n\neq 1$, $n\geq 1$, and quasi-def\/inite Toeplitz bi-moment functionals def\/ined up
to a~multiplicative nonzero constant.
This theorem can be proven as a~generalization to bi-orthogonal Laurent polynomials of a~similar result
in~\cite{CBGV}, for orthogonal Laurent polynomials on the unit circle.
For a~complete and independent proof, see~\cite{V}.
Thus to def\/ine the AL hierarchy vector f\/ields $T_{k}$, $k\in\mathbb{Z}$, it is enough to def\/ine them on
the bi-moments
\begin{gather}
\label{AL moments}
T_k(\mu_j)\equiv\frac{\partial\mu_j}{\partial t_k}=\mu_{j+k},
\qquad
T_{-k}(\mu_j)\equiv\frac{\partial\mu_j}{\partial s_k}=\mu_{j-k},
\qquad
\forall\, k\geq1,
\end{gather}
which, in the example of the bi-moment functional~\eqref{bimoment integral}, corresponds to deform the
weight~$w(z)$ as follows
\begin{gather}
\label{deformed weight}
w(z;t,s)=w(z)\exp\Bigg\{\sum_{j=1}^\infty\big(t_jz^j+s_jz^{-j}\big)\Bigg\}.
\end{gather}
Obviously $[T_k,T_l]=0$, $\forall\, k,l\in\mathbb{Z}$, if we def\/ine $T_0\mu_j=\mu_j$.
Then, all the objects introduced above become time dependent.
In particular $x_n(t,s)$ and $y_n(t,s)$ depend on~$t$,~$s$.
The Lax pair for the AL hierarchy is then obtained in Theorem~\ref{2-Toda} by ``dressing up'' the moment
equations~\eqref{AL moments} written in matrix form (see~\eqref{2-Toda evolution bi-moment matrix}).

Following an idea introduced by Haine and Semengue~\cite{HS} in the context of the semi-inf\/inite Toda
lattice, we def\/ine the following vector f\/ields on the bi-moments
\begin{gather}
\label{master symmetries}
V_k(\mu_j)=(j+k)\mu_{j+k},
\qquad
\forall\, k\in\mathbb{Z}.
\end{gather}
These vector f\/ields trivially satisfy the commutation relations
\begin{gather}
[V_k,V_l]=(l-k)V_{k+l},\label{commutation master sym}
\\
[V_k,T_l]=lT_{k+l},
\qquad
\forall\, k,l\in\mathbb{Z},\label{commutation master sym AL}
\end{gather}
from which it follows that
\begin{gather}
\label{commutation twice}
[[V_k,T_l],T_l]=l[T_{k+l},T_l]=0,
\qquad
\forall\, k,l\in\mathbb{Z}.
\end{gather}
Equations~\eqref{commutation master sym},~\eqref{commutation master sym AL} and~\eqref{commutation twice}
mean that the vector f\/ields $V_k$, $k\in\mathbb{Z}$, form a~centerless Virasoro algebra of master
symmetries, in the sense of Fuchssteiner~\cite{F}, for the AL hierarchy.
We remind the reader that master symmetries are generators for time dependent symmetries of the hierarchy
which are f\/irst degree polynomials in the time variables, that is
\begin{gather*}
X_{k,l}=V_{k}+t[V_k,T_l],
\qquad
k\in\mathbb{Z},
\end{gather*}
are time dependent symmetries of the vector f\/ield $T_l$ (run with time $t$) as one immediately checks that
\begin{gather*}
\frac{\partial X_{k,l}}{\partial t}+[T_l,X_{k,l}]=[V_k,T_l]+[T_l,V_k+t[V_k,T_l]]=0,
\end{gather*}
from the commutation relations~\eqref{commutation twice}.
Writing~\eqref{master symmetries} in matrix form (see~\eqref{master symmetries moment matrix}) and
``dressing up'' these equations, leads then in Theorem~\ref{Lax pair master symmetries} to the Lax pair
representation of the master symmetries on the CMV matrices $(A_1,A_2)$, which was our f\/irst goal and is
a~new result.

In Section~\ref{Section4}, we shall reach our second goal by translating the action of the master symmetries on
the tau-functions of the AL hierarchy.
One can show (see~\cite{AVM2, KMZ1, KMZ2}) that the general solution of the AL hierarchy can be expressed
in terms of the Toeplitz determinants
\begin{gather}
\label{Toeptau}
\tau_n(t,s)=\det \big(\mu_{k-l}(t,s)\big)_{0\leq k,l\leq n-1},
\end{gather}
as follows
\begin{gather*}
x_n(t,s)=\frac{S_n(-\tilde{\partial_t})\tau_n(t,s)}{\tau_n(t,s)},
\qquad
y_n(t,s)=\frac{S_n(-\tilde{\partial_s})\tau_n(t,s)}{\tau_n(t,s)}.
\end{gather*}
In this formula $S_n(t), t=(t_1,t_2,t_3,\ldots)$, are the so-called elementary Schur polynomials def\/ined
by the generating function
\begin{gather}
\label{Schur generating function}
\exp\Bigg(\sum_{k=1}^\infty t_k z^k\Bigg)=\sum_{n\in\mathbb{Z}}S_n(t_1,t_2,\ldots)z^n,
\end{gather}
and $S_n(-\tilde{\partial_t})=S_n\big({-}\frac{\partial}{\partial t_1}, -\frac{1}{2}\frac{\partial}{\partial
t_2}, -\frac{1}{3}\frac{\partial}{\partial t_3},\ldots\big)$, and similarly for $S_n(-\tilde{\partial_s})$.
The functions $\tau_n(t,s)$ are the tau-functions of the semi-inf\/inite AL hierarchy.
In the example of the bi-moment functional~\eqref{bimoment integral}, a~standard computation establishes
that
\begin{gather}
\label{Weyl form tau}
\tau_n(t,s)=\frac{1}{n!}\int_{(S^1)^n}\vert\Delta_n(z)\vert^2\prod_{k=1}^nw(z_k;t,s)\frac{\text{d}z_k}
{2\pi i z_k},
\end{gather}
with $w(z;t,s)$ the deformed weight introduced in~\eqref{deformed weight}, and $\Delta_n(z)$ the
Vandermonde determinant~\eqref{Vandermonde}.
Such integrals appear in combinatorics as well as in random matrix theory, see~\cite{AVM2, AVM3, AD, FW, R,
TW} and the references therein.
The special case $\tau_n(t,s;\eta,\theta)$~\eqref{tauint} considered at the beginning of this Introduction
corresponds to $w(z)=\chi_{]\eta,\theta[^c}(z)$, the characteristic function of the complement of an arc of
circle $]\eta,\theta[=\{z\in S^1: \eta<\text{arg}\, z<\theta\}$.

By a~simple computation, which will be recalled in Section~\ref{Section4}, one obtains that the
tau-functions~\eqref{Toeptau} admit the expansion
\begin{gather}
\label{tau Plucker}
\tau_n(t,s)=\sum_{\substack{0\leq i_0<\dots<i_{n-1}\\0\leq j_0<\dots<j_{n-1}}}
p_{\substack{i_0,\dots,i_{n-1}\\j_0,\dots,j_{n-1}}}
S_{i_{n-1}-(n-1),\dots,i_0}(t)S_{j_{n-1}-(n-1),\dots,j_0}(s),
\end{gather}
where
\begin{gather}
\label{Plucker}
p_{\substack{i_0,\dots,i_{n-1}\\j_0,\dots,j_{n-1}}}
=\det\big(\mu_{i_k-j_l}(0,0)\big)_{0\leq k,l\leq n-1},
\end{gather}
are the so-called Pl\"ucker coordinates, and $S_{i_1,\dots,i_k}(t)$ denote the Schur polynomials
\begin{gather}
\label{Schur polynomials}
S_{i_1,\dots,i_k}(t)=\det\big(S_{i_r+s-r}(t)\big)_{1\leq r,s\leq k}.
\end{gather}
In Theorem~\ref{Virasoro constraints theorem}, we will show that the induced action of the master
symmetries~\eqref{master symmetries} on the Pl\"ucker coordinates of the tau-function $\tau_n(t,s)$
translates into the centerless Virasoro algebra of partial dif\/ferential operators $L_k^{(n)}$,
$k\in\mathbb{Z}$, in the $(t,s)$ variables, that was introduced at the beginning of the Introduction,
a~result we announced without proof in~\cite{HV}.

For the convenience of the reader, we summarize below our main results, which will be established
respectively in Section~\ref{Section3} and Section~\ref{Section4} of the paper.
\begin{Theorem}
\label{full theorem}
The centerless Virasoro algebra $\{V_k, k\in\mathbb{Z}\}$, of master symmetries of the Ablowitz--Ladik
hierarchy which are defined on the bi-moments by~\eqref{master symmetries}, translates as follows on the
CMV matrices and the tau-functions of the hierarchy:

$1)$ On the CMV matrices $(A_1, A_2)$, the master symmetries admit the Lax pair representation
\begin{gather}
V_k(A_1)=\Big[A_1,\big(D_1A_1^{k+1}\big)_{--}+\big(A_1^{k+1}D_1^*\big)_{--}+k\big(A_1^k\big)_{--}
\Big],
\qquad
\forall\, k\in\mathbb{Z},
\label{Lax pair Vk A1}
\\
V_k(A_2)=\Big[\big(D_2A_2^{1-k}\big)_{--}+\big(A_2^{1-k}D_2^*\big)_{--}-k\big(A_2^{-k}\big)_{--}
,A_2\Big],
\qquad
\forall\, k\in\mathbb{Z},
\label{Lax pair Vk A2}
\end{gather}
where $A_{--}$ denotes the strictly lower triangular part of $A$, and $D_1$ and $(D_1^*)^T$ $($respectively
$D_2$ and $(D_2^*)^T)$ represent the operator of derivation $\text{\rm d}/\text{\rm d}z$ in the bases $(f_n(z))_{n\geq
0}$ and $\big(h_n^{-1}g_n(z^{-1})\big)_{n\geq 0}$ $($respectively $(g_n(z))_{n\geq 0}$ and
$\big(h_n^{-1}f_n(z^{-1})\big)_{n\geq 0})$, with $f_n(z)$, $g_n(z)$ the bi-orthogonal Laurent polynomials
satisfying~\eqref{Laurent bi-ortho} and $\mathcal{L}[f_m,g_n]=h_n\delta_{m,n}$.

$2)$ On the tau-functions $\tau_n(t,s)$, the master symmetries are given by a~centerless Virasoro algebra of partial
differential operators in the $(t,s)$ variables
\begin{gather*}
V_k\tau_n(t,s)=L_k^{(n)}\tau_n(t,s),
\qquad
\forall\, k\in\mathbb{Z},
\end{gather*}
with $L_k^{(n)}$ defined as in~\eqref{virplus},~\eqref{virzero} and~\eqref{virminus}.
\end{Theorem}

\section{Bi-orthogonal Laurent polynomials and CMV matrices}\label{Section2}

In this section, given $\mathcal{L}:\mathbb{C}\big[z,z^{-1}\big]\times \mathbb{C}\big[z,z^{-1}\big]\to \mathbb{C}$,
a~bi-moment functional as in~\eqref{bi-functional} which satisf\/ies the Toeplitz condition~\eqref{Toeplitz
condition} and is quasi-def\/inite~\eqref{quasi-definite}, we construct two sequences of bi-orthogonal
Laurent polynomials (in short L-polynomials), which can be thought of as a~Gram--Schmidt
bi-orthogonalization process applied to the ordered bases $\big\{1,z,z^{-1},z^2,z^{-2}, \ldots\big\}$ and
$\big\{1,z^{-1},z,z^{-2},z^2,\ldots\big\}$ of $\mathbb{C}\big[z,z^{-1}\big]$.
They will be called right and left bi-orthogonal L-po\-ly\-no\-mials respectively.
This is a~slight generalization of the Cantero, Moral and Vel\'azquez~\cite{CMV} construction\footnote{The
paper~\cite{CMV} deals with the case of a~sesquilinear quasi-def\/inite hermitian form on
$\mathbb{C}[z,z^{-1}]$, satisfying the Toeplitz condition.
Dropping the condition ``hermitian'' leads to bi-orthogonal L-polynomials, instead of orthogonal
L-polynomials.
For the applications we have in mind, see~\eqref{bimoment integral}, it is better to assume $\mathcal{L}$
bilinear rather than sesquilinear.}.

The two sequences of monic right and left bi-orthogonal L-polynomials we shall construct will be expressed
in terms of the sequence of monic bi-orthogonal polynomials $\big\{p_n^{(1)}(z),p_n^{(2)}(z)\big\}_{n\geq 0}$,
given by the well known formulae
\begin{gather*}
p_n^{(1)}(z)=\frac{1}{\tau_n}\det\left(
\begin{matrix}
\mu_{0,0}&\dots&\mu_{0,n-1}&1
\\
\mu_{1,0}&\dots&\mu_{1,n-1}&z
\\
\vdots&&\vdots&\vdots
\\
\mu_{n,0}&\dots&\mu_{n,n-1}&z^n
\end{matrix}
\right),
\\
p_n^{(2)}(z)=\frac{1}{\tau_n}\det\left(
\begin{matrix}
\mu_{0,0}&\mu_{0,1}&\dots&\mu_{0,n}
\\
\vdots&\vdots&&\vdots
\\
\mu_{n-1,0}&\mu_{n-1,1}&\dots&\mu_{n-1,n}
\\
1&z&\dots&z^n
\end{matrix}
\right),
\end{gather*}
with $\tau_n=\det\big(\mu_{kl}\big)_{0\leq k,l\leq n-1}$.
Denoting by $\{f_n,g_n\}_{n\geq 0}$ the sequence of monic right bi-orthogonal L-polynomials, multiplication
by $z$ in the bases $(f_n)_{n\geq 0}$ and $(g_n)_{n\geq 0}$ of $\mathbb{C}\big[z,z^{-1}\big]$ will be represented
by two pentadiagonal matrices $A_1$ and $A_2$, which we call the generalized CMV matrices (and similarly of
course for the sequence of left bi-orthogonal L-polynomials).
Moreover, the entries of $A_1$ and $A_2$ will have simple expressions in terms of the variables $x_n$ and
$y_n$ entering the Szeg\"o type recurrence relations~\eqref{Szegorecurrence}.

\subsection{Bi-orthogonal Laurent polynomials}

The following def\/inition is natural from our previous discussion.
We def\/ine the vector subspaces
\begin{gather*}
\mathbb{L}_{m,n}:=\left\langle z^m,z^{m+1},\dots,z^{n-1},z^n\right\rangle,
\qquad
\forall\, m,n\in\mathbb{Z},
\qquad
m\leq n,
\end{gather*}
and for $n\geq 0$
\begin{gather*}
\mathbb{L}_{2n}^+:=\mathbb{L}_{-n,n},
\qquad
\mathbb{L}_{2n+1}^+:=\mathbb{L}_{-n,n+1},
\qquad
\mathbb{L}_{2n}^-:=\mathbb{L}_{-n,n},
\qquad
\mathbb{L}_{2n+1}^-:=\mathbb{L}_{-n-1,n},
\end{gather*}
with the convention $\mathbb{L}_{-1}^+=\mathbb{L}_{-1}^-=\{0\}$.
\begin{Definition}
\label{Gram-Schmidt}
A sequence $\{f_n,g_n\}_{n\geq 0}$ in $\mathbb{C}[z,z^{-1}]$ is a~sequence of right (left) bi-orthogonal
L-polynomials with respect to $\mathcal{L}$ if
\begin{enumerate}\itemsep=0pt
\item[1)] $f_n, g_n\in\mathbb{L}_n^{+(-)}\setminus\mathbb{L}_{n-1}^{+(-)}$;
\item[2)] $\mathcal{L}[f_n,g_m]=h_n\delta_{n,m}$, with $h_n\neq 0$.
\end{enumerate}
\end{Definition}

\begin{Remark}
\label{equivcond-rl}
Similarly to orthogonal polynomials, condition $(2)$ in Def\/inition~\ref{Gram-Schmidt} can be replaced
equivalently by
\begin{gather*}
(3r) \ \left\{
\begin{array}{@{}lll}\mathcal{L}\big[f_{2n},z^k\big]=0,\qquad & \mathcal{L}\big[z^k,g_{2n}\big]=0
\qquad &
\text{if $-n+1\leq k\leq n$},
\vspace{0.5mm}\\
\mathcal{L}\big[f_{2n},z^{-n}\big]\neq0,\qquad & \mathcal{L}\big[z^{-n},g_{2n}\big]\neq0, &
\vspace{0.5mm}\\
\mathcal{L}\big[f_{2n+1},z^k\big]=0,\qquad & \mathcal{L}\big[z^k,g_{2n+1}\big]=0
\qquad &
\text{if $-n\leq k\leq n$},
\vspace{0.5mm}\\
\mathcal{L}\big[f_{2n+1},z^{n+1}\big]\neq0,\qquad & \mathcal{L}\big[z^{n+1},g_{2n+1}\big]\neq0, &
\end{array}
\right.
\end{gather*}
in the case of right bi-orthogonal L-polynomials.
For left bi-orthogonal L-polynomials the equiva\-lent condition is
\begin{gather*}
(3l) \ \left\{
\begin{array}{@{}llll}\mathcal{L}\big[f_{2n},z^k\big]=0,\qquad & \mathcal{L}\big[z^k,g_{2n}\big]=0
\qquad &
\text{if $-n\leq k\leq n-1$},
\vspace{0.5mm}\\
\mathcal{L}\big[f_{2n},z^{n}\big]\neq0,\qquad & \mathcal{L}\big[z^{n},g_{2n}\big]\neq0, &
\vspace{0.5mm}\\
\mathcal{L}\big[f_{2n+1},z^k\big]=0,\qquad & \mathcal{L}\big[z^k,g_{2n+1}\big]=0
\qquad &
\text{if $-n\leq k\leq n$},
\vspace{0.5mm}\\
\mathcal{L}\big[f_{2n+1},z^{-n-1}\big]\neq0,\qquad & \mathcal{L}\big[z^{-n-1},g_{2n+1}\big]\neq0. &
\end{array}
\right.
\end{gather*}
\end{Remark}
We start by proving that sequences of right and left bi-orthogonal L-polynomials for a~given Toeplitz
bi-moment functional $\mathcal{L}$ are closely related to each other.
\begin{Proposition}
\label{link left and right L-polynomials}
Let $f_{n}^*(z)=f_n\big(z^{-1}\big)$ and $g_{n}^*(z)=g_n\big(z^{-1}\big)$.
Then $\{f_n,g_n\}_{n\geq 0}$ is a~sequence of right bi-orthogonal L-polynomials with respect to
$\mathcal{L}$ if and only if $\{g_{n}^*,f_{n}^*\}_{n\geq 0}$ is a~sequence of left bi-orthogonal
L-polynomials with respect to $\mathcal{L}$.
\end{Proposition}
\begin{proof}
We have $f_{n}^*,g_{n}^*\in\mathbb{L}_n^-\setminus\mathbb{L}_{n-1}^-$ if and only if
$f_n, g_n\in\mathbb{L}_n^+\setminus\mathbb{L}_{n-1}^+$.
Using the Toeplitz condition~\eqref{Toeplitz condition}, the result then follows from
\begin{gather*}
\mathcal{L}\big[g_{m}^*(z),f_{n}^*(z)\big]=\mathcal{L}\big[g_{m}\big(z^{-1}\big),f_{n}\big(z^{-1}\big)\big]=\mathcal{L}
\big[f_{n}(z),g_{m}(z)\big].\tag*{\qed}
\end{gather*}
\renewcommand{\qed}{}
\end{proof}
Sequences of right or left bi-orthogonal L-polynomials with respect to $\mathcal{L}$ are also very closely
related to sequences of bi-orthogonal polynomials for $\mathcal{L}$.
This is proven in the next theorem.
\begin{Theorem}
\label{link right bi-orthogonal L-polynomials ans bi-orthogonal polynomials}
Let $\mathcal{L}$ be a~Toeplitz bi-moment functional and let $\{f_n,g_n\}_{n\geq 0}$ be a~sequence in
$\mathbb{C}\big[z,z^{-1}\big]$.
Let us define
\begin{gather}
p_{2n}^{(1)}(z)=z^ng_{2n}\big(z^{-1}\big),
\qquad
p_{2n+1}^{(1)}(z)=z^nf_{2n+1}(z),
\nonumber
\\
p_{2n}^{(2)}(z)=z^nf_{2n}\big(z^{-1}\big),
\qquad
p_{2n+1}^{(2)}(z)=z^ng_{2n+1}(z).
\label{link polynomials - laurent polynomials}
\end{gather}
The sequence $\{f_n,g_n\}_{n\geq 0}$ is a~sequence of right bi-orthogonal L-polynomials with respect to
$\mathcal{L}$ if and only if $\big\{p_n^{(1)},p_n^{(2)}\big\}_{n\geq 0}$ is a~sequence of bi-orthogonal polynomials
with respect to $\mathcal{L}$.
Furthermore we have
\begin{gather}
\label{ortho-pol-Lpol}
\mathcal{L}[f_n,g_n]=\mathcal{L}\big[p_{n}^{(1)},p_n^{(2)}\big].
\end{gather}
An analogous statement holds for sequences $\{f_n,g_n\}_{n\geq 0}$ of left bi-orthogonal L-polynomials, if
we define
\begin{gather}
\tilde{p}_{2n}^{(1)}(z)=z^nf_{2n}(z),
\qquad
\tilde{p}_{2n+1}^{(1)}(z)=z^ng_{2n+1}\big(z^{-1}\big),
\nonumber
\\
\tilde{p}_{2n}^{(2)}(z)=z^ng_{2n}(z),
\qquad
\tilde{p}_{2n+1}^{(2)}(z)=z^nf_{2n+1}\big(z^{-1}\big).
\label{link polynomials - llaurent polynomials}
\end{gather}
\end{Theorem}

\begin{proof}
For $n\geq 0$, we def\/ine $\mathbb{P}_n=\left\langle 1,z,\dots,z^n\right\rangle$ the vector subspace of
polynomials with degree less than or equal to $n$, and $\mathbb{P}_{-1}:=\{0\}$.
For $\big\{p_n^{(1)},p_n^{(2)}\big\}_{n\geq 0}$ def\/ined as in~\eqref{link polynomials - laurent polynomials} it
is trivial that
\begin{alignat*}{4}
& p_{2n}^{(1)},p_{2n}^{(2)}\in\mathbb{P}_{2n}\setminus\mathbb{P}_{2n-1}
\quad &&
\Leftrightarrow
\quad &&
g_{2n},f_{2n}\in\mathbb{L}_{2n}^+\setminus\mathbb{L}_{2n-1}^+, &
\\
& p_{2n+1}^{(1)},p_{2n+1}^{(2)}\in\mathbb{P}_{2n+1}\setminus\mathbb{P}_{2n}
\quad &&
\Leftrightarrow
\quad &&
f_{2n+1},g_{2n+1}\in\mathbb{L}_{2n+1}^+\setminus\mathbb{L}_{2n}^+.&
\end{alignat*}
\pagebreak

\noindent
Furthermore we have using the Toeplitz condition~\eqref{Toeplitz condition}
\begin{gather*}
\mathcal{L}\big[p_{2n+1}^{(1)}(z),z^{k}\big]=\mathcal{L}\big[z^nf_{2n+1}(z),z^{k}\big]=\mathcal{L}
\big[f_{2n+1}(z),z^{k-n}\big],
\end{gather*}
and similarly
\begin{gather*}
\mathcal{L}\big[p_{2n}^{(1)}(z),z^k\big]=\mathcal{L}\big[z^{n-k},g_{2n}(z)\big],
\qquad
\mathcal{L}\big[z^k,p_{2n+1}^{(2)}(z)\big]=\mathcal{L}\big[z^{k-n},g_{2n+1}(z)\big],
\\
\mathcal{L}\big[z^k,p_{2n}^{(2)}(z)\big]=\mathcal{L}\big[f_{2n}(z),z^{n-k}\big].
\end{gather*}
Consequently we have
\begin{alignat*}{6}
& \mathcal{L}\big[p_{2n+1}^{(1)}(z),z^k\big]=0,
\quad &&
0\leq k\leq2n
\quad &&
\Leftrightarrow
\quad &&
\mathcal{L}\big[f_{2n+1}(z),z^{k}\big]=0,
\quad &&
-n\leq k\leq n, &
\\
& \mathcal{L}\big[p_{2n}^{(1)}(z),z^k\big]=0,
\quad &&
0\leq k\leq2n-1
\quad &&
\Leftrightarrow
\quad &&
\mathcal{L}\big[z^k,g_{2n}(z)\big]=0,
\quad &&
-n+1\leq k\leq n, &
\\
& \mathcal{L}\big[z^k,p_{2n+1}^{(2)}(z)\big]=0,
\quad &&
0\leq k\leq2n
\quad &&
\Leftrightarrow
\quad &&
\mathcal{L}\big[z^k,g_{2n+1}(z)\big]=0,
\quad &&
-n\leq k\leq n, &
\\
& \mathcal{L}\big[z^k,p_{2n}^{(2)}(z)\big]=0,
\quad &&
0\leq k\leq2n-1
\quad &&
\Leftrightarrow
\quad &&
\mathcal{L}\big[f_{2n}(z),z^k\big]=0,
\quad &&
-n+1\leq k\leq n, &
\end{alignat*}
and
\begin{alignat*}{4}
& \mathcal{L}\big[p_{2n+1}^{(1)}(z),z^{2n+1}\big]\neq0
\quad &&
\Leftrightarrow
\quad &&
\mathcal{L}\big[f_{2n+1}(z),z^{n+1}\big]\neq0, &
\\
& \mathcal{L}\big[p_{2n}^{(1)}(z),z^{2n}\big]\neq0
\quad &&
\Leftrightarrow
\quad &&
\mathcal{L}\big[z^{-n},g_{2n}(z)\big]\neq0,&
\\
& \mathcal{L}\big[z^{2n+1},p_{2n+1}^{(2)}(z)\big]\neq0
\quad &&
\Leftrightarrow
\quad &&
\mathcal{L}\big[z^{n+1},g_{2n+1}(z)\big]\neq0, &
\\
& \mathcal{L}\big[z^{2n},p_{2n}^{(2)}(z)\big]\neq0
\quad &&
\Leftrightarrow
\quad &&
\mathcal{L}\big[f_{2n}(z),z^{-n}\big]\neq0.&
\end{alignat*}
Thus, according to Remark~\ref{equivcond-rl}, $\{f_n,g_n\}_{n\geq 0}$ is a~sequence of right
bi-orthogonal L-polynomials with respect to $\mathcal{L}$ if and only if $\big\{p_n^{(1)},p_n^{(2)}\big\}_{n\geq
0}$ is a~sequence of bi-orthogonal polynomials with respect to $\mathcal{L}$.
Equation~\eqref{ortho-pol-Lpol} follows immediately from the def\/inition~\eqref{link polynomials - laurent
polynomials} and the Toeplitz condition~\eqref{Toeplitz condition}.

The statement~\eqref{link polynomials - llaurent polynomials} for sequences of left bi-orthogonal
L-polynomials is an immediate consequence of the result for sequences of right bi-orthogonal L-polynomials
and Proposition~\ref{link left and right L-polynomials}.
This concludes the proof.
\end{proof}

We are now able to prove the existence and the unicity of bi-orthogonal L-polynomials with respect to
$\mathcal{L}$.
\begin{Corollary}
Consider a~Toeplitz bi-moment functional $\mathcal{L}$.
There exists a~sequence of right bi-orthogonal L-polynomials with respect to $\mathcal{L}$ and a~sequence
of left bi-orthogonal L-polynomials with respect to $\mathcal{L}$ if and only if $\mathcal{L}$ is
quasi-definite as defined in~\eqref{quasi-definite}.
Each L-polynomial in these sequences is uniquely determined up to an arbitrary non-zero factor.
\end{Corollary}
\begin{proof}
By virtue of Theorem~\ref{link right bi-orthogonal L-polynomials ans bi-orthogonal polynomials}, the
existence of a~sequence of right or left bi-orthogonal L-polynomials with respect to $\mathcal{L}$ is
equivalent to the existence of a~sequence of bi-orthogonal polynomials with respect to $\mathcal{L}$, which
are known to exist if and only $\mathcal{L}$ is quasi-def\/inite.
Since bi-orthogonal polynomials are uniquely determined up to an arbitrary non-zero factor, the same holds
for right and left bi-orthogonal L-polynomials.
\end{proof}

From now on we shall assume that $\{f_n,g_n\}_{n\geq 0}$ is a~sequence of \emph{monic} right bi-orthogonal
L-polynomials with respect to $\mathcal{L}$, i.e.\
the coef\/f\/icients of $z^{-n}$ in $f_{2n}$, $g_{2n}$ and $z^{n+1}$ in $f_{2n+1}$, $g_{2n+1}$ are equal to~$1$.
We denote by $\big\{p_n^{(1)},p_n^{(2)}\big\}_{n\geq 0}$ the associated sequence of monic bi-orthogonal polynomials
with respect to $\mathcal{L}$, as def\/ined by~\eqref{link polynomials - laurent polynomials}.

\subsection{Five term recurrence relations}

We now prove that bi-orthogonal L-polynomials with respect to a~quasi-def\/inite Toeplitz bi-moment
functional always satisfy f\/ive term recurrence relations.
This generalizes the result obtained in~\cite{CMV} for orthogonal L-polynomials associated with
a~quasi-def\/inite Toeplitz sesquilinear hermitian form.
The essential ingredient in the proof in~\cite{CMV} is the Toeplitz condition.
Consequently, it can immediately be translated to the case of bi-orthogonal L-polynomials.
\begin{Theorem}
\label{Five term recurrence relations}
Let $\{f_n,g_n\}_{n\geq 0}$ be a~sequence of monic right bi-orthogonal L-polynomials with respect to
$\mathcal{L}$, and $f^*_n(z)=f_n\big(z^{-1}\big)$, $g^*_n(z)=g_n\big(z^{-1}\big)$.
Then for $n\geq 0$ there exist five-term recurrence relations
\begin{gather*}
zf_n(z)=\sum_{i=n-2}^{n+2}\alpha_{n,i}f_i(z),
\qquad
zg_n(z)=\sum_{i=n-2}^{n+2}\beta_{n,i}g_i(z),
\\
zf_{n}^*(z)=\sum_{i=n-2}^{n+2}\alpha_{n,i}^*f_{i}^*(z),
\qquad
zg_{n}^*(z)=\sum_{i=n-2}^{n+2}\beta_{n,i}^*g_{i}^*(z),
\end{gather*}
where we use the convention $f_n(z)=g_n(z)=0$ if $n<0$, and
\begin{gather*}
\alpha_{n,i}^*=\frac{h_n}{h_i}\beta_{i,n},
\qquad
\beta_{n,i}^*=\frac{h_n}{h_i}\alpha_{i,n},
\end{gather*}
with $h_n=\mathcal{L}[f_n,g_n]$.
Moreover, we have for all $n\geq 0$
\begin{gather*}
\alpha_{2n-1,2n-3}=0,
\qquad
\alpha_{2n,2n+2}=0,
\qquad
\beta_{2n-1,2n-3}=0,
\qquad
\beta_{2n,2n+2}=0.
\end{gather*}
\end{Theorem}
\begin{proof}
As $f_n\in\mathbb{L}_{n}^+\setminus\mathbb{L}_{n-1}^+$, we have $zf_n(z)\in\mathbb{L}_{n+2}^+$.
This implies that $zf_n$ admits an expansion in terms of $f_0,\dots,f_{n+2}$
\begin{gather*}
zf_n(z)=\sum_{i=0}^{n+2}\alpha_{n,i}f_i(z),
\end{gather*}
with $\alpha_{n,i}\in\mathbb{C}$, $0\leq i\leq n+2$.
Consequently, by bi-orthogonality of the sequence $\{f_n,g_n\}_{n\geq 0}$ we have
\begin{gather*}
\mathcal{L}[zf_n,g_m]=\sum_{i=0}^{n+2}h_i\alpha_{n,i}\delta_{i,m}.
\end{gather*}
But we also have
\begin{gather*}
\mathcal{L}[zf_n,zg_k]=\mathcal{L}[f_n,g_k]=0,
\qquad
0\leq k\leq n-1,
\end{gather*}
and $\left\langle g_0,\dots,g_{n-3}\right\rangle\subset\left\langle zg_0,\dots,zg_{n-1}\right\rangle$.
It follows that
\begin{gather*}
\mathcal{L}[zf_n,g_k]=0,
\qquad
0\leq k\leq n-3.
\end{gather*}
Consequently we have $\alpha_{n,i}=0$ if $i<n-2$, and thus
\begin{gather*}
zf_n(z)=\sum_{i=n-2}^{n+2}\alpha_{n,i}f_i(z).
\end{gather*}
We prove that $\alpha_{2n,2n+2}=\alpha_{2n-1,2n-3}=0$.
We f\/irst prove that $\alpha_{2n,2n+2}=0$.
Indeed, we have $zf_{2n}(z)\in\left\langle z^{1-n},\dots,z^{1+n}\right\rangle$.
Consequently, using condition~$(3r)$ in Remark~\ref{equivcond-rl}, we have
$\mathcal{L}[zf_{2n},g_{2n+2}]=0$ and thus $\alpha_{2n,2n+2}=0$.
We also have $\alpha_{2n-1,2n-3}=0$.
Indeed, we have $\mathcal{L}[zf_{2n-1},g_{2n-3}]=\mathcal{L}[f_{2n-1},z^{-1}g_{2n-3}]$, and
$z^{-1}g_{2n-3}(z)\in\left\langle z^{1-n},\dots,z^{n-2}\right\rangle$.
From condition~$(3r)$ in Remark~\ref{equivcond-rl}, it follows that $\mathcal{L}[zf_{2n-1},g_{2n-3}]=0$.
A similar argument gives $\beta_{2n,2n+2}=\beta_{2n-1,2n-3}=0$.
The proof of the other recurrence relations is similar.

The coef\/f\/icients in the recurrence relations satisfy
\begin{gather*}
\alpha_{n,i}=\frac{\mathcal{L}[zf_n,g_i]}{\mathcal{L}[f_i,g_i]},
\qquad
\beta_{n,i}=\frac{\mathcal{L}[f_i,zg_n]}{\mathcal{L}[f_i,g_i]},
\\
\alpha_{n,i}^*=\frac{\mathcal{L}[g_{i}^*,zf_{n}^*]}{\mathcal{L}[g_{i}^*,f_{i}^*]},
\qquad
\beta_{n,i}^*=\frac{\mathcal{L}[zg_{n}^*,f_{i}^*]}{\mathcal{L}[g_{i}^*,f_{i}^*]}.
\end{gather*}
It follows from the def\/inition of $\{g_{n}^*,f_{n}^*\}_{n\geq 0}$ that
\begin{gather*}
\alpha_{n,i}^*=\frac{\mathcal{L}[g_{i}^*,zf_{n}^*]}{\mathcal{L}[g_{i}^*,f_{i}^*]}=\frac{\mathcal{L}
[f_n,zg_i]}{\mathcal{L}[f_i,g_i]}=\frac{\mathcal{L}[f_n,zg_i]}{\mathcal{L}[f_n,g_n]}\frac{\mathcal{L}
[f_n,g_n]}{\mathcal{L}[f_i,g_i]}=\beta_{i,n}\frac{h_n}{h_i}.
\end{gather*}
Similarly we have
\begin{gather*}
\beta_{n,i}^*=\frac{\mathcal{L}[zg_{n}^*,f_{i}^*]}{\mathcal{L}[g_{i}^*,f_{i}^*]}=\frac{\mathcal{L}
[zf_i,g_n]}{\mathcal{L}[f_i,g_i]}=\frac{\mathcal{L}[zf_i,g_n]}{\mathcal{L}[f_n,g_n]}\frac{\mathcal{L}
[f_n,g_n]}{\mathcal{L}[f_i,g_i]}=\alpha_{i,n}\frac{h_n}{h_i}.
\end{gather*}
This concludes the proof.
\end{proof}
\begin{Corollary}
\label{Corollary five term recurrence relations}
With the same notations as in Theorem~{\rm \ref{Five term recurrence relations}} we have
\begin{gather*}
z^{-1}f_n(z)=\sum_{i=n-2}^{n+2}\alpha_{n,i}^*f_i(z),
\qquad
z^{-1}g_n(z)=\sum_{i=n-2}^{n+2}\beta_{n,i}^*g_i(z),
\\
z^{-1}f_{n}^*(z)=\sum_{i=n-2}^{n+2}\alpha_{n,i}f_{i}^*(z),
\qquad
z^{-1}g_{n}^*(z)=\sum_{i=n-2}^{n+2}\beta_{n,i}g_{i}^*(z).
\end{gather*}
\end{Corollary}

Def\/ining the vectors
\begin{gather}
f(z)=\big(f_n(z)\big)_{n\geq0},
\qquad
g(z)=\big(g_n(z)\big)_{n\geq0},
\label{f,g}
\\
f^*(z)=f\big(z^{-1}\big)=\big(f_n^*(z)\big)_{n\geq0},
\qquad
g^*(z)=g\big(z^{-1}\big)=\big(g_n^*(z)\big)_{n\geq0},
\label{f^*,g^*}
\end{gather}
the f\/ive term recurrence relations obtained in Theorem~\ref{Five term recurrence relations} and
Corollary~\ref{Corollary five term recurrence relations} can be written in vector form
\begin{gather}
\label{vector form Five term recurrence relations}
\begin{cases}
zf(z)=A_1f(z),
\\
zg(z)=A_2g(z),
\\
z^{-1}f(z)=A_1^*f(z),
\\
z^{-1}g(z)=A_2^*g(z),
\end{cases}
\qquad
\begin{cases}
zf^*(z)=A_1^*f^*(z),
\\
zg^*(z)=A_2^*g^*(z),
\\
z^{-1}f^*(z)=A_1f^*(z),
\\
z^{-1}g^*(z)=A_2g^*(z),
\end{cases}
\end{gather}
with
\begin{gather*}
A_1=\big(\alpha_{i,j}\big)_{i,j\geq0},
\qquad
A_2=\big(\beta_{i,j}\big)_{i,j\geq0},
\end{gather*}
where $\alpha_{i,j}=\beta_{i,j}=0$ if $|i-j|>2$, and
\begin{gather}
\label{A star}
A_1^*=hA_2^Th^{-1},
\qquad
A_2^*=hA_1^Th^{-1},
\end{gather}
where $h=\diag(h_n)_{n\geq 0}$.
We call the matrices $A_1$, $A_2$ the (generalized) CMV matrices.
Clearly, from~\eqref{vector form Five term recurrence relations}, we have
\begin{gather}
\label{inverse A}
A^*_1=A_1^{-1},
\qquad
A_2^*=A_2^{-1}.
\end{gather}

\subsection{Explicit expression for the entries of the CMV matrices}

Explicit expressions for the entries of the CMV matrices can be found in terms of the variab\-les~$x_n$,~$y_n$
introduced in~\eqref{Szegovariables} entering the Szeg\"o type recurrence relations~\eqref{Szegorecurrence}.
\begin{Theorem}
\label{entries CMV matrices}
The non-zero entries of the CMV matrices $A_1$ and $A_2$ are
\begin{alignat*}{3}
& (A_1)_{2n-1,2n+1}=1, \qquad && (A_1)_{2n-1,2n-1}=-x_{2n}y_{2n-1}, &
\\
& (A_1)_{2n-1,2n}=-x_{2n+1}, \qquad && (A_1)_{2n-1,2n-2}=-x_{2n}(1-x_{2n-1}y_{2n-1}), &
\\
& (A_1)_{2n,2n+1}=y_{2n}, \qquad && (A_1)_{2n,2n-1}=y_{2n-1}(1-x_{2n}y_{2n}),&
\\
& (A_1)_{2n,2n}=-x_{2n+1}y_{2n}, \qquad &&
(A_1)_{2n,2n-2}=(1-x_{2n-1}y_{2n-1})(1-x_{2n}y_{2n}),
\end{alignat*}
and
\begin{alignat*}{3}
& (A_2)_{2n-1,2n+1}=1, \qquad && (A_2)_{2n-1,2n-1}=-x_{2n-1}y_{2n},&
\\
& (A_2)_{2n-1,2n}=-y_{2n+1}, \qquad && (A_2)_{2n-1,2n-2}=-y_{2n}(1-x_{2n-1}y_{2n-1}), &
\\
& (A_2)_{2n,2n+1}=x_{2n}, \qquad && (A_2)_{2n,2n-1}=x_{2n-1}(1-x_{2n}y_{2n}),&
\\
& (A_2)_{2n,2n}=-x_{2n}y_{2n+1}, \qquad && (A_2)_{2n,2n-2}=(1-x_{2n-1}y_{2n-1})(1-x_{2n}y_{2n}). &
\end{alignat*}
\end{Theorem}

\begin{proof}
(1) We have
\begin{gather*}
(A_1)_{2n-1,2n+1}=\frac{1}{h_{2n+1}}\mathcal{L}\big[zf_{2n-1}(z),g_{2n+1}(z)\big].
\end{gather*}
By virtue of Theorem~\ref{link right bi-orthogonal L-polynomials ans bi-orthogonal polynomials} we obtain
\begin{gather*}
(A_1)_{2n-1,2n+1}=\frac{1}{h_{2n+1}}\mathcal{L}\big[z^{2-n}p_{2n-1}^{(1)}(z),z^{-n}p_{2n+1}^{(2)}(z)\big]
=\frac{1}{h_{2n+1}}\mathcal{L}\big[z^{2}p_{2n-1}^{(1)}(z),p_{2n+1}^{(2)}(z)\big].
\end{gather*}
As $z^{2}p_{2n-1}^{(1)}(z)$ is a~monic polynomial of degree $2n+1$, using the bi-orthogonality of the
polynomials, we have
\begin{gather*}
(A_1)_{2n-1,2n+1}=\frac{1}{h_{2n+1}}\mathcal{L}\big[z^{2n+1},p_{2n+1}^{(2)}(z)\big]=1.
\end{gather*}

(2) We have
\begin{gather*}
(A_1)_{2n-1,2n}=\frac{1}{h_{2n}}\mathcal{L} [zf_{2n-1}(z),g_{2n}(z) ].
\end{gather*}
By virtue of Theorem~\ref{link right bi-orthogonal L-polynomials ans bi-orthogonal polynomials} we obtain
\begin{gather*}
(A_1)_{2n-1,2n}=\frac{1}{h_{2n}}\mathcal{L}\big[z^{2-n}p_{2n-1}^{(1)}(z),z^np_{2n}^{(1)}\big(z^{-1}\big)\big]
=\frac{1}{h_{2n}}\mathcal{L}\big[z^{2}p_{2n-1}^{(1)}(z),z^{2n}p_{2n}^{(1)}\big(z^{-1}\big)\big].
\end{gather*}
By using twice~\eqref{Szegorecurrence} we have
\begin{gather*}
z^2p_{2n-1}^{(1)}(z)=p_{2n+1}^{(1)}(z)-x_{2n+1}z^{2n}p_{2n}^{(2)}\big(z^{-1}\big)-x_{2n}z^{2n}p_{2n-1}^{(2)}\big(z^{-1}
\big),
\end{gather*}
and thus
\begin{gather*}
(A_1)_{2n-1,2n}=\frac{1}{h_{2n}}\mathcal{L}\big[p_{2n+1}^{(1)}(z),z^{2n}p_{2n}^{(1)}\big(z^{-1}
\big)\big]-\frac{x_{2n+1}}{h_{2n}}\mathcal{L}\big[p_{2n}^{(2)}(z^{-1}),p_{2n}^{(1)}\big(z^{-1}\big)\big]
\\
\phantom{(A_1)_{2n-1,2n}=}
{} -\frac{x_{2n}}{h_{2n}}\mathcal{L}\big[p_{2n-1}^{(2)}\big(z^{-1}\big),p_{2n}^{(1)}\big(z^{-1}\big)\big].
\end{gather*}
As $z^{2n}p_{2n}^{(1)}(z^{-1})$ is a~polynomial of degree $2n$, the f\/irst term is equal to $0$ by
bi-orthogonality.
The remaining terms give
\begin{gather*}
(A_1)_{2n-1,2n}=-\frac{x_{2n+1}}{h_{2n}}\mathcal{L}\big[p_{2n}^{(1)}(z),p_{2n}^{(2)}(z)\big]-\frac{x_{2n}}
{h_{2n}}\mathcal{L}\big[p_{2n}^{(1)}(z),p_{2n-1}^{(2)}(z)\big]
=-x_{2n+1}.
\end{gather*}

(3) We have
\begin{gather*}
(A_1)_{2n-1,2n-1}=\frac{1}{h_{2n-1}}\mathcal{L} [zf_{2n-1}(z),g_{2n-1}(z)].
\end{gather*}
By virtue of Theorem~\ref{link right bi-orthogonal L-polynomials ans bi-orthogonal polynomials} we obtain
\begin{gather*}
(A_1)_{2n-1,2n-1}=\frac{1}{h_{2n-1}}\mathcal{L}\big[z^{2-n}p_{2n-1}^{(1)}(z),z^{1-n}p_{2n-1}^{(2)}(z)\big]
=\frac{1}{h_{2n-1}}\mathcal{L}\big[zp_{2n-1}^{(1)}(z),p_{2n-1}^{(2)}(z)\big].
\end{gather*}
By using~\eqref{Szegorecurrence} and then~\eqref{Szegovariables} we have
\begin{gather*}
(A_1)_{2n-1,2n-1}=\frac{1}{h_{2n-1}}\mathcal{L}\big[p_{2n}^{(1)}(z)-x_{2n}z^{2n-1}p_{2n-1}^{(2)}\big(z^{-1}
\big),p_{2n-1}^{(2)}(z)\big]
\\
\phantom{(A_1)_{2n-1,2n-1}}{}
=-\frac{x_{2n}}{h_{2n-1}}\mathcal{L}\big[z^{2n-1}p_{2n-1}^{(2)}\big(z^{-1}\big),p_{2n-1}^{(2)}(z)\big]
\\
\phantom{(A_1)_{2n-1,2n-1}}{}
=-\frac{x_{2n}}{h_{2n-1}}\mathcal{L}\big[y_{2n-1}z^{2n-1},p_{2n-1}^{(2)}(z)\big]
=-x_{2n}y_{2n-1}.
\end{gather*}

(4) We have
\begin{gather*}
(A_1)_{2n-1,2n-2}=\frac{1}{h_{2n-2}}\mathcal{L} [zf_{2n-1}(z),g_{2n-2}(z) ].
\end{gather*}
By virtue of Theorem~\ref{link right bi-orthogonal L-polynomials ans bi-orthogonal polynomials} we obtain
\begin{gather*}
(A_1)_{2n-1,2n-2}=\frac{1}{h_{2n-2}}\mathcal{L}\big[z^{2-n}p_{2n-1}^{(1)}(z),z^{n-1}p_{2n-2}^{(1)}\big(z^{-1}\big)\big]
\\
\phantom{(A_1)_{2n-1,2n-2}}
{}=\frac{1}{h_{2n-2}}\mathcal{L}\big[zp_{2n-1}^{(1)}(z),z^{2n-2}p_{2n-2}^{(1)}\big(z^{-1}\big)\big].
\end{gather*}
Using~\eqref{Szegorecurrence} we obtain
\begin{gather*}
(A_1)_{2n-1,2n-2}
=\frac{1}{h_{2n-2}}\mathcal{L}\big[p_{2n}^{(1)}(z)-x_{2n}z^{2n-1}p_{2n-1}^{(2)}\big(z^{-1}
\big),z^{2n-2}p_{2n-2}^{(1)}\big(z^{-1}\big)\big]
\\
\phantom{(A_1)_{2n-1,2n-2}}{}
=\frac{1}{h_{2n-2}}\mathcal{L}\big[p_{2n}^{(1)}(z),z^{2n-2}p_{2n-2}^{(1)}\big(z^{-1}\big)\big]-\frac{x_{2n}}
{h_{2n-2}}\mathcal{L}\big[zp_{2n-1}^{(2)}\big(z^{-1}\big),p_{2n-2}^{(1)}\big(z^{-1}\big)\big].
\end{gather*}
The f\/irst term is equal to $0$ as $z^{2n-2}p_{2n-2}^{(1)}\big(z^{-1}\big)$ is a~polynomial of degree $2n-2$.
Consequently, using~\eqref{condition h}, we have
\begin{gather*}
(A_1)_{2n-1,2n-2}=-\frac{x_{2n}}{h_{2n-2}}\mathcal{L}\big[zp_{2n-2}^{(1)}(z),p_{2n-1}^{(2)}(z)\big]
=-\frac{x_{2n}}{h_{2n-2}}\mathcal{L}\big[z^{2n-1},p_{2n-1}^{(2)}(z)\big]
\\
\phantom{(A_1)_{2n-1,2n-2}}{}
=-\frac{h_{2n-1}}{h_{2n-2}}x_{2n}
=-(1-x_{2n-1}y_{2n-1})x_{2n}.
\end{gather*}

(5) The other relations are proven in a~similar way.
This f\/inishes the proof.
\end{proof}

\section{The AL hierarchy and a~Lax pair for its master symmetries}\label{Section3}

In this section we ``dress up'' the equations def\/ining the Ablowitz--Ladik hierarchy~\eqref{AL moments}
and its master symmetries~\eqref{master symmetries} on the bi-moments.
This leads to Lax pair representations both for the hierarchy and its master symmetries on the CMV matrices.
In all this section we shall denote the time variables $(t,s)=(t_1,t_2,\ldots, s_1,s_2,\ldots)$ of the AL
hierarchy by $(t_k)_{k\in\mathbb{Z}}$, with $t_{-k}=s_{k}$, $k\geq 1$, and $T_0$ def\/ined as in the
Introduction (see below~\eqref{deformed weight}).
It is only in the next section that the notation~$(t,s)$ will be more convenient.

\subsection{The Ablowitz--Ladik hierarchy}

Let
\begin{gather}
\label{chi}
\chi(z)=\big(1,z,z^{-1},z^2,z^{-2},\dots\big)^T,
\end{gather}
and let $\mathcal{L}$ be a~quasi-def\/inite bi-moment functional satisfying the Toeplitz condition.
We introduce two matrices $S_1$ and $S_2$ by writing the vectors $f(z)$, $g(z)$~\eqref{f,g} of monic right
bi-orthogonal L-polynomials with respect to $\mathcal{L}$ as follows
\begin{gather}\label{vector form f,g}
f(z)=S_1\chi(z),
\qquad
g(z)=h\big(S_2^T\big)^{-1}\chi(z),
\end{gather}
with $h=\diag(h_n)_{n\geq 0}$ and $h_n=\mathcal{L}[f_n,g_n]$.
With this def\/inition, $S_1$ is a~lower triangular matrix with all diagonal elements equal to~$1$, and
$S_2$ is an upper triangular matrix such that $h^{-1}S_2$ has all diagonal elements equal to~$1$.

Associated to $\mathcal{L}$ we also def\/ine the semi-inf\/inite bi-moment matrix
\begin{gather}
M=\left(
\begin{matrix}
\mu_{0,0}&\mu_{0,1}&\mu_{0,-1}&\dots
\\
\mu_{1,0}&\mu_{1,1}&\mu_{1,-1}&\dots
\\
\mu_{-1,0}&\mu_{-1,1}&\mu_{-1,-1}&\dots
\\
\vdots&\vdots&\vdots&\ddots
\end{matrix}
\right),
\label{L-bi-moment matrix}
\end{gather}
with $\mu_{m,n}$ as in~\eqref{bimoments first form},~\eqref{bimoments second form}.
The bi-moment matrix $M$ can be written in terms of the vec\-tor~$\chi (z)$ in~\eqref{chi}
\begin{gather*}
M=\big(\mathcal{L}\big[\big(\chi(z)\big)_m,\big(\chi(z)\big)_n\big]\big)_{0\leq m,n<\infty}.
\end{gather*}
The existence of a~sequence of right bi-orthogonal L-polynomials for $\mathcal{L}$ is equivalent to the
existence of a~factorisation of the bi-moment matrix $M$ in a~product of a~lower triangular matrix and an
upper triangular matrix with non-zero diagonal elements.
\begin{Proposition}
\label{factorization bi-moment matrix}
The bi-moment matrix $M$ factorizes in a~product of a~lower triangular matrix and an upper triangular matrix
\begin{gather*}
M=S_1^{-1}S_2.
\end{gather*}
\end{Proposition}
\begin{proof}
By bi-orthogonality of the sequence $\{f_n,g_n\}_{n\geq 0}$, we have
\begin{gather*}
\mathcal{L}[f_m,g_n]=h_m\delta_{m,n}.
\end{gather*}
This can be written in matrix form
\begin{gather*}
h=\big(\mathcal{L}[f_m,g_n]\big)_{0\leq m,n<\infty}.
\end{gather*}
Using the expressions~\eqref{vector form f,g} we obtain
\begin{gather*}
h=\big(\mathcal{L}\big[\big(S_1\chi(z)\big)_m,\big(h\big(S_2^T\big)^{-1}
\chi(z)\big)_n\big]\big)_{0\leq m,n\leq\infty}=S_1M S_2^{-1}h.
\end{gather*}
Consequently we have
\begin{gather*}
M=S_1^{-1}S_2,
\end{gather*}
which establishes the result.
\end{proof}

We def\/ine the semi-inf\/inite shift matrix $\Lambda$ by
\begin{gather}
\label{Lambda}
\Lambda\chi(z)=z\chi(z).
\end{gather}
We have
\begin{gather}
\label{Lambda matrix}
\Lambda=\left(
\begin{matrix}
0&1&0&0&0&0&\dots
\\
0&0&0&1&0&0&\dots
\\
1&0&0&0&0&0&\dots
\\
0&0&0&0&0&1&\dots
\\
0&0&1&0&0&0&\dots
\\
\vdots&\vdots&\vdots&\vdots&\vdots&\vdots&\ddots
\end{matrix}
\right),
\end{gather}
and $\Lambda^{-1}=\Lambda^{T}$.
We leave to the reader to check that, because of the Toeplitz property satisf\/ied by the bi-moments
in~\eqref{L-bi-moment matrix}, we have the commutation relation
\begin{gather}
\label{commutation Lambda M}
[\Lambda,M]=0.
\end{gather}

The CMV matrices can be obtained by ``dressing up'' the shift $\Lambda$.
\begin{Proposition}
We have
\begin{alignat}{3}
& A_1=S_1\Lambda S_1^{-1},
\qquad &&
A_2=h\big(S_2^T\big)^{-1}\Lambda S_2^Th^{-1},&
\label{Expression A}
\\
& A_1^{-1}=S_2\Lambda^TS_2^{-1},
\qquad &&
A_2^{-1}=h\big(S_1^{-1}\big)^T\Lambda^TS_1^Th^{-1},&
\label{Expression A^{-1}}
\end{alignat}
with $S_1$ and $S_2$ defined in~\eqref{vector form f,g}.
\end{Proposition}
\begin{proof}
We have
\begin{gather*}
A_1f(z)=zf(z)=zS_1\chi(z)=S_1\Lambda\chi(z)=S_1\Lambda S_1^{-1}f(z).
\end{gather*}
It follows that
\begin{gather*}
A_1=S_1\Lambda S_1^{-1}.
\end{gather*}
The proof for $A_2$ is similar.
The factorisations in~\eqref{Expression A^{-1}} follow from~\eqref{A star},~\eqref{inverse A}
and~\eqref{Expression A}.
\end{proof}

Remember that because $\mathcal{L}:\mathbb{C}[z,z^{-1}]\times\mathbb{C}[z,z^{-1}]\to\mathbb{C}$ is
a~Toeplitz bi-moment functional, the bi-moments $\mu_{m,n}=\mathcal{L}[z^m,z^n]$ only depend on the
dif\/ference $m-n$ and can be written as in~\eqref{bimoments second form} $\mu_{m,n}:=\mu_{m-n}$.

The Ablowitz--Ladik hierarchy is def\/ined on the space of bi-moments by the vector f\/ields
\begin{gather}
\label{Toeplitz vector fields bi-moments}
\boxed{T_k\mu_j\equiv\frac{\partial\mu_j}{\partial t_k}=\mu_{j+k},
\qquad
\forall\, k\in\mathbb{Z},}
\end{gather}
where we have put $s_k=t_{-k}$ in~\eqref{AL moments}.
Obviously, these vector f\/ields satisfy the commutation relations
\begin{gather*}
[T_k,T_l]=0,
\qquad
\forall\, k,l\in\mathbb{Z}.
\end{gather*}

It follows from the def\/inition of $\Lambda$ in~\eqref{Lambda} and~\eqref{Toeplitz vector fields
bi-moments} that the time evolution of the bi-moment matrix $M$ is given by the equations
\begin{gather}
\label{2-Toda evolution bi-moment matrix}
\boxed{\frac{\partial M}{\partial t_k}=\Lambda^kM,
\qquad
\forall\, k\in\mathbb{Z}.}
\end{gather}
Equations~\eqref{Toeplitz vector fields bi-moments} and~\eqref{2-Toda evolution bi-moment matrix} are two
equivalent formulations of the Ablowitz--Ladik vector f\/ields at the level of the bi-moments.

For a~square matrix $A$, we def\/ine
\begin{itemize}\itemsep=0pt
\item $A_0$ the diagonal part of $A$;
\item $A_-$ (resp.\ $A_+$) the lower (resp.\ upper) triangular part of $A$;
\item $A_{--}$ (resp.\ $A_{++}$) the strictly lower (resp.\ strictly upper) triangular part of $A$.
\end{itemize}
We establish the following lemma, based on the factorisation of the moment matrix $M$ in
Proposition~\ref{factorization bi-moment matrix} in a~product of a~lower triangular and an upper triangular
matrix.
\begin{Lemma}
\label{2-Toda evolution S}
We have for $k\in\mathbb{Z}$
\begin{gather}
\frac{\partial S_1}{\partial t_k}S_1^{-1}=-\big(A_1^k\big)_{--},
\label{derivative S1}
\\
\big(S_2^Th^{-1}\big)^{-1}\frac{\partial\big(S_2^Th^{-1}\big)}{\partial t_k}=\big(A_2^{-k}\big)_{--}.
\label{derivative S2}
\end{gather}
\end{Lemma}
\begin{proof}
On the one hand, we have using Proposition~\ref{factorization bi-moment matrix}
\begin{gather*}
\frac{\partial M}{\partial t_k}=-S_1^{-1}\frac{\partial S_1}{\partial t_k}S_1^{-1}S_2+S_1^{-1}
\frac{\partial S_2}{\partial t_k}.
\end{gather*}
On the other hand, from equation~\eqref{2-Toda evolution bi-moment matrix} we have
\begin{gather*}
\frac{\partial M}{\partial t_k}=\Lambda^kM=\Lambda^kS_1^{-1}S_2.
\end{gather*}
As $A_1=S_1 \Lambda S_1^{-1}$, we obtain
\begin{gather*}
A_1^k=-\frac{\partial S_1}{\partial t_k}S_1^{-1}+\frac{\partial S_2}{\partial t_k}S_2^{-1}.
\end{gather*}
Since $\frac{\partial S_1}{\partial t_k}$ is strictly lower triangular, the f\/irst term in the right hand
side of this equation is strictly lower triangular.
The second term is upper triangular.
Consequently, taking the strictly lower triangular part of both sides of the equation yields
\begin{gather*}
\frac{\partial S_1}{\partial t_k}S_1^{-1}=-\big(A_1^k\big)_{--},
\end{gather*}
which establishes~\eqref{derivative S1}.

To establish the other formula, we write $M=\big(S_1^{-1}h\big)\big(h^{-1}S_2\big)$ which gives
\begin{gather*}
\frac{\partial M}{\partial t_k}=\frac{\partial\big(S_1^{-1}h\big)}{\partial t_k}\big(h^{-1}S_2\big)+\big(S_1^{-1}
h\big)\frac{\partial\big(h^{-1}S_2\big)}{\partial t_k}.
\end{gather*}
Using the commutation relation~\eqref{commutation Lambda M} and~\eqref{2-Toda evolution bi-moment matrix},
we also have
\begin{gather*}
\frac{\partial M}{\partial t_k}=M\Lambda^k=\big(S_1^{-1}h\big)\big(h^{-1}S_2\big)\Lambda^k.
\end{gather*}
As $A_2=\big(S_2^Th^{-1}\big)^{-1}\Lambda\big(S_2^Th^{-1}\big)$, we obtain after some algebra
\begin{gather*}
A_2^{-k}=\frac{\partial\big(S_1^{-1}h\big)^T}{\partial t_k}\big(\big(S_1^{-1}h\big)^T\big)^{-1}+\big(S_2^Th^{-1}\big)^{-1}
\frac{\partial\big(S_2^Th^{-1}\big)}{\partial t_k}.
\end{gather*}
Since $\big(S_1^{-1}h\big)^T$ is upper triangular, the f\/irst term in the right hand side of this equation is
upper triangular.
As $S_2^Th^{-1}$ is lower triangular with all diagonal entries equal to $1$, the second term is strictly
lower triangular.
Consequently, taking the strictly lower triangular part of both sides of the equation yields
\begin{gather*}
\big(S_2^Th^{-1}\big)^{-1}\frac{\partial\big(S_2^Th^{-1}\big)}{\partial t_k}=\big(A_2^{-k}\big)_{--},
\end{gather*}
which establishes~\eqref{derivative S2}, completing the proof.
\end{proof}

We are now able to obtain a~Lax pair representation for the Ablowitz--Ladik hierarchy.
\begin{Theorem}
\label{2-Toda}
The ``dressed up'' form of the moment equation~\eqref{2-Toda evolution bi-moment matrix} gives the
following Lax pair representation for the Ablowitz--Ladik hierarchy on the semi-infinite CMV matrices
$(A_1,A_2)$
\begin{gather}
\label{Toda equations}
\boxed{\frac{\partial A_1}{\partial t_k}=\big[A_1,\big(A_1^k\big)_{--}\big],
\qquad
\frac{\partial A_2}{\partial t_k}=\big[A_2,\big(A_2^{-k}\big)_{--}\big],
\qquad
\forall\, k\in\mathbb{Z}.}
\end{gather}
\end{Theorem}
\begin{proof}
As $A_1=S_1\Lambda S_1^{-1}$ and $A_2=\big(S_2^Th^{-1}\big)^{-1}\Lambda\big(S_2^Th^{-1}\big)$, we have
\begin{gather*}
\frac{\partial A_1}{\partial t_k}=\left[\frac{\partial S_1}{\partial t_k}S_1^{-1},A_1\right]
\qquad
\text{and}
\qquad
\frac{\partial A_2}{\partial t_k}=\left[A_2,\big(S_2^Th^{-1}\big)^{-1}\frac{\partial\big(S_2^Th^{-1}\big)}
{\partial t_k}\right].
\end{gather*}
By Lemma~\ref{2-Toda evolution S} we obtain
\begin{gather*}
\frac{\partial A_1}{\partial t_k}=\big[{-}\big(A_1^k\big)_{--},A_1\big]
\qquad
\text{and}
\qquad
\frac{\partial A_2}{\partial t_k}=\big[A_2,\big(A_2^{-k}\big)_{--}\big],
\end{gather*}
which establishes~\eqref{Toda equations}, concluding the proof.
\end{proof}
\begin{Remark}
\label{Lax xy}
Looking back at the explicit expressions for the entries of the CMV matrices in Theorem~\ref{entries CMV
matrices}, the reader will observe that the entries of $A_2$ are obtained from those of $A_1$ by exchanging
the roles of the variables $x_n$ and $y_n$.
Also $A_1$ contains as entries $-x_{2n+1}$ and $y_{2n}$ and thus $A_2$ contains as entries $x_{2n}$ and
$-y_{2n+1}$, $n\geq 0$ (remember that $x_0=y_0=1$).
Thus the pair of Lax equations in~\eqref{Toda equations} completely determines the Ablowitz--Ladik
hierarchy in terms of the variables $x_n$ and $y_n$.
\end{Remark}

Using the explicit expressions in terms of the variables $x_n$ and $y_n$ for the entries of the CMV
matrices obtained in Theorem~\ref{entries CMV matrices}, and Theorem~\ref{2-Toda}, one easily computes the
equations for the vector f\/ields $T_1$ and $T_{-1}$
\begin{gather*}
\begin{split}
& \frac{\partial x_n}{\partial t_1}=(1-x_ny_n)x_{n+1},
\qquad
\frac{\partial x_n}{\partial t_{-1}}=-(1-x_ny_n)x_{n-1},
\\
& \frac{\partial y_n}{\partial t_1}=-(1-x_ny_n)y_{n-1},
\qquad
\frac{\partial y_n}{\partial t_{-1}}=(1-x_ny_n)y_{n+1}.
\end{split}
\end{gather*}
After the rescaling $x_n\to e^{-2t}x_n$, $y_n\to e^{2t}y_n$, the vector f\/ield $T_1-T_{-1}$ reduces to the
Ablowitz--Ladik equations as written in~\eqref{ALequations}.
In this paper, we won't discuss the Hamiltonian structure of the AL hierarchy in terms of the CMV matrices
$A_1$ and $A_2$.
One can show that for $k\geq 1$
\begin{gather*}
\frac{\partial x_n}{\partial t_k}=(1-x_n y_n)\frac{\partial H_k^{(1)}}{\partial y_n},
\qquad
\frac{\partial x_n}{\partial t_{-k}}=(1-x_n y_n)\frac{\partial H_k^{(2)}}{\partial y_n},
\\
\frac{\partial y_n}{\partial t_k}=-(1-x_n y_n)\frac{\partial H_k^{(1)}}{\partial x_n},
\qquad
\frac{\partial y_n}{\partial t_{-k}}=-(1-x_n y_n)\frac{\partial H_k^{(2)}}{\partial x_n},
\end{gather*}
where $H_k^{(1)}=-\frac{1}{k}\, \text{Tr}\, A_1^k, H_k^{(2)}=\frac{1}{k}\,\text{Tr}\, A_2^k$ and $\text{Tr}$ denotes the formal trace,
see~\cite{V} for a~proof inspired by~\cite{AVM2} in the context of Hessenberg matrices.

\subsection{A Lax pair for the master symmetries}

In this section we translate the action of the master symmetries vector f\/ields $V_k$, $k\in\mathbb{Z}$,
def\/ined on the bi-moments by~\eqref{master symmetries}, on the CMV matrices $(A_1,A_2)$.

We f\/irst decompose the vector f\/ields $V_k$ as follows
\begin{gather}
\label{V V script}
V_k=kT_k+\mathcal{V}_k,
\end{gather}
where $T_k$ are the Ablowitz--Ladik vector f\/ields~\eqref{Toeplitz vector fields bi-moments}.
At the level of the bi-moments, the vector f\/ields $\mathcal{V}_k$ are given by
\begin{gather}
\label{V script}
\boxed{\mathcal{V}_k\mu_j\equiv\frac{\text{d}}{\text{d}u_k}\mu_j=j\mu_{j+k},
\qquad
j,k\in\mathbb{Z}.}
\end{gather}
These vector f\/ields satisfy the following commutation relations
\begin{gather*}
[\mathcal{V}_k,\mathcal{V}_l]=(l-k)\mathcal{V}_{k+l},
\qquad
[\mathcal{V}_k,T_l]=lT_{k+l}.
\end{gather*}
It follows that
\begin{gather*}
[[\mathcal{V}_k,T_l],T_l]=0,
\qquad
\forall\, k,l\in\mathbb{Z}.
\end{gather*}
Consequently, like the vector f\/ields $V_k$, the vector f\/ields $\mathcal{V}_k$, $k\in\mathbb{Z}$, form
a~Virasoro algebra of master symmetries for the Ablowitz--Ladik hierarchy.

The dif\/ferentiation of $\chi(z)$ with respect to $z$ is def\/ined by
\begin{gather}
\label{delta}
\frac{\text{d}}{\text{d}z}\chi(z)=\delta\chi(z),
\end{gather}
where
\begin{gather}
\label{Delta}
\delta=\Delta\Lambda^T,
\qquad
\text{with}
\qquad
\Delta=\diag(0,1,-1,2,-2,\ldots),
\end{gather}
and $\Lambda$ is as in~\eqref{Lambda matrix}.

Remembering the notation~\eqref{bimoments second form},~\eqref{V script} writes
\begin{gather*}
\frac{\text{d}}{\text{d}u_k}\mu_{m,n}=(m-n)\mu_{m+k,n},
\end{gather*}
which is equivalent to the following equation on the bi-moment matrix $M$
\begin{gather}
\label{master symmetries moment matrix}
\boxed{\frac{\text{d}M}{\text{d}u_k}=\Delta\Lambda^kM-\Lambda^kM\Delta=\big[\Delta,\Lambda^kM\big].}
\end{gather}

Remember from~\eqref{vector form f,g} that
\begin{gather}\label{vector form f,g*}
f(z)=S_1\chi(z),
\qquad
g(z)=h\big(S_2^T\big)^{-1}\chi(z),
\end{gather}
and, according to~\eqref{vector form Five term recurrence relations} and~\eqref{A star}, these vectors
satisfy
\begin{gather}
A_1f(z)=zf(z),
\qquad
A_1^T\big(h^{-1}g^*(z)\big)=z\big(h^{-1}g^*(z)\big),
\label{Five term recurrence f,g*}
\\
A_2g(z)=zg(z),
\qquad
A_2^T\big(h^{-1}f^*(z)\big)=z\big(h^{-1}f^*(z)\big).
\label{five term recurrence g,f*}
\end{gather}
We def\/ine the semi-inf\/inite matrices $D_1$, $D_1^*$ and $D_2$, $D_2^*$ by the relations
\begin{gather}
\frac{\text{d}}{\text{d}z}f(z)=D_1f(z),
\qquad
\frac{\text{d}}{\text{d}z}\big(h^{-1}g^*(z)\big)=(D_1^*)^T\big(h^{-1}g^*(z)\big),
\label{D D* matrices}
\\
\frac{\text{d}}{\text{d}z}g(z)=D_2g(z),
\qquad
\frac{\text{d}}{\text{d}z}\big(h^{-1}f^*(z)\big)=(D_2^*)^T\big(h^{-1}f^*(z)\big).
\label{D_2 D_2^* matrices}
\end{gather}
These matrices can be ``dressed up'' as explained in the next lemma.
\begin{Lemma}
\label{factorisation D}
We have
\begin{gather}
D_1=S_1\Delta\Lambda^TS_1^{-1},
\qquad
D_1^*=-S_2\Lambda^T\Delta S_2^{-1},
\label{dressing D_1}
\\
D_2=\big(S_2^Th^{-1}\big)^{-1}\Delta\Lambda^T\big(S_2^Th^{-1}\big),
\qquad
D_2^*=-\big(S_1^Th^{-1}\big)^{-1}\Lambda^T\Delta\big(S_1^Th^{-1}\big),
\label{dressing D_2}
\end{gather}
with $\Delta$ as in~\eqref{Delta}.
\end{Lemma}
\begin{proof}
Using~\eqref{vector form f,g*} and~\eqref{D D* matrices}, we have
\begin{gather*}
D_1f(z)=\frac{\text{d}}{\text{d}z}f(z)=S_1\frac{\text{d}}{\text{d}z}\chi(z).
\end{gather*}
By def\/inition of $\delta$ in~\eqref{delta} and~\eqref{Delta}, we get
\begin{gather*}
D_1f(z)=S_1\delta S_1^{-1}f(z)=S_1\Delta\Lambda^TS_1^{-1}f(z).
\end{gather*}
This proves the f\/irst formula in~\eqref{dressing D_1}.

Using~\eqref{vector form f,g*} and remembering from~\eqref{f^*,g^*} that $g^*(z)=g\big(z^{-1}\big)$, we have
\begin{gather*}
\frac{\text{d}}{\text{d}z}g^*(z)=h\big(S_2^T\big)^{-1}\frac{\text{d}}{\text{d}z}\chi(z^{-1})=-h\big(S_2^T\big)^{-1}z^{-2}
\left(\frac{\text{d}}{\text{d}u}\chi(u)\right)\bigg|_{u=z^{-1}},
\end{gather*}
which gives, using~\eqref{delta},~\eqref{Delta},~\eqref{vector form f,g*} and remembering the
def\/inition~\eqref{Lambda} of the shift mat\-rix~$\Lambda$,
\begin{gather*}
\frac{\text{d}}{\text{d}z}g^*(z)=-h\big(S_2^T\big)^{-1}\delta z^{-2}\chi(z^{-1})
=-h\big(S_2^T\big)^{-1}\Delta\Lambda^T\Lambda^2\chi(z^{-1})
\\
\phantom{\frac{\text{d}}{\text{d}z}g^*(z)}{}
=-h\big(S_2^T\big)^{-1}\Delta\Lambda\big(h\big(S_2^T\big)^{-1}\big)^{-1}g^*(z).
\end{gather*}
Consequently, using the def\/inition~\eqref{D D* matrices} of $D_1^*$
\begin{gather*}
(D_1^*)^T\big(h^{-1}g^*(z)\big)=\frac{\text{d}}{\text{d}z}\big(h^{-1}g^*(z)\big)=-\big(S_2^T\big)^{-1}
\Delta\Lambda S_2^T\big(h^{-1}g^*(z)\big).
\end{gather*}
This proves the second formula in~\eqref{dressing D_1}.

The proof of~\eqref{dressing D_2} is identical to the proof of~\eqref{dressing D_1} using~\eqref{vector
form f,g*} and the def\/initions of $D_2$ and $D_2^*$ in~\eqref{D_2 D_2^* matrices}.
This establishes the lemma.
\end{proof}
\begin{Lemma}
\label{master}
We have for $k\in\mathbb{Z}$
\begin{gather}
\frac{\text{\rm d}S_1}{\text{\rm d}u_k}S_1^{-1}=-\big(D_1A_1^{k+1}\big)_{--}-\big(A_1^{k+1}D_1^*\big)_{--},
\label{master S_1}
\\
\big(S_2^Th^{-1}\big)^{-1}\frac{\text{\rm d}\big(S_2^Th^{-1}\big)}{\text{\rm d}u_k}=-\big(D_2A_2^{1-k}\big)_{--}-\big(A_2^{1-k}
D_2^*\big)_{--}.
\label{master S_2}
\end{gather}
\end{Lemma}
\begin{proof}
By substituting the factorisation $M=S_1^{-1}S_2$ of the moment matrix into~\eqref{master symmetries
moment matrix}, we obtain
\begin{gather*}
-S_1^{-1}\frac{\text{d}S_1}{\text{d}u_k}S_1^{-1}S_2+S_1^{-1}\frac{\text{d}S_2}{\text{d}u_k}=\Delta\Lambda^kS_1^{-1}
S_2-\Lambda^kS_1^{-1}S_2\Delta.
\end{gather*}
Multiplying this equation on the left by $S_1$ and on the right by $S_2^{-1}$, we get
\begin{gather}
\label{Term 1 and 2}
-\frac{\text{d}S_1}{\text{d}u_k}S_1^{-1}+\frac{\text{d}S_2}{\text{d}u_k}S_2^{-1}
=\underbrace{S_1\Delta\Lambda^kS_1^{-1}}_{\text{Term1}}
-\underbrace{S_1\Lambda^kS_1^{-1}S_2\Delta S_2^{-1}}_{\text{Term2}}.
\end{gather}
Using the factorisation of $A_1$ given in~\eqref{Expression A} and the factorisation of $D_1$
in~\eqref{dressing D_1}, {Term1} gives
\begin{gather*}
\text{Term1}=S_1\Delta\Lambda^T\Lambda^{k+1}S_1^{-1}=\big(S_1\Delta\Lambda^TS_1^{-1}
\big)\big(S_1\Lambda^{k+1}S_1^{-1}\big)=D_1A_1^{k+1}.
\end{gather*}
Similarly, Term2 gives
\begin{gather*}
\text{Term2}=A_1^kS_2\Delta S_2^{-1}=A_1^{k+1}A_1^{-1}S_2\Delta S_2^{-1}.
\end{gather*}
Using the factorisation of $A_1^{-1}$ in~\eqref{Expression A^{-1}} we get
\begin{gather*}
\text{Term2}=A_1^{k+1}\big(S_2\Lambda^TS_2^{-1}\big)S_2\Delta S_2^{-1}=A_1^{k+1}
\big(S_2\Lambda^T\Delta S_2^{-1}\big)=-A_1^{k+1}D_1^*,
\end{gather*}
where we have used the expression of $D_1^*$ in Lemma~\ref{factorisation D}.
Substituting these results in~\eqref{Term 1 and 2}, we obtain
\begin{gather*}
-\frac{\text{d}S_1}{\text{d}u_k}S_1^{-1}+\frac{\text{d}S_2}{\text{d}u_k}S_2^{-1}=D_1A_1^{k+1}+A_1^{k+1}D_1^*.
\end{gather*}
The f\/irst term in the left-hand side is strictly lower triangular, while the second term in the left-hand
side is upper triangular.
Consequently, taking the strictly lower triangular part in both sides, we obtain
\begin{gather*}
\frac{\text{d}S_1}{\text{d}u_k}S_1^{-1}=-\big(D_1A_1^{k+1}\big)_{--}-\big(A_1^{k+1}D_1^*\big)_{--},
\end{gather*}
which establishes~\eqref{master S_1}.

To establish the other formula, we substitute the factorisation $M=\big(S_1^{-1}h\big)\big(h^{-1}S_2\big)$ into
equation~\eqref{master symmetries moment matrix} rewritten as
\begin{gather*}
\frac{\text{d}M}{\text{d}u_k}=\big[\Delta,M\Lambda^k\big],
\end{gather*}
which follows from the commutation relation~\eqref{commutation Lambda M}.
This gives
\begin{gather*}
\frac{\text{d}\big(S_1^{-1}h\big)}{\text{d}u_k}\big(h^{-1}S_2\big)+\big(S_1^{-1}h\big)\frac{\text{d}\big(h^{-1}S_2\big)}{\text{d}u_k}=\Delta\big(S_1^{-1}
h\big)\big(h^{-1}S_2\big)\Lambda^k-\big(S_1^{-1}h\big)\big(h^{-1}S_2\big)\Lambda^k\Delta.
\end{gather*}
Multiplying this equation on the left by $\big(S_1^{-1}h\big)^{-1}$ and on the right by $\big(h^{-1}S_2\big)^{-1}$, we
get
\begin{gather}
\big(S_1^{-1}h\big)^{-1}\frac{\text{d}\big(S_1^{-1}h\big)}{\text{d}u_k}+\frac{\text{d}\big(h^{-1}S_2\big)}{\text{d}u_k}\big(h^{-1}S_2\big)^{-1}
\nonumber
\\
\qquad{}
=\underbrace{\big(S_1^{-1}h\big)^{-1}\Delta\big(S_1^{-1}h\big)\big(h^{-1}S_2\big)\Lambda^k\big(h^{-1}S_2\big)^{-1}}_{\text{Term1}}
-\underbrace{\big(h^{-1}S_2\big)\Lambda^k\Delta\big(h^{-1}S_2\big)^{-1}}_{\text{Term2}}.
\label{Term 1 Term 2 A_2}
\end{gather}
Using the factorisation of $A_2$ in~\eqref{Expression A} and the factorisation of $D_2$ in~\eqref{dressing D_2},
{Term2} gives
\begin{gather*}
\text{Term2}=\big(h^{-1}S_2\big)\Lambda^{k-1}\Lambda\Delta\big(h^{-1}S_2\big)^{-1}
\\
\phantom{\text{Term2}}{}
=\big(h^{-1}S_2\big)\Lambda^{k-1}\big(h^{-1}S_2\big)^{-1}\big(h^{-1}S_2\big)\Lambda\Delta\big(h^{-1}
S_2\big)^{-1}=\big(A_2^T\big)^{1-k}D_2^T.
\end{gather*}
Similarly, using the factorisation of $A_2$ in~\eqref{Expression A}, gives
\begin{gather*}
\text{Term1}=\big(S_1^{-1}h\big)^{-1}\Delta\big(S_1^{-1}h\big)\big(A_2^T\big)^{-k}
=\big(S_1^{-1}h\big)^{-1}\Delta\big(S_1^{-1}h\big)\big(A_2^T\big)^{-1}\big(A_2^T\big)^{1-k}.
\end{gather*}
Using the factorisation of $A_2^{-1}$ in~\eqref{Expression A^{-1}} and the factorisation of $D_2^*$
in~\eqref{dressing D_2}, we get
\begin{gather*}
\text{Term1}=\big(h^{-1}S_1\big)\Delta\Lambda\big(h^{-1}S_1\big)^{-1}\big(A_2^T\big)^{1-k}
=-(D_2^*)^T\big(A_2^T\big)^{1-k}.
\end{gather*}
Substituting these results in the transpose of~\eqref{Term 1 Term 2 A_2}, we obtain
\begin{gather*}
\frac{\text{d}\big(S_1^{-1}h\big)^T}{du_k}\big(\big(S_1^{-1}h\big)^T)^{-1}+\big(S_2^Th^{-1}\big)^{-1}\frac{\text{d}\big(S_2^Th^{-1}\big)}{\text{d}u_k}
=-D_2A_2^{1-k}-A_2^{1-k}D_2^*.
\end{gather*}
Since $\big(S_1^{-1}h\big)^T$ is upper triangular and $S_2^Th^{-1}$ is lower triangular with diagonal elements
equal to~1, by taking the strictly lower part of both sides of this equation, we obtain~\eqref{master S_2}.
This concludes the proof of the lemma.
\end{proof}

We are now able to obtain a~Lax pair representation for the master symmetries vector f\/ields~$\mathcal{V}_k$, $k\in\mathbb{Z}$.
\begin{Theorem}
\label{Lax pair master symmetries}
The ``dressed up'' form of the moment equation~\eqref{master symmetries moment matrix} gives the following
Lax pair representation for the master symmetries vector fields $\mathcal{V}_k$ on the semi-infinite
CMV matrices $(A_1,A_2)$
\begin{gather}
\label{Lax pair master symmetries 1}
\boxed{\left.
\begin{array}{@{}ll}
\displaystyle \frac{\text{\rm d}}{\text{\rm d}u_k}A_1=\Big[A_1,\big(D_1A_1^{k+1}\big)_{--}+\big(A_1^{k+1}D_1^*\big)_{--}\Big],
\qquad
\forall\, k\in\mathbb{Z},
\vspace{1mm}\\
\displaystyle \frac{\text{\rm d}}{\text{\rm d}u_k}A_2=\Big[\big(D_2A_2^{1-k}\big)_{--}+\big(A_2^{1-k}D_2^*\big)_{--},A_2\Big],
\qquad
\forall\, k\in\mathbb{Z},
\end{array}
\right.}
\end{gather}
or equivalently
\begin{gather*}
\boxed{\left.
\begin{array}{@{}ll}
\displaystyle\frac{\text{\rm d}}{\text{\rm d}u_k}A_1=A_1^{k+1}+\Big[\big(D_1A_1^{k+1}\big)_{+}-\big(A_1^{k+1}D_1^*\big)_{--}
,A_1\Big],
\qquad
\forall\, k\in\mathbb{Z},
\vspace{1mm}\\
\displaystyle \frac{\text{\rm d}}{\text{\rm d}u_k}A_2=A_2^{1-k}+\Big[A_2,\big(A_2^{1-k}D_2^*)_{+}-\big(D_2A_2^{1-k}\big)_{--}\Big],
\qquad
\forall\, k\in\mathbb{Z}.
\end{array}
\right.}
\end{gather*}
\end{Theorem}
\begin{proof}
As $A_1=S_1\Lambda S_1^{-1}$ and $A_2=\big(S_2^Th^{-1}\big)^{-1}\Lambda\big(S_2^Th^{-1}\big)$, we have{\samepage
\begin{gather*}
\frac{\text{d}A_1}{\text{d}u_k}=\left[\frac{\text{d}S_1}{\text{d}u_k}S_1^{-1},A_1\right]
\qquad
\text{and}
\qquad
\frac{\text{d}A_2}{\text{d}u_k}=\left[A_2,\big(S_2^Th^{-1}\big)^{-1}\frac{\text{d}\big(S_2^Th^{-1}\big)}{\text{d}u_k}\right].
\end{gather*}
Using~\eqref{master S_1} and~\eqref{master S_2} in Lemma~\ref{master}, we obtain~\eqref{Lax pair master
symmetries 1}.}

From~\eqref{Five term recurrence f,g*},~\eqref{D D* matrices} and from~\eqref{five term recurrence
g,f*},~\eqref{D_2 D_2^* matrices}, we deduce that $[A_1,D_1]=1$ and $[D_2^*,A_2]=1$.
From these commutation relations, one readily obtains that
\begin{gather*}
\big[A_1,\big(D_1A_1^{k+1}\big)_{+}\big]+\big[A_1,\big(D_1A_1^{k+1}\big)_{--}\big]=A_1^{k+1},
\\
\big[\big(A_2^{1-k}D_2^*)_{+},A_2\big]+\big[\big(A_2^{1-k}D_2^*\big)_{--},A_2\big]=A_2^{1-k},
\end{gather*}
which gives the equivalent formulation for the representation of the master symmetries on the CMV matrices
$(A_1,A_2)$.
This concludes the proof.
\end{proof}

We notice that as a~consequence of the Lax pair representation~\eqref{Toda equations} for the AL hierarchy
in Theorem~\ref{2-Toda}, the relation between the vector f\/ields $V_k$ and $\mathcal{V}_k$ in~\eqref{V V
script} and the Lax pair representation~\eqref{Lax pair master symmetries 1} of $\mathcal{V}_k$ in
Theorem~\ref{Lax pair master symmetries}, we have established the Lax pair representation~\eqref{Lax pair
Vk A1},~\eqref{Lax pair Vk A2} of the vector f\/ields $V_k$ as announced in Theorem~\ref{full theorem} in
the Introduction.

We emphasize that Theorem~\ref{Lax pair master symmetries} exhibits a~\emph{full} centerless Virasoro
algebra of master symmetries for the AL hierarchy.
This result stands in contrast with the Toda lattice and Korteweg--de~Vries hierarchies which possess only
half of a~Virasoro algebra of master symmetries~$\mathcal{V}_k$, $k\geq -1$, satisfying $[\mathcal{V}_k,
\mathcal{V}_l]=(l-k)\mathcal{V}_{k+l}$, $k,l\geq -1$, see~\cite{AVM1, D, FG, Fe, GH, MZ}.

Using the explicit form of the CMV matrices $(A_1,A_2)$ in Theorem~\ref{entries CMV matrices}, and Theorem~\ref{Lax pair master symmetries}, remembering Remark~\ref{Lax xy}, one can compute the f\/irst few
master symmetries vector f\/ields $\mathcal{V}_{-2}$, $\mathcal{V}_{-1}$, $\mathcal{V}_0$, $\mathcal{V}_1$ in
terms of the variables $x_n$, $y_n$:
\begin{gather*}
\mathcal{V}_{-2}(x_n)=(n-4)x_{n-2}(1-x_{n-1}y_{n-1})(1-x_ny_n)
\\
\phantom{\mathcal{V}_{-2}(x_n)=}{}
-x_{n-1}(1-x_ny_n)\big((n-4)x_{n-1}y_n+(n-1)x_ny_{n+1}\big)
\\
\phantom{\mathcal{V}_{-2}(x_n)=}{}
-2x_{n-1}(1-x_ny_n)\sum_{k=1}^{n}y_kx_{k-1}+x_n\sum_{k=1}^{n}y_k^2x_{k-1}^2
\\
\phantom{\mathcal{V}_{-2}(x_n)=}{}
-2x_n\sum_{k=2}^{n}y_kx_{k-2}+2x_n\sum_{k=2}^{n}y_ky_{k-1}x_{k-1}x_{k-2},
\\
\mathcal{V}_{-2}(y_n)=-ny_{n+2}(1-x_ny_n)(1-x_{n+1}y_{n+1})
\\
\phantom{\mathcal{V}_{-2}(y_n)=}{}
+y_{n+1}(1-x_ny_n)\big(nx_{n}y_{n+1}+(n-1)x_{n-1}y_{n}\big)
\\
\phantom{\mathcal{V}_{-2}(y_n)=}{}
+2y_{n+1}(1-x_ny_n)\sum_{k=1}^{n}y_kx_{k-1}-y_n\sum_{k=1}^{n}y_k^2x_{k-1}^2
\\
\phantom{\mathcal{V}_{-2}(y_n)=}{}
+2y_n\sum_{k=2}^{n}y_kx_{k-2}-2y_n\sum_{k=2}^{n}y_ky_{k-1}x_{k-1}x_{k-2},
\\
\mathcal{V}_{-1}(x_n)=(n-2)x_{n-1}(1-x_ny_n)-x_n\sum_{k=1}^{n}y_kx_{k-1},
\\
\mathcal{V}_{-1}(y_n)=-ny_{n+1}(1-x_ny_n)+y_n\sum_{k=1}^{n}y_kx_{k-1},
\\
\mathcal{V}_0(x_n)=nx_{n},
\\
\mathcal{V}_0(y_n)=-ny_{n},
\\
\mathcal{V}_{1}(x_n)=nx_{n+1}(1-x_ny_n)-x_n\sum_{k=1}^{n}x_ky_{k-1},
\\
\mathcal{V}_{1}(y_n)=-(n-2)y_{n-1}(1-x_ny_n)+y_n\sum_{k=1}^{n}x_ky_{k-1}.
\end{gather*}

\section{The action of the master symmetries on the tau-functions}\label{Section4}

As we recalled in the Introduction in formula~\eqref{Toeptau}, the tau-functions of the semi-inf\/inite AL
hierarchy are given by
\begin{gather}
\label{Toeptaubis}
\tau_n(t,s)=\det \big(\mu_{k-l}(t,s)\big)_{0\leq k,l<n}.
\end{gather}
It immediately follows from the generating function of the elementary Schur polynomials~\eqref{Schur
generating function} that
\begin{gather}
\label{diff elementary Schur}
\frac{\partial}{\partial t_k}S_n(t)=S_{n-k}(t),
\end{gather}
which shows that the formal solution of the AL hierarchy~\eqref{AL moments} on the moments is
\begin{gather}
\label{moments-Schur}
\mu_j(t,s)=\sum_{m,n=0}^{\infty}S_m(t)S_n(s)\mu_{j+m-n}(0,0),
\qquad
\forall\, j\in\mathbb{Z}.
\end{gather}
The expansion~\eqref{tau Plucker} of the tau-functions in terms of the Pl\"ucker
coordinates~\eqref{Plucker} and the Schur polynomials~\eqref{Schur polynomials} easily follows.
Indeed, by substituting~\eqref{moments-Schur} into~\eqref{Toeptaubis}, we have
\begin{gather*}
\tau_n(t,s)=\sum_{\substack{0\leq i_0,i_1,\dots,i_{n-1}\\0\leq j_0,j_1,\dots,j_{n-1}}}
\det\big[\mu_{k-l+i_k-j_l}(0,0)\big]_{0\leq k,l<n}S_{i_0}(t)\cdots S_{i_{n-1}}(t)S_{j_0}(s)\cdots S_{j_{n-1}}(s).
\end{gather*}
Relabeling the indices as follows $i_k\mapsto i_k-k$, $j_l\mapsto j_l-l$, we get
\begin{gather}
\tau_n(t,s)=\sum_{\substack{0\leq i_0,\dots,i_{n-1}\\0\leq j_0,\dots,j_{n-1}}}
\det\big[\mu_{i_k-j_l}(0,0)\big]_{0\leq k,l<n}S_{i_0}(t)S_{i_1-1}(t)\cdots S_{i_{n-1}-(n-1)}(t)\nonumber
\\
\phantom{\tau_n(t,s)=}{}
\times S_{j_0}(s)S_{j_1-1}(s)\cdots S_{j_{n-1}-(n-1)}(s)\nonumber
\\
\phantom{\tau_n(t,s)}
=\sum_{\substack{0\leq i_0<\cdots<i_{n-1}\\0\leq j_0<\cdots<j_{n-1}}}
\sum_{\sigma_1,\sigma_2\in S_n}(-1)^{\sigma_1}(-1)^{\sigma_2}\det\big[\mu_{i_k-j_l}(0,0)\big]_{0\leq k,l<n}
S_{i_{\sigma_1(0)}}(t)\nonumber
\\
\phantom{\tau_n(t,s)=}{}
\times S_{i_{\sigma_1(1)}-1}(t)\cdots S_{i_{\sigma_1(n-1)}-(n-1)}
(t)S_{j_{\sigma_2(0)}}(s)S_{j_{\sigma_2(1)}-1}(s)\cdots S_{j_{\sigma_2(n-1)}-(n-1)}(s)\nonumber
\\
\phantom{\tau_n(t,s)}{}
=\sum_{\substack{0\leq i_0<\dots<i_{n-1}\\0\leq j_0<\dots<j_{n-1}}}
p_{\substack{i_0,\dots,i_{n-1}\\j_0,\dots,j_{n-1}}}
S_{i_{n-1}-(n-1),\dots,i_0}(t)S_{j_{n-1}-(n-1),\dots,j_0}(s),
\label{tau Plucker proof}
\end{gather}
with $(-1)^{\sigma}$ the sign of the permutation $\sigma$.

The aim of this section is to establish the second part of Theorem~\ref{full theorem}. 
\begin{Theorem}
\label{Virasoro constraints theorem}
For all $k\in\mathbb{Z}$, we have
\begin{gather}
\label{master tau}
L^{(n)}_k\tau_n(t,s)=\sum_{\substack{0\leq i_0<\dots<i_{n-1}\\0\leq j_0<\dots<j_{n-1}}}
V_k\Big(p_{\substack{i_0,\dots,i_{n-1}\\j_0,\dots,j_{n-1}}}\Big)
S_{i_{n-1}-(n-1),\dots,i_0}(t)S_{j_{n-1}-(n-1),\dots,j_0}(s),
\end{gather}
with $L^{(n)}_k$, $k\in \mathbb{Z}$, defined as in~\eqref{virplus},~\eqref{virzero},~\eqref{virminus}, and
$V_k \Big(p_{\substack{i_0,\dots,i_{n-1}\\j_0,\dots,j_{n-1}}}\Big)$
the Lie derivative of the Pl\"ucker coordinates~\eqref{Plucker} in the direction
of the master symmetries $V_k$ of the AL hierarchy, as defined in~\eqref{master symmetries}.
\end{Theorem}

This theorem is the key to the quick derivation of the various ``Virasoro-type'' constraints sa\-tisf\/ied by
\emph{special} tau-functions of the AL hierarchy.
As an illustration we establish the following result.
\begin{Corollary}
\label{unitary matrix model}
The partition function of the unitary matrix model
\begin{gather*}
\tau_n(t,s)
=\int_{U(n)}\exp\left\{\sum_{j=1}^{\infty}\big(t_j\text{\rm Tr}\, U^j+s_j\text{\rm Tr}\, U^{-j}\big)\right\}\text{\rm d}U,
\end{gather*}
where $U(n)$ is the group of unitary $n\times n$ matrices and $\text{\rm d}U$ is the standard Haar measure, normalized
so that the total volume is $1$, satisfies the Virasoro constraints
\begin{gather*}
L_k^{(n)}\tau_n(t,s)=0,
\qquad
\forall\, k\in\mathbb{Z},
\end{gather*}
with $L_k^{(n)}$ defined as in~\eqref{virplus},~\eqref{virzero} and~\eqref{virminus}.
\end{Corollary}
\begin{proof}
By using Weyl's integral formula, one has that
\begin{gather*}
\tau_n(t,s)=\frac{1}{n!}\int_{(S^1)^n}\vert\Delta_n(z)\vert^2\prod_{k=1}^n\exp\left\{\sum_{j=1}
^\infty\big(t_jz_k^j+s_jz_k^{-j}\big)\right\}\frac{\text{d}z_k}{2\pi i z_k},
\end{gather*}
is a~tau-function of the AL hierarchy as in~\eqref{Weyl form tau}, with $w(z)=1$ in the deformed
weight~\eqref{deformed weight}.
Thus the initial moments (at time $(t,s)=(0,0))$ are given by
\begin{gather*}
\mu_{j}(0,0)=\oint_{S^1}z^j\frac{\text{d}z}{2\pi iz}=\delta_{j,0},
\end{gather*}
with $\delta_{j,k}$ the usual Kronecker symbol.
By the def\/inition~\eqref{master symmetries} of the master symmetries $V_k$, it follows that
\begin{gather*}
V_k(\mu_j)_{\vert(t,s)=(0,0)}=(j+k)\mu_{j+k}(0,0)=(j+k)\delta_{j+k,0}=0,
\end{gather*}
which, using the def\/inition of the Pl\"ucker coordinates~\eqref{Plucker} and formula~\eqref{master tau},
establishes the result.
\end{proof}
\begin{Remark}
\label{Virasoro unitary matrix model}
After~\cite{HV} was completed, we found out that Corollary~\ref{unitary matrix model}, which can be seen as
a~particular case of our result recalled in~\eqref{virtauint}, had already been obtained by Bowick, Morozov
and Shevitz~\cite{BMS}, using the Lagrangian approach~\cite{MM} to derive Virasoro constraints.
However, these authors didn't notice the commutation relations~\eqref{Virasoro L} of the centerless
Virasoro algebra.
In contrast with Corollary~\ref{unitary matrix model}, the partition function of the Hermitian matrix model
(which is a~tau-function of the Toda lattice hierarchy) and the partition function of 2d-quantum gravity
(which is a~tau-function of the KdV hierarchy) are characterized by Virasoro constraints $L_k
\tau(t)=0$, $k\geq-1$, corresponding to ``half of'' a~Virasoro algebra, see~\cite{AVM1,DVV,FKN,HH,KS,M,MM} for
the explicit form of the operators~$L_k$ in those cases.
\end{Remark}

Actually, in the proof of Theorem~\ref{Virasoro constraints theorem}, we shall need to know that the
operators $L_k^{(n)}$, $k\in \mathbb{Z}$, satisfy the commutation relations of the centerless Virasoro
algebra.
For the convenience of the reader we repeat the proof given in~\cite{HV}.
Consider the complex Lie algebra $\mathcal{A}$ given by the direct sum of two commuting copies of the
Heisenberg algebra with bases $\{\hbar_a,a_j\,|\,j\in\mathbb{Z}\}$ and $\{\hbar_b,b_j\,|\,j\in\mathbb{Z}\}$ and
def\/ining commutation relations
\begin{alignat}{3}
& [\hbar_a,a_j]=0,
\qquad &&
[a_j,a_k]=j\delta_{j,-k}\hbar_a,& \nonumber
\\
& [\hbar_b,b_j]=0,
\qquad &&
[b_j,b_k]=j\delta_{j,-k}\hbar_b, &
\nonumber
\\
& [\hbar_a,\hbar_b]=0,
\qquad&&
[a_j,b_k]=0,
\qquad
[\hbar_a,b_j]=0,
\qquad
[\hbar_b,a_j]=0,& \label{Heisenberg algebra}
\end{alignat}
with $j,k\in\mathbb{Z}$.
Let $\mathcal{B}$ be the space of formal power series in the variables $t_1, t_2, \dots$
and $s_1, s_2, \dots$,
and consider the following representation of~$\mathcal{A}$ in~$\mathcal{B}$
\begin{gather}
a_j=\frac{\partial}{\partial t_j}
,
\qquad
a_{-j}=jt_j
,
\qquad
b_j=\frac{\partial}{\partial s_j}
,
\qquad
b_{-j}=js_j,\nonumber
\\
a_0=b_0=\mu
,
\qquad
\hbar_a=\hbar_b=1,
\label{oscillator representation Heisenberg}
\end{gather}
for $j>0$, and $\mu\in\mathbb{C}$.
Def\/ine the operators
\begin{gather}
\label{AB Virasoro operators}
A_k^{(n)}=\frac{1}{2}\sum_{j\in\mathbb{Z}}:a_{-j}a_{j+k}:
,
\qquad
B_k^{(n)}=\frac{1}{2}\sum_{j\in\mathbb{Z}}:b_{-j}b_{j+k}:,
\end{gather}
where $k\in\mathbb{Z}$, $a_j$, $b_j$ are as in~\eqref{oscillator representation Heisenberg} with $\mu=n$, and
where the colons indicate normal ordering, def\/ined by
\begin{gather*}
:a_ja_k:=
\begin{cases}a_ja_k
\quad &
\text{if}
\ \
j\leq k,
\\
a_ka_j
\quad &
\text{if}
\ \
j>k,
\end{cases}
\end{gather*}
and a~similar def\/inition for $:b_jb_k:$, obtained by changing the $a$'s in $b$'s in the former.
Expanding the expressions in~\eqref{AB Virasoro operators} we obtain for $k>0$
\begin{gather*}
A_0^{(n)}=\sum_{j>0}jt_j\frac{\partial}{\partial t_j}+\frac{n^2}{2},
\\
A_k^{(n)}=\frac{1}{2}\sum_{0<j<k}\frac{\partial^2}{\partial t_j\partial t_{k-j}}+\sum_{j>k}(j-k)t_{j-k}
\frac{\partial}{\partial t_j}+n\frac{\partial}{\partial t_k},
\\
A_{-k}^{(n)}=\frac{1}{2}\sum_{0<j<k}j(k-j)t_jt_{k-j}+\sum_{j>k}jt_{j}\frac{\partial}{\partial t_{j-k}}
+nkt_k,
\end{gather*}
and similar expressions for $B_k^{(n)}$, by changing the $t$-variables in $s$-variables.
Using these notations, we can rewrite~\eqref{virplus},~\eqref{virzero} and~\eqref{virminus} as follows
\begin{gather}
L_k^{(n)}=A_k^{(n)}-B_{-k}^{(n)}+\frac{1}{2}\sum_{j=1}^{k-1}(a_j-b_{-j})(a_{k-j}-b_{j-k}),
\qquad
k\geq1,\nonumber
\\
L_0^{(n)}=A_0^{(n)}-B_0^{(n)},
\label{L Virasoro in terms of AB}
\\
L_{-k}^{(n)}=A_{-k}^{(n)}-B_k^{(n)}-\frac{1}{2}\sum_{j=1}^{k-1}(a_{-j}-b_{j})(a_{j-k}-b_{k-j}),
\qquad
k\geq1.\nonumber
\end{gather}
As shown in~\cite{KR} (see Lecture~2) the operators $A_k^{(n)}$, $k\in\mathbb{Z}$, provide a~representation
of the Virasoro algebra in $\mathcal{B}$ with central charge $c=1$, that is
\begin{gather}
\label{Virasoro A}
\big[A_k^{(n)},A_l^{(n)}\big]=(k-l)A_{k+l}^{(n)}+\delta_{k,-l}\frac{k^3-k}{12},
\end{gather}
for $k,l\in\mathbb{Z}$.
Similarly, the operators $B_k^{(n)}$ satisfy the commutation relations
\begin{gather}
\label{Virasoro B}
\big[B_k^{(n)},B_l^{(n)}\big]=(k-l)B_{k+l}^{(n)}+\delta_{k,-l}\frac{k^3-k}{12},
\end{gather}
for $k,l\in\mathbb{Z}$.
Furthermore we have for $k,l\in\mathbb{Z}$
\begin{gather}
\label{commutator Heisenberg and Virasoro operators}
\big[a_k,A_l^{(n)}\big]=ka_{k+l},
\qquad
\big[b_k,B_l^{(n)}\big]=kb_{k+l},
\qquad
\big[a_k,B_l^{(n)}\big]=0,
\qquad
\big[b_k,A_l^{(n)}\big]=0.
\end{gather}
\begin{Proposition}
The operators $L_{k}^{(n)}$ defined as in~\eqref{virplus},~\eqref{virzero},~\eqref{virminus} satisfy the
commutation relations of the centerless Virasoro algebra
\begin{gather}
\label{virc}
\big[L_k^{(n)},L_l^{(n)}\big]=(k-l)L_{k+l}^{(n)},
\qquad
\forall\, k,l\in\mathbb{Z}.
\end{gather}
\end{Proposition}
\begin{proof}
We give the proof for $k,l\geq 0$, the other cases being similar.
As $\big[A_{i}^{(n)},B_{j}^{(n)}\big]=0$, $i,j\in\mathbb{Z}$, we have using~\eqref{Heisenberg
algebra},~\eqref{Virasoro A},~\eqref{Virasoro B} and~\eqref{commutator Heisenberg and Virasoro operators}
\begin{gather*}
\big[L_k^{(n)},L_l^{(n)}\big]=(k-l)\big(A_{k+l}^{(n)}-B_{-k-l}^{(n)}\big)-\frac{1}{2}\sum_{j=1}^{l-1}j(a_{j+k}
-b_{-j-k})(a_{l-j}-b_{j-l})
\\
\qquad
{}-\frac{1}{2}\sum_{j=1}^{l-1}(l-j)(a_{j}-b_{-j})(a_{k+l-j}-b_{j-k-l})+\frac{1}{2}\sum_{j=1}^{k-1}j(a_{j+l}
-b_{-j-l})(a_{k-j}-b_{j-k})
\\
\qquad
{}+\frac{1}{2}\sum_{j=1}^{k-1}(k-j)(a_{j}-b_{-j})(a_{k+l-j}-b_{j-k-l}).
\end{gather*}
Relabeling the indices in the sums, we have
\begin{gather*}
\big[L_k^{(n)},L_l^{(n)}\big]=(k-l)\big(A_{k+l}^{(n)}-B_{-k-l}^{(n)}\big)
{}-\frac{1}{2}\sum_{j=k+1}^{k+l-1}(j-k)(a_{j}-b_{-j})(a_{k+l-j}-b_{j-k-l})
\\
\qquad
{}-\frac{1}{2}\sum_{j=1}^{l-1}(l-j)(a_{j}-b_{-j})(a_{k+l-j}-b_{j-k-l})
+\frac{1}{2}\!\sum_{j=l+1}^{k+l-1}(j-l)(a_{j}-b_{-j})(a_{k+l-j}-b_{j-k-l})
\\
\qquad
{}+\frac{1}{2}\sum_{j=1}^{k-1}(k-j)(a_{j}-b_{-j})(a_{k+l-j}-b_{j-k-l})
=(k-l)L_{k+l}^{(n)}.
\end{gather*}
This concludes the proof.
\end{proof}

The plan of the rest of the section is as follows.
After some algebraic preliminaries, we shall translate the master symmetries on the Pl\"ucker coordinates
$p_{\substack{i_0,\dots,i_{n-1}\\j_0,\dots,j_{n-1}}}$.
Next we shall compute the action of the Virasoro operators on the products
$S_{i_{n-1}-(n-1),\dots,i_0}(t)S_{j_{n-1}-(n-1),\dots,j_0}(s)$ of Schur polynomials.
Finally we shall end with the proof of Theorem~\ref{Virasoro constraints theorem}.

\subsection{Some algebraic lemmas}

We shall need the following lemmas.
In order to formulate them, we introduce some notations.
Given $n$ vectors $x_1,\dots,x_n\in\mathbb{R}^n$, we shall denote by $|x_1x_2\dots x_n|$ the determinant of
the $n\times n$ matrix formed with the columns $x_i$.
Also, given two vectors $x$ and $y$, $x\wedge y$ denotes the usual wedge product, with components $(x\wedge
y)_{rs}=x_r y_s-x_s y_r$.
Finally, for an $n\times n$ matrix $A$, $A_r$ will denote the $r$th column of A, and $A_r^T$ the $r$th
column of the transposed matrix, and $\tr(A)$ will mean the trace of $A$.
With these conventions, we have the following lemma.
\begin{Lemma}[Haine--Semengue~\cite{HS}]\label{algebraic lemma H.S.}
Let $A$ and $B$ be $n\times n$ matrices, with $A$ invertible.
Then
\begin{gather*}
(i)\quad\sum_{r=1}^{n}|A_1\dots A_{r-1}B_rA_{r+1}\dots A_n|=(\det A)\tr(BA^{-1}),
\\
(ii)\quad\sum_{1\leq r<s\leq n}|A_1\dots A_{r-1}B_rA_{r+1}\dots A_{s-1}B_sA_{s+1}\dots A_n|
\\
\hphantom{(ii)\quad} \qquad{}
=(\det A)\sum_{1\leq r<s\leq n}\big(\big(BA^{-1}\big)_r\wedge\big(BA^{-1}\big)_s\big)_{rs}.
\end{gather*}
\end{Lemma}

\begin{proof}
$(i)$ Let $A$, $B$ be $n\times n$ matrices, with $A$ invertible.
As $A$ is invertible, its columns form a~basis of $\mathbb{C}^n$ and thus we have
\begin{gather}
B_r=Ac^{(r)}=\sum_jc_j^{(r)}A_j,
\label{B basis A}
\end{gather}
for a~certain $c^{(r)}\in\mathbb{C}^n$, whose components are $c_j^{(r)}=\big(A^{-1}B\big)_{jr}$.
It then follows that
\begin{gather*}
\sum_{r=1}^{n}|A_1\dots A_{r-1}B_rA_{r+1}\dots A_n|=\sum_{r=1}^{n}\Big|A_1\dots A_{r-1}\Big(\sum_jc_j^{(r)}
A_j\Big)A_{r+1}\dots A_n\Big|
\\
\qquad{}
=\det A\sum_{r=1}^{n}c_r^{(r)}
=(\det A)\tr\big(BA^{-1}\big).
\end{gather*}

$(ii)$ Using~\eqref{B basis A}, we have
\begin{gather*}
\sum_{1\leq r<s\leq n}|A_1\dots A_{r-1}B_rA_{r+1}\dots A_{s-1}B_sA_{s+1}\dots A_n|
\\
\qquad
=\sum_{1\leq r<s\leq n}\Big|A_1\dots A_{r-1}\Big(\sum_jc_j^{(r)}A_j\Big)A_{r+1}\dots A_{s-1}
\Big(\sum_jc_j^{(s)}A_j\Big)A_{s+1}\dots A_n\Big|
\\
\qquad{}
=\sum_{1\leq r<s\leq n}\Big|A_1\dots A_{r-1}\big(c_r^{(r)}A_r+c_s^{(r)}A_s\big)A_{r+1}\dots A_{s-1}
\big(c_r^{(s)}A_r+c_s^{(s)}A_s\big)A_{s+1}\dots A_n\Big|
\\
\qquad{}
=\det A\sum_{1\leq r<s\leq n}\big(c_r^{(r)}c_s^{(s)}-c_s^{(r)}c_r^{(s)}\big)
=\det A\sum_{1\leq r<s\leq n}\big((A^{-1}B)_{r}\wedge(A^{-1}B)_{s}\big)_{rs}.
\end{gather*}
We thus obtain
\begin{gather*}
\sum_{1\leq r<s\leq n}|A_1\dots A_{r-1}B_rA_{r+1}\dots A_{s-1}B_sA_{s+1}\dots A_n|
\\
\qquad{}
=\det A\sum_{1\leq r<s\leq n}\big(\big(BA^{-1}\big)_{r}\wedge\big(BA^{-1}\big)_{s}\big)_{rs},
\end{gather*}
where we have used the fact that for $X$, $Y$ two $n\times n$ matrices, we have
\begin{gather}
\sum_{1\leq r<s\leq n}\big((XY)_r\wedge(XY)_s\big)_{rs}=\sum_{1\leq r<s\leq n}
\big((YX)_r\wedge(YX)_s\big)_{rs}.
\label{commutator wedge}
\end{gather}
This concludes the proof of the lemma.
\end{proof}

We will also need a~transposed version of this lemma.
\begin{Lemma} \label{Lemma version H.S. transposed}
With the same conditions as in Lemma~{\rm \ref{algebraic lemma H.S.}}, we have
\begin{gather*}
(i)\quad\sum_{r=1}^{n}\big|A^T_1\dots A^T_{r-1}(B)^T_rA^T_{r+1}\dots A^T_n\big|
=\sum_{r=1}^{n}|A_1\dots A_{r-1}B_rA_{r+1}\dots A_n|,
\\
(ii)\quad\sum_{1\leq r<s\leq n}\big|A^T_1\dots A^T_{r-1}B^T_rA^T_{r+1}\dots A^T_{s-1}B^T_sA^T_{s+1}\dots A^T_n\big|
\\
\hphantom{(ii)\quad}\qquad
=\sum_{1\leq r<s\leq n}|A_1\dots A_{r-1}B_rA_{r+1}\dots A_{s-1}B_sA_{s+1}\dots A_n|.
\end{gather*}
\end{Lemma}
\begin{proof}
Both formulas are direct consequences of Lemma~\ref{algebraic lemma H.S.}, by observing that for $X$, $Y$ two
$n\times n$ matrices, we have~\eqref{commutator wedge} and
\begin{gather*}
\big(X^T_r\wedge X^T_s\big)_{rs}=(X_r\wedge X_s)_{rs}.\tag*{\qed}
\end{gather*}
\renewcommand{\qed}{}
\end{proof}

We give two consequences of this lemma.
First we particularize the preceding lemma to the Pl\"ucker coordinates, and then we particularize it to
the Schur polynomials.
\begin{Lemma}
\label{technical lemma}
For $m\in\mathbb{Z}$ we have
\begin{gather*}
(i)\quad
\sum_{l=1}^{n}p_{\substack{i_0,\dots,i_{n-1}\\j_0,\dots,j_{n-l}-m,\dots,j_{n-1}}}
=\sum_{l=1}^{n}p_{\substack{i_0,\dots,i_{n-l}+m,\dots,i_{n-1}\\j_0,\dots,j_{n-1}}},
\\
(ii)\quad
\sum_{1\leq r<s\leq n}p_{\substack{i_0,\dots,i_{n-1}\\j_0,\dots,j_{n-s}-m,\dots,j_{n-r}-m,\dots,j_{n-1}}}
=\sum_{1\leq r<s\leq n}p_{\substack{i_0,\dots,i_{n-s}+m,\dots,i_{n-r}+m,\dots,i_{n-1}\\j_0,\dots,j_{n-1}}}.
\end{gather*}
\end{Lemma}
\begin{proof}
Def\/ine the $n\times n$ matrices
\begin{gather*}
A=(\mu_{i_k-j_l})_{0\leq k,l\leq n-1}
\qquad
\text{and}
\qquad
B(m)=(\mu_{i_k-j_l+m})_{0\leq k,l\leq n-1}.
\end{gather*}
We then have
\begin{gather*}
\sum_{l=1}^{n}p_{\substack{i_0,\dots,i_{n-1}\\j_0,\dots,j_{n-l}-m,\dots,j_{n-1}}}
=\sum_{l=1}^{n}\big|A_1\dots A_{n-l-1}\big(B(m)\big)_{n-l}A_{n-l+1}
\dots A_{n-1}\big|
\\
\qquad
=\sum_{l=1}^{n}\big|A^T_1\dots A^T_{n-l-1}\big(B(m)\big)^T_{n-l}A^T_{n-l+1}\dots A^T_{n-1}\big|\nonumber
=\sum_{l=1}^{n}p_{\substack{i_0,\dots,i_{n-l}+m,\dots,i_{n-1}\\j_0,\dots,j_{n-1}}},
\end{gather*}
where we have used Lemma~\ref{Lemma version H.S. transposed}$(i)$ in the second equality.
This proves~$(i)$. The proof of~$(ii)$ is similar.
\end{proof}
\begin{Lemma}
\label{determinants of Schur polynomials}
The following holds
\begin{gather*}
(i)
\quad
\sum_{l=1}^{n}S_{i_{n-1}-(n-1),\dots,i_{n-l}-(n-l)-1,\dots,i_0}(t)
\\
\qquad\qquad
=\det\left(
\begin{matrix}
S_{i_{n-1}-n}(t)&S_{i_{n-1}-(n-2)}(t)&\cdots&S_{i_{n-1}}(t)
\\
S_{i_{n-2}-n}(t)&S_{i_{n-2}-(n-2)}(t)&\cdots&S_{i_{n-2}}(t)
\\
\vdots&\vdots&&\vdots
\\
S_{i_{0}-n}(t)&S_{i_{0}-(n-2)}(t)&\cdots&S_{i_{0}}(t)
\end{matrix}
\right),
\\[1.2mm]
(ii)
\quad
\sum_{l=1}^{n-1}S_{i_{n-1}-(n-1),\dots,i_{n-l}-(n-l)+1,\dots,i_1-1}(t)\nonumber
\\
\qquad\qquad
=\det\left(
\begin{matrix}
S_{i_{n-1}-(n-1)}(t)&\cdots&S_{i_{n-1}-2}(t)&S_{i_{n-1}}(t)
\\
S_{i_{n-2}-(n-1)}(t)&\cdots&S_{i_{n-2}-2}(t)&S_{i_{n-2}}(t)
\\
\vdots&&\vdots&\vdots
\\
S_{i_{1}-(n-1)}(t)&\cdots&S_{i_{1}-2}(t)&S_{i_{1}}(t)
\end{matrix}
\right).
\end{gather*}
\end{Lemma}
\begin{proof}
We prove $(i)$.
Def\/ine the $n\times n$ matrices
\begin{gather*}
A=\big(S_{i_{n-k}-(n-k)+l-k}(t)\big)_{1\leq k,l\leq n},
\qquad
B(m)=\big(S_{i_{n-k}-(n-k)+l-k+m}(t)\big)_{1\leq k,l\leq n}.
\end{gather*}
We have $S_{i_{n-1}-(n-1),\dots,i_0}(t)=\det A$.
It then follows that
\begin{gather*}
\sum_{l=1}^{n}S_{i_{n-1}-(n-1),\dots,i_{n-l}-(n-l)-1,\dots,i_0}(t)=\sum_{l=1}^{n}\big|A_1^T\dots A_{n-l-1}
^T\big(B(-1)\big)_{n-l}^T A_{n-l+1}^T\dots A_{n-1}^T\big|.
\end{gather*}
Using Lemma \ref{Lemma version H.S. transposed}$(i)$ we get
\begin{gather*}
\sum_{l=1}^{n}S_{i_{n-1}-(n-1),\dots,i_{n-l}-(n-l)-1,\dots,i_0}(t)=\sum_{l=1}^{n}\big|A_1\dots A_{n-l-1}
\big(B(-1)\big)_{n-l}A_{n-l+1}\dots A_{n-1}\big|.
\end{gather*}
In the right-hand side, in the $l^{\rm th}$ term, the $l^{\rm th}$ and $(l-1)^{\rm th}$ columns coincide in the
determinant, provided that $l\neq 1$.
Consequently, only the f\/irst term of the right-hand side gives a~non zero contribution.
This proves~$(i)$.
The proof of~$(ii)$ is similar.
\end{proof}

\subsection{Expression of the master symmetries on the Pl\"ucker coordinates}

We now translate the master symmetries on Pl\"ucker coordinates.
\begin{Lemma}
\label{master symmetries on Plucker coordinates}
Let $V_kp_{\substack{i_0,\dots,i_{n-1}\\j_0,\dots,j_{n-1}}}$ denote
the Lie derivative of the Pl\"ucker coordinates in the direction of the vector fields~$V_k$.
Then for $k\in\mathbb{Z}$,
\begin{gather}
V_kp_{\substack{i_0,\dots,i_{n-1}\\j_0,\dots,j_{n-1}}}
=\sum_{l=0}^{n-1}(i_l+k)p_{\substack{i_0,\dots,i_{l-1},i_{l}+k,i_{l+1},\dots,i_{n-1}\\j_0,\dots,j_{n-1}}}
-\sum_{l=0}^{n-1}j_{l}p_{\substack{i_0,\dots,i_{n-1}\\j_0,\dots,j_{l-1},j_{l}-k,j_{l+1},\dots,j_{n-1}}}
\nonumber
\\
\phantom{V_kp_{\substack{i_0,\dots,i_{n-1}\\j_0,\dots,j_{n-1}}}}{}
=\sum_{l=0}^{n-1}i_lp_{\substack{i_0,\dots,i_{l-1},i_{l}+k,i_{l+1},\dots,i_{n-1}\\j_0,\dots,j_{n-1}}}
-\sum_{l=0}^{n-1}(j_{l}-k)p_{\substack{i_0,\dots,i_{n-1}\\j_0,\dots,j_{l-1},j_{l}-k,j_{l+1},\dots,j_{n-1}}}.
\label{equation master symmetries on Plucker coordinates}
\end{gather}
\end{Lemma}

\begin{proof}
Fix $0\leq i_0<i_1<\dots<i_{n-1}$ and $0\leq j_0<j_1<\dots<j_{n-1}$.
We introduce the $n\times n$ matrices
\begin{gather*}
A=\big(\mu_{i_k-j_l}(0,0)\big)_{0\leq k,l\leq n-1},
\qquad
B(m)=\big(\mu_{i_k-j_l+m}(0,0)\big)_{0\leq k,l\leq n-1},
\end{gather*}
as well as the diagonal matrix $D=\diag(j_0,\dots,j_{n-1})$.
We notice that $p_{\substack{i_0,\dots,i_{n-1}\\j_0,\dots,j_{n-1}}}=\det A$,
by def\/inition of the Pl\"ucker coordinates.
From the def\/inition of $V_k$ and using Leibniz's rule we f\/ind for $k\in \mathbb{Z}$
\begin{gather*}
V_kp_{\substack{i_0,\dots,i_{n-1}\\j_0,\dots,j_{n-1}}}
\\
\qquad{}=\sum_{l=0}^{n-1}\left|
\begin{matrix}
\mu_{i_0-j_0}&\mu_{i_{0}-j_1}&\dots&\mu_{i_{0}-j_{n-1}}
\\
\vdots&\vdots&&\vdots
\\
\mu_{i_{l-1}-j_0}&\mu_{i_{l-1}-j_1}&\dots&\mu_{i_{l-1}-j_{n-1}}
\\
(i_l-j_0+k)\mu_{i_{l}-j_0+k}&(i_l-j_1+k)\mu_{i_{l}-j_1+k}&\dots&(i_l-j_{n-1}+k)\mu_{i_{l}-j_{n-1}+k}
\\
\mu_{i_{l-1}-j_0}&\mu_{i_{l-1}-j_1}&\dots&\mu_{i_{l-1}-j_{n-1}}
\\
\vdots&\vdots&&\vdots
\\
\mu_{i_{n-1}-j_0}&\mu_{i_{n-1}-j_1}&\dots&\mu_{i_{n-1}-j_{n-1}}
\end{matrix}
\right|,
\end{gather*}
or equivalently,
\begin{gather*}
V_kp_{\substack{i_0,\dots,i_{n-1}\\j_0,\dots,j_{n-1}}}
=\sum_{l=0}^{n-1}(i_l+k)p_{\substack{i_0,\dots,i_{l-1},i_l+k,i_{l+1},\dots,i_{n-1}\\j_0,\dots,j_{n-1}}}
-\sum_{l=1}^{n}\big|A_1^T\dots A_{l-1}^T\big(B(k)D\big)_l^T A_{l+1}
^T\dots A_n^T\big|.
\end{gather*}
Using Lemma \ref{Lemma version H.S. transposed}$(i)$ we obtain
\begin{gather*}
\begin{split}
& V_kp_{\substack{i_0,\dots,i_{n-1}\\j_0,\dots,j_{n-1}}}
=\sum_{l=0}^{n-1}(i_l+k)p_{\substack{i_0,\dots,i_{l-1},i_l+k,i_{l+1},\dots,i_{n-1}\\j_0,\dots,j_{n-1}}}
-\sum_{l=1}^{n}\big|A_1\dots A_{l-1}\big(B(k)D\big)_l A_{l+1}\dots A_n\big|
\\
& \phantom{V_kp_{\substack{i_0,\dots,i_{n-1}\\j_0,\dots,j_{n-1}}}}{}
=\sum_{l=0}^{n-1}(i_l+k)p_{\substack{i_0,\dots,i_{l-1},i_l+k,i_{l+1},\dots,i_{n-1}\\j_0,\dots,j_{n-1}}}
-\sum_{l=1}^{n}j_{l-1}\big|A_1\dots A_{l-1}\big(B(k)\big)_l A_{l+1}\dots A_n\big|.
\end{split}
\end{gather*}
This gives the f\/irst equality in~\eqref{equation master symmetries on Plucker coordinates}.
The second equality in~\eqref{equation master symmetries on Plucker coordinates} can be derived from the
f\/irst one by using Lemma~\ref{technical lemma}$(i)$.
\end{proof}

\subsection[Action of the Virasoro operators $L_k^{(n)}$ on the Schur polynomials]{Action of the Virasoro
operators $\boldsymbol{L_k^{(n)}}$ on the Schur polynomials}

Next we shall compute the action of the Virasoro operators on the products of Schur polynomials
$S_{i_{n-1}-(n-1),\dots,i_0}(t)S_{j_{n-1}-(n-1),\dots,j_0}(s)$.
We have the following lemma.
\begin{Lemma}
\label{Virasoro operators on the products of Schur polynomials}
\begin{gather}
(i)\quad L_0^{(n)}S_{i_{n-1}-(n-1),\dots,i_0}(t)S_{j_{n-1}-(n-1),\dots,j_0}(s)\nonumber
\\
\qquad\qquad
=\sum_{l=0}^{n-1}(i_l-j_l)S_{i_{n-1}-(n-1),\dots,i_0}(t)S_{j_{n-1}-(n-1),\dots,j_0}(s),\nonumber
\\
(ii)\quad L_1^{(n)}S_{i_{n-1}-(n-1),\dots,i_0}(t)S_{j_{n-1}-(n-1),\dots,j_0}(s)\nonumber
\\
\qquad\qquad
=\sum_{l=1}^{n}i_{n-l}S_{i_{n-1}-(n-1),\dots,i_{n-l}-(n-l)-1,\dots,i_0}(t)S_{j_{n-1}
-(n-1),\dots,j_0}(s)\nonumber
\\
\qquad\qquad\phantom{=}
{}-\sum_{l=1}^{n}(j_{n-l}+1)S_{i_{n-1}-(n-1),\dots,i_0}(t)S_{j_{n-1}-(n-1),\dots,j_{n-l}
-(n-l)+1,\dots,j_0}(s),\nonumber
\\
(iii)\quad L_2^{(n)}S_{i_{n-1}-(n-1),\dots,i_0}(t)S_{j_{n-1}-(n-1),\dots,j_0}(s)\nonumber
\\
\qquad\qquad
=\sum_{l=1}^{n}i_{n-l}S_{i_{n-1}-(n-1),\dots,i_{n-l}-(n-l)-2,\dots,i_0}(t)S_{j_{n-1}
-(n-1),\dots,j_0}(s)\nonumber
\\
\qquad\qquad\phantom{=}
{}+\sum_{1\leq k<l\leq n}S_{i_{n-1}-(n-1),\dots,i_{n-k}-(n-k)-1,\dots,i_{n-l}
-(n-l)-1,\dots,i_0}(t)S_{j_{n-1}-(n-1),\dots,j_0}(s)\nonumber
\\
\qquad\qquad\phantom{=}
{}-\sum_{l=1}^{n}(j_{n-l}+2)S_{i_{n-1}-(n-1),\dots,i_0}(t)S_{j_{n-1}-(n-1),\dots,j_{n-l}
-(n-l)+2,\dots,j_0}(s)\nonumber
\\
\qquad\qquad\phantom{=}
{}-\sum_{1\leq k<l\leq n}S_{i_{n-1}-(n-1),\dots,i_0}(t)S_{j_{n-1}-(n-1),\dots,j_{n-k}
-(n-k)+1,\dots,j_{n-l}-(n-l)+1,\dots,j_0}(s)\nonumber
\\
\qquad\qquad\phantom{=}
{}+s_{1}\sum_{l=1}^{n}S_{i_{n-1}-(n-1),\dots,i_0}(t)S_{j_{n-1}-(n-1),\dots,j_{n-l}
-(n-l)+1,\dots,j_0}(s)\nonumber
\\
\qquad\qquad\phantom{=}
{}-s_{1}\sum_{l=1}^{n}S_{i_{n-1}-(n-1),\dots,i_{n-l}-(n-l)-1,\dots,i_0}(t)S_{j_{n-1}
-(n-1),\dots,j_0}(s).\nonumber
\end{gather}
\end{Lemma}
\begin{proof}
By using Leibniz's rule and~\eqref{diff elementary Schur} we have for $j\geq 1$,
\begin{gather}
\frac{\partial}{\partial t_j}S_{i_{n-1}-(n-1),\dots,i_0}(t)=\sum_{l=1}^{n}S_{i_{n-1}-(n-1),\dots,i_{n-l}
-(n-l)-j,\dots,i_0}(t),
\label{first derivative}
\end{gather}
and
\begin{gather}
\frac{\partial^2}{\partial t_1^2}S_{i_{n-1}-(n-1),\dots,i_0}(t)
=\sum_{l=1}^{n}S_{i_{n-1}-(n-1),\dots,i_{n-l}-(n-l)-2,\dots,i_0}(t)
\nonumber
\\
\qquad
{}+2\sum_{1\leq r<s\leq n}S_{i_{n-1}-(n-1),\dots,i_{n-r}-(n-r)-1,\dots,i_{n-s}-(n-s)-1,\dots,i_0}(t).
\label{second derivative}
\end{gather}
Def\/ine the following $n\times n$ matrices
\begin{gather*}
A(t):=\left(
\begin{matrix}
S_{i_{n-1}-(n-1)}(t)&\dots&S_{i_{n-1}}(t)
\\
\vdots&&\vdots
\\
S_{i_0-(n-1)}(t)&\dots&S_{i_0}(t)
\end{matrix}
\right),
\\
B(j,t):=\left(
\begin{matrix}
S_{i_{n-1}-(n-1)-j}(t)&\dots&S_{i_{n-1}-j}(t)
\\
\vdots&&\vdots
\\
S_{i_0-(n-1)-j}(t)&\dots&S_{i_0-j}(t)
\end{matrix}
\right),
\end{gather*}
and $D=\diag(n-1,n-2,\dots,0)$.
We shall denote $\hat{A}(s)$ and $\hat{B}(j,s)$ the same matrices with $t\rightarrow s$ and
$(i_0,\dots,i_{n-1})\rightarrow(j_0,\dots,j_{n-1})$.
From the def\/inition~\eqref{Schur generating function} of the elementary Schur polynomials it follows
easily that for $j\geq 0$,
\begin{gather}
\sum_{k=1}^{\infty}kt_k\frac{\partial}{\partial t_{k+j}}S_i(t)=(i-j)S_{i-j}(t),\nonumber
\\
\sum_{k=j+1}^{\infty}kt_k\frac{\partial}{\partial t_{k-j}}S_i(t)=(i+j)S_{i+j}(t)-\sum_{1\leq l\leq j}
lt_lS_{i+j-l}(t).\nonumber
\end{gather}
Consequently, by f\/irst using Leibniz's rule and then Lemma~\ref{algebraic lemma H.S.}$(i)$ we have for
$j\geq 0$
\begin{gather}
\sum_{k=1}^{\infty}kt_k\frac{\partial}{\partial t_{k+j}}S_{i_{n-1}-(n-1),\dots,i_0}(t)=\sum_{l=1}^{n}
(i_{n-l}-j)S_{i_{n-1}-(n-1),\dots,i_{n-l}-(n-l)-j,\dots,i_0}(t)
\nonumber
\\
\hphantom{\sum_{k=1}^{\infty}kt_k\frac{\partial}{\partial t_{k+j}}S_{i_{n-1}-(n-1),\dots,i_0}(t)=}
{}-\big(\det A(t)\big)\tr\big(A(t)^{-1}B(j,t)D\big),
\label{M j}
\\
\sum_{k=j+1}^{\infty}kt_k\frac{\partial}{\partial t_{k-j}}S_{i_{n-1}-(n-1),\dots,i_0}(t)=\sum_{l=1}^{n}
(i_{n-l}+j)S_{i_{n-1}-(n-1),\dots,i_{n-l}-(n-l)+j,\dots,i_0}(t)
\nonumber
\\
\hphantom{\sum_{k=j+1}^{\infty}kt_k\frac{\partial}{\partial t_{k-j}}S_{i_{n-1}-(n-1),\dots,i_0}(t)=}{}
-\big(\det A(t)\big)\tr\big(A(t)^{-1}B(-j,t)D\big)
\nonumber
\\
\hphantom{\sum_{k=j+1}^{\infty}kt_k\frac{\partial}{\partial t_{k-j}}S_{i_{n-1}-(n-1),\dots,i_0}(t)=}{}
-\sum_{m=1}^{j}mt_m\big(\det A(t)\big)\tr\big(A(t)^{-1}B(m-j,t)\big).
\label{M -j}
\end{gather}
We are now ready to prove the lemma.

$(i)$ From~\eqref{L Virasoro in terms of AB}, we have $L_0^{(n)}=A_0^{(n)}-B_{0}^{(n)}$.
Using~\eqref{M j} with $j=0$, we obtain
\begin{gather*}
A_0^{(n)}S_{i_{n-1}-(n-1),\dots,i_0}(t)
=\sum_{k=1}^{\infty}kt_k\frac{\partial}{\partial t_k}S_{i_{n-1}-(n-1),\dots,i_0}(t)+\frac{n^2}{2}S_{i_{n-1}
-(n-1),\dots,i_0}(t)
\\
\qquad
=\sum_{l=1}^{n}i_{n-l}S_{i_{n-1}-(n-1),\dots,i_0}(t)-\big(\!\det A(t)\big)\tr\big(A(t)^{-1}
B(0,t)D\big)+\frac{n^2}{2}S_{i_{n-1}-(n-1),\dots,i_0}(t).
\end{gather*}
We have $B(0,t)=A(t)$, and thus
\begin{gather*}
\big(\det A(t)\big)\tr\big(A(t)^{-1}B(0,t)D\big)=\big(\det A(t)\big)\tr(D)=\frac{n(n-1)}{2}S_{i_{n-1}
-(n-1),\dots,i_0}(t).
\end{gather*}
Consequently, we get
\begin{gather*}
A_0^{(n)}S_{i_{n-1}-(n-1),\dots,i_0}(t)=\Bigg[\sum_{l=1}^{n}i_{n-l}+\frac{n}{2}\Bigg]S_{i_{n-1}
-(n-1),\dots,i_0}(t).
\end{gather*}
Similarly, we get
\begin{gather*}
B_0^{(n)}S_{j_{n-1}-(n-1),\dots,j_0}(s)=\left[\sum_{l=1}^{n}j_{n-l}+\frac{n}{2}\right]S_{j_{n-1}
-(n-1),\dots,j_0}(s).
\end{gather*}
Combining both equations, we obtain $(i)$.

$(ii)$ From~\eqref{L Virasoro in terms of AB}, we have $L_1^{(n)}=A_1^{(n)}-B_{-1}^{(n)}$.
We compute, using~\eqref{first derivative} and~\eqref{M j}
\begin{gather*}
A_1^{(n)}S_{i_{n-1}-(n-1),\dots,i_0}(t)=\Bigg[\sum_{j=1}^{\infty}jt_{j}\frac{\partial}{\partial t_{j+1}}
+n\frac{\partial}{\partial t_1}\Bigg]S_{i_{n-1}-(n-1),\dots,i_0}(t)
\\
\qquad{}\!
=\sum_{l=1}^{n}\big(i_{n-l}+n-1\big)S_{i_{n-1}-(n-1),\dots,i_{n-l}-(n-l)-1,\dots,i_0}
(t)-\!\big(\!\det A(t)\big)\tr\!\big(A(t)^{-1}B(1,t)D\big).
\end{gather*}
By virtue of Lemma~\ref{algebraic lemma H.S.}$(i)$, we have
\begin{gather*}
\big(\det A(t)\big)\tr\big(A(t)^{-1}B(1,t)D\big)=(n-1)\big|\big(B(1,t)\big)_1A_2(t)\dots A_n(t)\big|.
\end{gather*}
But by virtue of Lemma~\ref{determinants of Schur polynomials}$(i)$, this gives
\begin{gather*}
\big(\det A(t)\big)\tr\big(A(t)^{-1}B(1,t)D\big)=(n-1)\sum_{l=1}^{n}S_{i_{n-1}-(n-1),\dots,i_{n-l}
-(n-l)-1,\dots,i_0}(t).
\end{gather*}
Hence, we obtain
\begin{gather}
A_1^{(n)}S_{i_{n-1}-(n-1),\dots,i_0}(t)=\sum_{l=1}^{n}i_{n-l}S_{i_{n-1}-(n-1),\dots,i_{n-l}
-(n-l)-1,\dots,i_0}(t).
\label{A1 S}
\end{gather}
Similarly, we have using~\eqref{M -j}
\begin{gather*}
B_{-1}^{(n)}S_{j_{n-1}-(n-1),\dots,j_0}(s)=\left[\sum_{j=2}^{\infty}js_{j}\frac{\partial}{\partial s_{j-1}}
+ns_1\right]S_{j_{n-1}-(n-1),\dots,j_0}(s)
\\
\qquad{}
=\sum_{l=1}^{n}(j_{n-l}+1)S_{j_{n-1}-(n-1),\dots,j_{n-l}-(n-l)+1,\dots,j_0}(s)-\big(\det\hat{A}
(s)\big)\tr\big(\hat{A}(s)^{-1}\hat{B}(-1,s)D\big)
\\
\qquad
\phantom{=}
{}-s_1\big(\det\hat{A}(s)\big)\tr\big(\hat{A}(s)^{-1}\hat{B}(0,s)\big)+ns_1S_{j_{n-1}
-(n-1),\dots,j_0}(s).
\end{gather*}
We have using Lemma~\ref{algebraic lemma H.S.}$(i)$
\begin{gather}
\big(\det\hat{A}(s)\big)\tr\big(\hat{A}(s)^{-1}\hat{B}(-1,s)D\big)=0,\nonumber
\end{gather}
and, obviously, we also have
\begin{gather}
\big(\det\hat{A}(s)\big)\tr\big(\hat{A}(s)^{-1}\hat{B}(0,s)\big)=nS_{j_{n-1}-(n-1),\dots,j_0}(s).\nonumber
\end{gather}
Consequently we obtain
\begin{gather}
\label{B-1 S}
B_{-1}^{(n)}S_{j_{n-1}-(n-1),\dots,j_0}(s)=\sum_{l=1}^{n}(j_{n-l}+1)S_{j_{n-1}-(n-1),\dots,j_{n-l}
-(n-l)+1,\dots,j_0}(s).
\end{gather}
Subtracting~\eqref{A1 S} and~\eqref{B-1 S} gives $(ii)$.

$(iii)$ From~\eqref{L Virasoro in terms of AB}, we have
\begin{gather*}
L_2^{(n)}=A_2^{(n)}-B_{-2}^{(n)}+\frac{1}{2}\left(\frac{\partial}{\partial t_1}-s_1\right)^2.
\end{gather*}
We study separately the contributions of the three terms in the operator $L_2^{(n)}$ on the product of
Schur functions.
We start with the contribution of $A_2^{(n)}$.
We compute, using~\eqref{first derivative},~\eqref{second derivative} and~\eqref{M j}
\begin{gather*}
A_2^{(n)}S_{i_{n-1}-(n-1),\dots,i_0}(t)
=\left[\frac{1}{2}\frac{\partial^2}{\partial t_1^2}+\sum_{j=1}^{\infty}jt_{j}\frac{\partial}
{\partial t_{j+2}}+n\frac{\partial}{\partial t_2}\right]S_{i_{n-1}-(n-1),\dots,i_0}(t)
\\
\hphantom{A_2^{(n)}S_{i_{n-1}-(n-1),\dots,i_0}(t)}{}
=\sum_{l=1}^{n}\left(i_{n-l}+n-\frac{3}{2}\right)S_{i_{n-1}-(n-1),\dots,i_{n-l}-(n-l)-2,\dots,i_0}(t)
\\
\hphantom{A_2^{(n)}S_{i_{n-1}-(n-1),\dots,i_0}(t)=}{}
+\sum_{1\leq k<l\leq n}S_{i_{n-1}-(n-1),\dots,i_{n-k}-(n-k)-1,\dots,i_{n-l}-(n-l)-1,\dots,i_0}(t)
\\
\hphantom{A_2^{(n)}S_{i_{n-1}-(n-1),\dots,i_0}(t)=}{}
-\big(\det A(t)\big)\tr\big(A(t)^{-1}B(2,t)D\big).
\end{gather*}
The last term in this equation gives by developing the trace
\begin{gather*}
\big(\det A(t)\big)\tr\big(A(t)^{-1}B(2,t)D\big)
\\
\qquad
=\big(\det A(t)\big)\Big[(n-1)\big(A(t)^{-1}B(2,t)\big)_{11}+(n-2)\big(A(t)^{-1}B(2,t)\big)_{22}\Big].
\end{gather*}
We have
\begin{gather*}
\big(\det A(t)\big)\tr\big(A(t)^{-1}B(2,t)\big)=\big(\det A(t)\big)\big[\big(A(t)^{-1}B(2,t)\big)_{11}
+\big(A(t)^{-1}B(2,t)\big)_{22}\big],
\end{gather*}
and by a~short computation
\begin{gather*}
\big(A(t)^{-1}B(2,t)\big)_{22}=-\sum_{1\leq k<l\leq n}\big(\big(A(t)^{-1}B(1,t)\big)_k\wedge\big(A(t)^{-1}
B(1,t)\big)_l\big)_{kl}.
\end{gather*}
Consequently we have
\begin{gather*}
\big(\det A(t)\big)\tr\big(A(t)^{-1}B(2,t)D\big)=(n-1)\big(\det A(t)\big)\tr\big(A(t)^{-1}B(2,t)\big)
\\
\qquad
{}+\big(\det A(t)\big)\sum_{1\leq k<l\leq n}\Big(\big(A(t)^{-1}B(1,t)\big)_k\wedge\big(A(t)^{-1}
B(1,t)\big)_l\Big)_{kl}.
\end{gather*}
Using Lemma~\ref{algebraic lemma H.S.}, we obtain
\begin{gather*}
\big(\det A(t)\big)\tr\big(A(t)^{-1}B(2,t)D\big)=(n-1)\sum_{l=1}^{n}S_{i_{n-1}-(n-1),\dots,i_{n-l}
-(n-l)-2,\dots,i_0}(t)
\\
\qquad
+\sum_{1\leq k<l\leq n}S_{i_{n-1}-(n-1),\dots,i_{n-k}-(n-k)-1,\dots,i_{n-l}-(n-l)-1,\dots,i_0}(t).
\end{gather*}
Hence, we get
\begin{gather}
\label{A2 S}
A_2^{(n)}S_{i_{n-1}-(n-1),\dots,i_0}(t)=\sum_{l=1}^{n}\left(i_{n-l}-\frac{1}{2}\right)S_{i_{n-1}
-(n-1),\dots,i_{n-l}-(n-l)-2,\dots,i_0}(t).
\end{gather}

We now turn to the contribution of $B_{-2}^{(n)}$.
We have using~\eqref{M -j}
\begin{gather*}
B_{-2}^{(n)}S_{j_{n-1}-(n-1),\dots,j_0}(s)
=\left[\frac{1}{2}s_1^2+\sum_{j=3}^{\infty}js_{j}\frac{\partial}{\partial s_{j-2}}+2ns_2\right]S_{j_{n-1}
-(n-1),\dots,j_0}(s)
\\
\qquad
=\left[\frac{1}{2}s_1^2+2ns_2\right]S_{j_{n-1}-(n-1),\dots,j_0}(s)+\sum_{l=1}^{n}\big(j_{n-l}+2\big)S_{j_{n-1}
-(n-1),\dots,j_{n-l}-(n-l)+2,\dots,j_0}(s)
\\
\qquad
\phantom{=}
{}-\big(\det\hat{A}(s)\big)\tr\big(\hat{A}(s)^{-1}\hat{B}(-2,s)D\big)-\sum_{m=1}^{2}
ms_m\big(\det\hat{A}(s)\big)\tr\big(\hat{A}(s)^{-1}\hat{B}(m-2,s)\big).
\end{gather*}
By a~similar argument as above, we have
\begin{gather*}
\big(\det\hat{A}(s)\big)\tr\big(\hat{A}(s)^{-1}\hat{B}(-2,s)D\big)
\\
\qquad
=-\big(\det\hat{A}(s)\big)\sum_{1\leq k<l\leq n}\big(\big(\hat{A}(s)^{-1}\hat{B}
(-1,s)\big)_k\wedge\big(\hat{A}(s)^{-1}\hat{B}(-1,s)\big)_l\big)_{kl},
\end{gather*}
and thus using Lemma~\ref{algebraic lemma H.S.}$(ii)$, we obtain
\begin{gather*}
\big(\det\hat{A}(s)\big)\tr\big(\hat{A}(s)^{-1}\hat{B}(-2,s)D\big)
\\
\qquad
=-\sum_{1\leq k<l\leq n}S_{j_{n-1}-(n-1),\dots,j_{n-k}-(n-k)+1,\dots,j_{n-l}-(n-l)+1,\dots,j_0}(s).
\end{gather*}
We also have, using Lemma~\ref{algebraic lemma H.S.}$(i)$,
\begin{gather*}
\sum_{m=1}^{2}ms_m\big(\det\hat{A}(s)\big)\tr\big(\hat{A}(s)^{-1}\hat{B}(m-2,s)\big)
\\
\qquad
=s_1\sum_{l=1}^{n}S_{j_{n-1}-(n-1),\dots,j_{n-l}-(n-l)+1,\dots,j_0}(s)+2ns_2S_{j_{n-1}-(n-1),\dots,j_0}(s).
\end{gather*}
Consequently, we have
\begin{gather}
B_{-2}^{(n)}S_{j_{n-1}-(n-1),\dots,j_0}(s)
=\sum_{l=1}^{n}\big(j_{n-l}+2\big)S_{j_{n-1}-(n-1),\dots,j_{n-l}-(n-l)+2,\dots,j_0}(s)
\nonumber
\\
\qquad
{}+\sum_{1\leq k<l\leq n}S_{j_{n-1}-(n-1),\dots,j_{n-k}-(n-k)+1,\dots,j_{n-l}-(n-l)+1,\dots,j_0}(s)
\nonumber
\\
\qquad
{}-s_1\sum_{l=1}^{n}S_{j_{n-1}-(n-1),\dots,j_{n-l}-(n-l)+1,\dots,j_0}(s)+\frac{1}{2}s_1^2S_{j_{n-1}-(n-1),\dots,j_0}(s).
\label{B-2 S}
\end{gather}

Finally, we turn to the contribution of the term $\frac{1}{2}\big(\frac{\partial}{\partial t_1}-s_1\big)^2$.
We have using~\eqref{first derivative} and~\eqref{second derivative}
\begin{gather}
\frac{1}{2}\left[\frac{\partial}{\partial t_1}-s_1\right]^2S_{i_{n-1}-(n-1),\dots,i_0}(t)\nonumber
=\frac{1}{2}\left[\frac{\partial^2}{\partial t_1^2}-2s_1\frac{\partial}{\partial t_1}+s_1^2\right]S_{i_{n-1}
-(n-1),\dots,i_0}(t)\nonumber
\\
\qquad{}
=\frac{1}{2}\sum_{l=1}^{n}S_{i_{n-1}-(n-1),\dots,i_{n-l}-(n-l)-2,\dots,i_0}(t)\nonumber
\\
\qquad
\phantom{=}
{}+\sum_{1\leq k<l\leq n}S_{i_{n-1}-(n-1),\dots,i_{n-k}-(n-k)-1,\dots,i_{n-l}-(n-l)-1,\dots,i_0}(t)\nonumber
\\
\qquad
\phantom{=}
{}-s_1\sum_{l=1}^{n}S_{i_{n-1}-(n-1),\dots,i_{n-l}-(n-l)-1,\dots,i_0}(t)+\frac{1}{2}
s_1^2S_{i_{n-1}-(n-1),\dots,i_0}(t).
\label{last contribution L2}
\end{gather}

Combining~\eqref{A2 S},~\eqref{B-2 S} and~\eqref{last contribution L2}, we obtain $(iii)$.
\end{proof}
\begin{Remark}
We observe that by def\/inition of the operators $L_k^{(n)}$ we have
\begin{gather*}
L_{-k}^{(n)}S_{i_{n-1}-(n-1),\dots,i_0}(t)S_{j_{n-1}-(n-1),\dots,j_0}(s)\nonumber
\\
\qquad
=-L_{k}^{(n)}S_{i_{n-1}-(n-1),\dots,i_0}(t)S_{j_{n-1}-(n-1),\dots,j_0}(s)
\Big|_{\substack{t\leftrightarrow s\\(i_0,\dots,i_{n-1})
\leftrightarrow(j_0,\dots,j_{n-1})}}.
\end{gather*}
\end{Remark}

\subsection{Proof of the main theorem}

We now turn to the last part of this section.
We will prove Theorem~\ref{Virasoro constraints theorem}.
We f\/irst prove the following lemma.
\begin{Lemma}\label{corollary technical lemma}
\begin{gather}
\sum_{l=1}^{n}\sum_{\substack{0\leq i_0<\dots<i_{n-1}\\0\leq j_0<\dots<j_{n-1}}}
p_{\substack{i_0,\dots,i_{n-1}\\j_0,\dots,j_{n-1}}}
S_{i_{n-1}-(n-1),\dots,i_0}(t)S_{j_{n-1}-(n-1),\dots,j_{n-l}-(n-l)+1,\dots,j_0}(s)\nonumber
\\
\qquad\quad{}
+\sum_{\substack{0\leq i_0<\dots<i_{n-1}\\0<j_1<\dots<j_{n-1}}}
p_{\substack{i_0,\dots,i_{n-1}\\-1,j_1,\dots,j_{n-1}}}
S_{i_{n-1}-(n-1),\dots,i_0}(t)S_{j_{n-1}-(n-1),\dots,j_{1}-1,0}(s)\nonumber
\\
\qquad
=\sum_{l=1}^{n}\sum_{\substack{0\leq i_0<\dots<i_{n-1}\\0\leq j_0<\dots<j_{n-1}}}
p_{\substack{i_0,\dots,i_{n-1}\\j_0,\dots,j_{n-1}}}
S_{i_{n-1}-(n-1),\dots,i_{n-l}-(n-l)-1,\dots,i_0}(t)S_{j_{n-1}-(n-1),\dots,j_0}(s).\!\!\!
\label{equation lemma main theorem}
\end{gather}
\end{Lemma}
\begin{proof}
For simplicity, we will use the notations
\begin{gather}
\label{simple notations}
\mathcal{S}_i(t)=S_{i_{n-1}-(n-1),\dots,i_0}(t),
\qquad
\mathcal{S}_j(s)=S_{j_{n-1}-(n-1),\dots,j_0}(s),
\end{gather}
when no 'special' shift on the indices of the Schur functions occur.
Relabeling each term in the f\/irst sum of the left-hand side of~\eqref{equation lemma main theorem} in the
following way $j_{n-l}\mapsto j_{n-l}-1$ gives
\begin{gather*}
\sum_{l=1}^{n}\sum_{\substack{0\leq i_0<\dots<i_{n-1}\\0\leq j_0<\dots<j_{n-1}}}
p_{\substack{i_0,\dots,i_{n-1}\\j_0,\dots,j_{n-1}}}
\mathcal{S}_i(t)S_{j_{n-1}-(n-1),\dots,j_{n-l}-(n-l)+1,\dots,j_0}(s)
\\
\qquad
=\sum_{l=1}^{n}\sum_{\substack{0\leq i_0<\dots<i_{n-1}\\0\leq j_0<\dots<j_{n-l}-1<\dots<j_{n-1}}}
p_{\substack{i_0,\dots,i_{n-1}\\j_0,\dots,j_{n-l}-1,\dots,j_{n-1}}}
\mathcal{S}_i(t)\mathcal{S}_j(s).
\end{gather*}
On the one hand, for a~f\/ixed $1\leq l\leq n-1$, if $j_{n-l}=j_{n-l-1}+1$, then
$p_{\substack{i_0,\dots,i_{n-1}\\j_0,\dots,j_{n-l}-1,\dots,j_{n-1}}}=0$.
On the other hand, for a~f\/ixed $2\leq l\leq n$, if $j_{n-l}=j_{n-l+1}$, then
$S_{j_{n-1}-(n-1),\dots,j_{n-l}-(n-l),\dots,j_0}(s)=0$.
Therefore
\begin{gather*}
\sum_{l=1}^{n}\sum_{\substack{0\leq i_0<\dots<i_{n-1}\\0\leq j_0<\dots<j_{n-1}}}
p_{\substack{i_0,\dots,i_{n-1}\\j_0,\dots,j_{n-1}}}
\mathcal{S}_i(t)S_{j_{n-1}-(n-1),\dots,j_{n-l}-(n-l)+1,\dots,j_0}(s)
\\
\qquad{}
=\sum_{l=1}^{n}\sum_{\substack{0\leq i_0<\dots<i_{n-1}\\0\leq j_0<\dots<j_{n-1}}}
p_{\substack{i_0,\dots,i_{n-1}\\j_0,\dots,j_{n-l}-1,\dots,j_{n-1}}}
\mathcal{S}_i(t)\mathcal{S}_j(s)
\\
\qquad
\phantom{=}
{}-\sum_{\substack{0\leq i_0<\dots<i_{n-1}\\0<j_1<\dots<j_{n-1}}}
p_{\substack{i_0,\dots,i_{n-1}\\-1,j_1,\dots,j_{n-1}}}
\mathcal{S}_i(t)S_{j_{n-1}-(n-1),\dots,j_1-1,0}(s).
\end{gather*}
Consequently, the left-hand side of~\eqref{equation lemma main theorem} is equal to
\begin{gather}
\sum_{l=1}^{n}\sum_{\substack{0\leq i_0<\dots<i_{n-1}\\0\leq j_0<\dots<j_{n-1}}}
p_{\substack{i_0,\dots,i_{n-1}\\j_0,\dots,j_{n-1}}}
\mathcal{S}_i(t)S_{j_{n-1}-(n-1),\dots,j_{n-l}-(n-l)+1,\dots,j_0}(s)\nonumber
\\
\qquad\phantom{=}
{}+\sum_{\substack{0\leq i_0<\dots<i_{n-1}\\0<j_1<\dots<j_{n-1}}}
p_{\substack{i_0,\dots,i_{n-1}\\-1,j_1,\dots,j_{n-1}}}
\mathcal{S}_i(t)S_{j_{n-1}-(n-1),\dots,j_{1}-1,0}(s)\nonumber
\\
\qquad{}
=\sum_{l=1}^{n}\sum_{\substack{0\leq i_0<\dots<i_{n-1}\\0\leq j_0<\dots<j_{n-1}}}
p_{\substack{i_0,\dots,i_{n-1}\\j_0,\dots,j_{n-l}-1,\dots,j_{n-1}}}
\mathcal{S}_i(t)\mathcal{S}_j(s).
\label{left hand side}
\end{gather}
Similarly, one can show that the right-hand side of~\eqref{equation lemma main theorem} is equal to
\begin{gather}
\sum_{l=1}^{n}\sum_{\substack{0\leq i_0<\dots<i_{n-1}\\0\leq j_0<\dots<j_{n-1}}}
p_{\substack{i_0,\dots,i_{n-1}\\j_0,\dots,j_{n-1}}}
S_{i_{n-1}-(n-1),\dots,i_{n-l}-(n-l)-1,\dots,i_0}(t)\mathcal{S}_j(s)\nonumber
\\
\qquad
=\sum_{l=1}^{n}\sum_{\substack{0\leq i_0<\dots<i_{n-1}\\0\leq j_0<\dots<j_{n-1}}}
p_{\substack{i_0,\dots,i_{n-l}+1,\dots,i_{n-1}\\j_0,\dots,j_{n-1}}}
\mathcal{S}_i(t)\mathcal{S}_j(s).
\label{right hand side}
\end{gather}
By virtue of Lemma~\ref{technical lemma}$(i)$,~\eqref{left hand side} and~\eqref{right hand side} are equal.
\end{proof}
\begin{proof}
[Proof of Theorem~\ref{Virasoro constraints theorem}] We will prove the theorem for $k\geq 0$.
The case $k<0$ is similar.
Using the Pl\"ucker expansion~\eqref{tau Plucker proof} of $\tau_n(t)$, and Lemmas~\ref{master symmetries
on Plucker coordinates} and~\ref{Virasoro operators on the products of Schur polynomials} we have for
$k=0,1$, using the notations~\eqref{simple notations},
\begin{gather*}
V_k\tau_n(s,t)=\sum_{\substack{0\leq i_0<\dots<i_{n-1}\\0\leq j_0<\dots<j_{n-1}}}
V_k p_{\substack{i_0,\dots,i_{n-1}\\j_0,\dots,j_{n-1}}}
\mathcal{S}_i(t)\mathcal{S}_j(s)
\\
\phantom{V_k\tau_n(s,t)}{}
=\sum_{\substack{0\leq i_0<\dots<i_{n-1}\\0\leq j_0<\dots<j_{n-1}}}
p_{\substack{i_0,\dots,i_{n-1}\\j_0,\dots,j_{n-1}}}
L_k^{(n)}\mathcal{S}_i(t)\mathcal{S}_j(s)
=L_k^{(n)}\tau_n(s,t),
\end{gather*}
where, in the second equality, we have performed some relabeling of the indices as in the proof of
Lemma~\ref{corollary technical lemma}.
We will f\/inish the proof with the case $k=2$, for which we provide some more details, but f\/irst we
prove the theorem for general $k\geq 3$.
We proceed by induction.
Assume the theorem holds for some $k\geq 2$.
We will establish it for $k+1$.
The argument follows from the commutation relations~\eqref{virc} and~\eqref{commutation master sym}.
We have
\begin{gather*}
(k-1)V_{k+1}\tau_{n}(s,t)=\sum_{\substack{0\leq i_0<\dots<i_{n-1}\\0\leq j_0<\dots<j_{n-1}}}
[V_1,V_k]p_{\substack{i_0,\dots,i_{n-1}\\j_0,\dots,j_{n-1}}}
\mathcal{S}_i(t)\mathcal{S}_j(s)
\\
\hphantom{(k-1)V_{k+1}\tau_{n}(s,t)}{}
=\sum_{\substack{0\leq i_0<\dots<i_{n-1}\\0\leq j_0<\dots<j_{n-1}}}
p_{\substack{i_0,\dots,i_{n-1}\\j_0,\dots,j_{n-1}}}
\big[L_k^{(n)},L_1^{(n)}\big]\mathcal{S}_i(t)\mathcal{S}_j(s)
=(k-1)L_{k+1}^{(n)}\tau_{n}(s,t),
\end{gather*}
where in the second equality we have used the induction hypothesis.

We now provide some details for the case $k=2$.
Using Lemmas~\ref{Virasoro operators on the products of Schur polynomials} and~\ref{corollary technical lemma} we have
\begin{gather}
L_2^{(n)}\tau_n(s,t)=T_1+T_2+T_3+T_4\nonumber
\\
\phantom{L_2^{(n)}\tau_n(s,t)=}
{}-s_1\sum_{\substack{0\leq i_0<\dots<i_{n-1}\\0<j_1<\dots<j_{n-1}}}
p_{\substack{i_0,\dots,i_{n-1}\\-1,j_1,\dots,j_{n-1}}}
\mathcal{S}_i(t)S_{j_{n-1}-(n-1),\dots,j_1-1,0}(s),
\label{main proof equation 1}
\end{gather}
with
\begin{gather*}
T_1:=\sum_{\substack{0\leq i_0<\dots<i_{n-1}\\0\leq j_0<\dots<j_{n-1}}}
p_{\substack{i_0,\dots,i_{n-1}\\j_0,\dots,j_{n-1}}}
\sum_{l=1}^{n}i_{n-l}S_{i_{n-1}-(n-1),\dots,i_{n-l}-(n-l)-2,\dots,i_0}(t)\mathcal{S}
_j(s),
\\[-1ex]
T_2:=-\sum_{\substack{0\leq i_0<\dots<i_{n-1}\\0\leq j_0<\dots<j_{n-1}}}
p_{\substack{i_0,\dots,i_{n-1}\\j_0,\dots,j_{n-1}}}
\sum_{l=1}^{n}(j_{n-l}+2)\mathcal{S}_i(t)S_{j_{n-1}-(n-1),\dots,j_{n-l}
-(n-l)+2,\dots,j_0}(s),
\\[-1ex]
T_3:=\sum_{\substack{0\leq i_0<\dots<i_{n-1}\\0\leq j_0<\dots<j_{n-1}}}
p_{\substack{i_0,\dots,i_{n-1}\\j_0,\dots,j_{n-1}}}
\sum_{1\leq k<l\leq n}S_{i_{n-1}-(n-1),\dots,i_{n-k}-(n-k)-1,\dots,i_{n-l}
-(n-l)-1,\dots,i_0}(t)\mathcal{S}_j(s),
\\[-1ex]
T_4:=-\sum_{\substack{0\leq i_0<\dots<i_{n-1}\\0\leq j_0<\dots<j_{n-1}}}
p_{\substack{i_0,\dots,i_{n-1}\\j_0,\dots,j_{n-1}}}
\\[-1ex]
\phantom{T_4:=}{}
\times\sum_{1\leq k<l\leq n}\mathcal{S}_i(t)S_{j_{n-1}-(n-1),\dots,j_{n-k}
-(n-k)+1,\dots,j_{n-l}-(n-l)+1,\dots,j_0}(s).
\end{gather*}
We will consider separately the four terms $T_1$, $T_2$, $T_3$, $T_4$.
By arguments similar to those used in the proof of Lemma~\ref{corollary technical lemma}, and using the
fact that $S_{i_{n-1}-(n-1),\dots,i_0}(t)=0$ if $i_k<0$ for some $0\leq k\leq n-1$, we get for~$T_1$
\begin{gather*}
T_1=\sum_{l=1}^{n}\sum_{\substack{0\leq i_0<\dots<i_{n-1}\\0\leq j_0<\dots<j_{n-1}}}(i_{n-l}+2)
p_{\substack{i_0,\dots,i_{n-l}+2,\dots,i_{n-1}\\j_0,\dots,j_{n-1}}}
\mathcal{S}_i(t)\mathcal{S}_j(s)
\\[-1ex]
\phantom{T_1=}
{}+\sum_{l=1}^{n-1}
\sum_{\substack{-1\leq i_0-1<\dots<i_{n-l-1}-1\\=i_{n-l}<\dots<i_{n-1}\\0\leq j_0<\dots<j_{n-1}}}(i_{n-l}+2)
p_{\substack{i_0,\dots,i_{n-l}+2,\dots,i_{n-1}\\j_0,\dots,j_{n-1}}}
\mathcal{S}_i(t)\mathcal{S}_j(s)
\\[-1ex]
\phantom{T_1=}
{}-\sum_{l=2}^{n}
\sum_{\substack{-1\leq i_0-1<\dots<i_{n-l-1}-1\\<i_{n-l}+1=i_{n-l+1}<\dots<i_{n-1}\\0\leq j_0<\dots<j_{n-1}}}(i_{n-l}+2)
p_{\substack{i_0,\dots,i_{n-l}+2,\dots,i_{n-1}\\j_0,\dots,j_{n-1}}}
\mathcal{S}_i(t)\mathcal{S}_j(s).
\end{gather*}
The two last terms in this expression annihilate, i.e.
\begin{gather}
0=\sum_{l=1}^{n-1}
\sum_{\substack{-1\leq i_0-1<\dots<i_{n-l-1}-1\\=i_{n-l}<\dots<i_{n-1}\\0\leq j_0<\dots<j_{n-1}}}(i_{n-l}+2)
p_{\substack{i_0,\dots,i_{n-l}+2,\dots,i_{n-1}\\j_0,\dots,j_{n-1}}}
\mathcal{S}_i(t)\mathcal{S}_j(s)\nonumber
\\[-1ex]
\phantom{0=}
{}-\sum_{l=2}^{n}
\sum_{\substack{-1\leq i_0-1<\dots<i_{n-l-1}-1\\<i_{n-l}+1=i_{n-l+1}<\dots<i_{n-1}\\0\leq j_0<\dots<j_{n-1}}}(i_{n-l}+2)
p_{\substack{i_0,\dots,i_{n-l}+2,\dots,i_{n-1}\\j_0,\dots,j_{n-1}}}
\mathcal{S}_i(t)\mathcal{S}_j(s).
\label{last two terms annihilate}
\end{gather}
Indeed, we have for $1\leq l\leq n-1$
\begin{gather*}
\sum_{\substack{-1\leq i_0-1<\dots<i_{n-l-1}-1\\=i_{n-l}<\dots<i_{n-1}\\0\leq j_0<\dots<j_{n-1}}}(i_{n-l}+2)
p_{\substack{i_0,\dots,i_{n-l}+2,\dots,i_{n-1}\\j_0,\dots,j_{n-1}}}
\mathcal{S}_i(t)\mathcal{S}_j(s)
\\[-1ex]
\qquad
=\sum_{\substack{-1\leq k_0-1<\dots<k_{n-l-1}\\=k_{n-l}-1<\dots<k_{n-1}\\0\leq j_0<\dots<j_{n-1}}}(k_{n-l-1}+2)
p_{\substack{k_0,\dots,k_{n-l-2},k_{n-l},k_{n-l-1}+2,k_{n-l+1},\dots,k_{n-1}\\j_0,\dots,j_{n-1}}}
\\[-1ex]
\qquad
\phantom{=}
\times
S_{k_{n-1}-(n-1),\dots,k_{n-l+1}-(n-l+1),k_{n-l-1}-(n-l),k_{n-l}-(n-l-1),k_{n-l-2}-(n-l-2),\dots,k_0}
(t)\mathcal{S}_j(s),
\end{gather*}
where we have made the relabeling $i_{n-l-1}\mapsto k_{n-l}$, $i_{n-l}\mapsto k_{n-l-1}$, and $i_m\mapsto
k_m$ if $m\neq n-l-1, n-l$.
As the Pl\"ucker coordinates and the Schur functions are determinants, we have, permuting lines in the
determinants,
\begin{gather*}
p_{\substack{k_0,\dots,k_{n-l-2},k_{n-l},k_{n-l-1}+2,k_{n-l+1},\dots,k_{n-1}\\j_0,\dots,j_{n-1}}}
=-p_{\substack{k_0,\dots,k_{n-l-1}+2,\dots,k_{n-1}\\j_0,\dots,j_{n-1}}},
\end{gather*}
and
\begin{gather*}
S_{k_{n-1}-(n-1),\dots,k_{n-l+1}-(n-l+1),k_{n-l-1}-(n-l),k_{n-l}-(n-l-1),k_{n-l-2}-(n-l-2),\dots,k_0}
(t)=-\mathcal{S}_k(t),
\end{gather*}
and hence
\begin{gather*}
\sum_{\substack{-1\leq i_0-1<\dots<i_{n-l-1}-1=i_{n-l}<\dots<i_{n-1}\\0\leq j_0<\dots<j_{n-1}}}(i_{n-l}+2)
p_{\substack{i_0,\dots,i_{n-l}+2,\dots,i_{n-1}\\j_0,\dots,j_{n-1}}}
\mathcal{S}_i(t)\mathcal{S}_j(s)
\\
\qquad{} =\sum_{\substack{-1\leq k_0-1<\dots<k_{n-l-1}=k_{n-l}-1<\dots<k_{n-1}\\0\leq j_0<\dots<j_{n-1}}}(k_{n-l-1}+2)
p_{\substack{k_0,\dots,k_{n-l-1}+2,\dots,k_{n-1}\\j_0,\dots,j_{n-1}}}
\mathcal{S}_k(t)\mathcal{S}_j(s).
\end{gather*}
Summing this expression for $1\leq l\leq n-1$, and relabeling $l\mapsto l-1$ we get~\eqref{last two terms
annihilate}.
Consequently we obtain
\begin{gather}
T_1=\sum_{l=1}^{n}\sum_{\substack{0\leq i_0<\dots<i_{n-1}\\0\leq j_0<\dots<j_{n-1}}}(i_{n-l}+2)
p_{\substack{i_0,\dots,i_{n-l}+2,\dots,i_{n-1}\\j_0,\dots,j_{n-1}}}
\mathcal{S}_i(t)\mathcal{S}_j(s).
\label{Term 1}
\end{gather}
By similar arguments, we have
\begin{gather}
T_2=-\sum_{l=1}^{n}\sum_{\substack{0\leq i_0<\dots<i_{n-1}\\0\leq j_0<\dots<j_{n-1}}}j_{n-l}
p_{\substack{i_0,\dots,i_{n-1}\\j_0,\dots,j_{n-l}-2,\dots,j_{n-1}}}
\mathcal{S}_i(t)\mathcal{S}_j(s)\nonumber
\\
\phantom{T_2=}
{}+\sum_{\substack{0\leq i_0<\dots<i_{n-1}\\0\leq j_1<\dots<j_{n-1}}}
p_{\substack{i_0,\dots,i_{n-1}\\-1,j_1,\dots,j_{n-1}}}\mathcal{S}_i(t)S_{j_{n-1}-(n-1),\dots,j_1-1,1}(s),
\label{Term 2}
\\
T_3=\sum_{1\leq k<l\leq n}\sum_{\substack{0\leq i_0<\dots<i_{n-1}\\0\leq j_0<\dots<j_{n-1}}}
p_{\substack{i_0,\dots,i_{n-l}+1,\dots,i_{n-k}+1,\dots,i_{n-1}
\\
j_0,\dots,j_{n-1}}}\mathcal{S}_i(t)\mathcal{S}_j(s),
\label{Term 3}
\\
T_4=-\sum_{1\leq k<l\leq n}\sum_{\substack{0\leq i_0<\dots<i_{n-1}\\0\leq j_0<\dots<j_{n-1}}}
p_{\substack{i_0,\dots,i_{n-1}\\j_0,\dots,j_{n-l}-1,\dots,j_{n-k}-1,\dots,j_{n-1}}}
\mathcal{S}_i(t)\mathcal{S}_j(s)
\nonumber
\\
\phantom{T_4=}
{}+\sum_{1\leq k\leq n-1}\sum_{\substack{0\leq i_0<\dots<i_{n-1}\\0<j_1<\dots<j_{n-1}}}
p_{\substack{i_0,\dots,i_{n-1}\\-1,j_1,\dots,j_{n-k}-1,\dots,j_{n-1}}}
\mathcal{S}_i(t)S_{j_{n-1}-(n-1),\dots,j_1-1,0}(s).
\label{Term 4}
\end{gather}
Substituting~\eqref{Term 1},~\eqref{Term 2},~\eqref{Term 3} and~\eqref{Term 4} in~\eqref{main proof
equation 1}, using Lemma~\ref{technical lemma}$(ii)$ and Lemma~\ref{master symmetries on Plucker coordinates}
we obtain
\begin{gather*}
L_2^{(n)}\tau_n(s,t)
=\sum_{\substack{0\leq i_0<\dots<i_{n-1}\\0\leq j_0<\dots<j_{n-1}}}
V_2p_{\substack{i_0,\dots,i_{n-1}\\j_0,\dots,j_{n-1}}}\mathcal{S}_i(t)\mathcal{S}_j(s)
\\
\phantom{L_2^{(n)}\tau_n(s,t)=}
{}+\sum_{k=1}^{n-1}\sum_{\substack{0\leq i_0<\dots<i_{n-1}\\0<j_1<\dots<j_{n-1}}}
p_{\substack{i_0,\dots,i_{n-1}\\-1,j_1,\dots,j_{n-k}-1,\dots,j_{n-1}}}\mathcal{S}_i(t)S_{j_{n-1}-(n-1),\dots,j_1-1,0}(s)
\\
\phantom{L_2^{(n)}\tau_n(s,t)=}
{}-s_1\sum_{\substack{0\leq i_0<\dots<i_{n-1}\\0<j_1<\dots<j_{n-1}}}
p_{\substack{i_0,\dots,i_{n-1}\\-1,j_1,\dots,j_{n-1}}}\mathcal{S}_i(t)S_{j_{n-1}-(n-1),\dots,j_1-1,0}(s)
\\
\phantom{L_2^{(n)}\tau_n(s,t)=}
{}+\sum_{\substack{0\leq i_0<\dots<i_{n-1}\\0\leq j_1<\dots<j_{n-1}}}
p_{\substack{i_0,\dots,i_{n-1}\\-1,j_1,\dots,j_{n-1}}}\mathcal{S}_i(t)S_{j_{n-1}-(n-1),\dots,j_1-1,1}(s).
\end{gather*}
We prove that the last three terms in this expression annihilate
\begin{gather}
0=\sum_{k=1}^{n-1}\sum_{\substack{0\leq i_0<\dots<i_{n-1}\\0<j_1<\dots<j_{n-1}}}
p_{\substack{i_0,\dots,i_{n-1}\\-1,j_1,\dots,j_{n-k}-1,\dots,j_{n-1}}}
\mathcal{S}_i(t)S_{j_{n-1}-(n-1),\dots,j_1-1,0}(s)
\nonumber
\\
\phantom{0=}
{}-s_1\sum_{\substack{0\leq i_0<\dots<i_{n-1}\\0<j_1<\dots<j_{n-1}}}
p_{\substack{i_0,\dots,i_{n-1}\\-1,j_1,\dots,j_{n-1}}}
\mathcal{S}_i(t)S_{j_{n-1}-(n-1),\dots,j_1-1,0}(s)
\nonumber
\\
\phantom{0=}
{}+\sum_{\substack{0\leq i_0<\dots<i_{n-1}\\0\leq j_1<\dots<j_{n-1}}}
p_{\substack{i_0,\dots,i_{n-1}\\-1,j_1,\dots,j_{n-1}}}
\mathcal{S}_i(t)S_{j_{n-1}-(n-1),\dots,j_1-1,1}(s),
\label{proof main theorem equation 2}
\end{gather}
and hence
\begin{gather}
L_2^{(n)}\tau_n(s,t)
=\sum_{\substack{0\leq i_0<\dots<i_{n-1}\\0\leq j_0<\dots<j_{n-1}}}
V_2p_{\substack{i_0,\dots,i_{n-1}\\j_0,\dots,j_{n-1}}}\mathcal{S}_i(t)\mathcal{S}_j(s).
\label{proof main theorem equation 3}
\end{gather}
Indeed, developing the determinant $S_{j_{n-1}-(n-1),\dots,j_1-1,1}(s)$ with respect to the last line,
\mbox{using} the fact that the f\/irst elementary Schur polynomials are $S_0(s)=1$ and $S_1(s)=s_1$, and
Lemma~\ref{determinants of Schur polynomials}$(ii)$, we have
\begin{gather*}
\sum_{\substack{0\leq i_0<\dots<i_{n-1}\\0\leq j_1<\dots<j_{n-1}}}
p_{\substack{i_0,\dots,i_{n-1}\\-1,j_1,\dots,j_{n-1}}}
\mathcal{S}_i(t)S_{j_{n-1}-(n-1),\dots,j_1-1,1}(s)
\\
\qquad
=s_1\sum_{\substack{0\leq i_0<\dots<i_{n-1}\\0\leq j_1<\dots<j_{n-1}}}
p_{\substack{i_0,\dots,i_{n-1}\\-1,j_1,\dots,j_{n-1}}}
\mathcal{S}_i(t)S_{j_{n-1}-(n-1),\dots,j_1-1}(s)
\\
\qquad
\phantom{=}
{}-\sum_{\substack{0\leq i_0<\dots<i_{n-1}\\0\leq j_1<\dots<j_{n-1}}}
p_{\substack{i_0,\dots,i_{n-1}\\-1,j_1,\dots,j_{n-1}}}
\mathcal{S}_i(t)\sum_{l=1}^{n-1}S_{j_{n-1}-(n-1),\dots,j_{n-l}
-(n-l)+1,\dots,j_1-1}(s).
\end{gather*}
By an argument similar to that of the proof of Lemma~\ref{corollary technical lemma}, we get
\begin{gather*}
\sum_{\substack{0\leq i_0<\dots<i_{n-1}\\0\leq j_1<\dots<j_{n-1}}}
p_{\substack{i_0,\dots,i_{n-1}\\-1,j_1,\dots,j_{n-1}}}
\mathcal{S}_i(t)S_{j_{n-1}-(n-1),\dots,j_1-1,1}(s)
\\
\qquad
=s_1\sum_{\substack{0\leq i_0<\dots<i_{n-1}\\0\leq j_1<\dots<j_{n-1}}}
p_{\substack{i_0,\dots,i_{n-1}\\-1,j_1,\dots,j_{n-1}}}
\mathcal{S}_i(t)S_{j_{n-1}-(n-1),\dots,j_1-1}(s)
\\
\qquad
\phantom{=}
{}-\sum_{l=1}^{n-1}\sum_{\substack{0\leq i_0<\dots<i_{n-1}\\0\leq j_1<\dots<j_{n-1}}}
p_{\substack{i_0,\dots,i_{n-1}\\-1,j_1,\dots,j_{n-l}-1,\dots,j_{n-1}}}
\mathcal{S}_i(t)S_{j_{n-1}-(n-1),\dots,j_1-1}(s).
\end{gather*}
Noticing that $S_{j_{n-1}-(n-1),\dots,j_1-1}(s)=0$ when $j_1=0$, and
\begin{gather*}
S_{j_{n-1}-(n-1),\dots,j_1-1}(s)=S_{j_{n-1}-(n-1),\dots,j_1-1,0}(s),
\end{gather*}
when $j_1>0$, we get~\eqref{proof main theorem equation 2}, and hence~\eqref{proof main theorem equation 3}.
This proves the case $k=2$ and f\/inishes the proof.
\end{proof}

It would be nice to have a~proof of Theorem~\ref{Virasoro constraints theorem} using the vertex operators
techniques deve\-loped by the Kyoto school~\cite{DKJM}, but this remains a~challenge for the future!

\subsection*{Acknowledgements}

The authors thank the referees for their useful comments on this work and
for drawing attention to the references~\cite{KM, KMZ1, KMZ2}.
The f\/irst author acknowledges the partial support of the Belgian Interuniversity Attraction Poles P06/02
and P07/18.
During part of this research, the second author was a~Research Fellow of the Belgian National Science
Foundation (FNRS), whose support is also gratefully acknowledged.

\LastPageEnding


\addcontentsline{toc}{section}{References}
\begin{thebibliography}{99}
\footnotesize\itemsep=0pt

\bibitem{AL1}
Ablowitz M.J., Ladik J.F., Nonlinear dif\/ferential-dif\/ference equations,
\href{http://dx.doi.org/10.1063/1.522558}{\textit{J.~Math. Phys.}} \textbf{16} (1975), 598--603.

\bibitem{AL2}
Ablowitz M.J., Ladik J.F., Nonlinear dif\/ferential-dif\/ference equations and
  {F}ourier analysis, \href{http://dx.doi.org/10.1063/1.523009}{\textit{J.~Math. Phys.}} \textbf{17} (1976), 1011--1018.

\bibitem{ASVM}
Adler M., Shiota T., van Moerbeke P., Random matrices, {V}irasoro algebras, and
  noncommutative {KP}, \href{http://dx.doi.org/10.1215/S0012-7094-98-09417-0}{\textit{Duke Math.~J.}} \textbf{94} (1998), 379--431,
  \href{http://arxiv.org/abs/solv-int/9812006}{solv-int/9812006}.

\bibitem{AVM1}
Adler M., van Moerbeke P., Matrix integrals, {T}oda symmetries, {V}irasoro
  constraints, and orthogonal polynomials, \href{http://dx.doi.org/10.1215/S0012-7094-95-08029-6}{\textit{Duke Math.~J.}} \textbf{80}
  (1995), 863--911, \href{http://arxiv.org/abs/solv-int/9706010}{solv-int/9706010}.

\bibitem{AVM2}
Adler M., van Moerbeke P., Integrals over classical groups, random
  permutations, {T}oda and {T}oeplitz lattices, \href{http://dx.doi.org/10.1002/1097-0312(200102)54:2<153::AID-CPA2>3.0.CO;2-5}{\textit{Comm. Pure Appl. Math.}}
  \textbf{54} (2001), 153--205, \href{http://arxiv.org/abs/math.CO/9912143}{math.CO/9912143}.

\bibitem{AVM3}
Adler M., van Moerbeke P., Recursion relations for unitary integrals,
  combinatorics and the {T}oeplitz lattice, \textit{Comm. Math. Phys.}
  \textbf{237} (2003), 397--440, \href{http://arxiv.org/abs/math-ph/0201063}{math-ph/0201063}.

\bibitem{AD}
Aldous D., Diaconis P., Longest increasing subsequences: from patience sorting
  to the {B}aik--{D}eift--{J}ohansson theorem, \href{http://dx.doi.org/10.1090/S0273-0979-99-00796-X}{\textit{Bull. Amer. Math. Soc.}}
  \textbf{36} (1999), 413--432.

\bibitem{BMS}
Bowick M.J., Morozov A., Shevitz D., Reduced unitary matrix models and the
  hierarchy of {$\tau$}-functions, \href{http://dx.doi.org/10.1016/0550-3213(91)90365-5}{\textit{Nuclear Phys.~B}} \textbf{354}
  (1991), 496--530.

\bibitem{C}
Cafasso M., Matrix biorthogonal polynomials on the unit circle and non-abelian
  {A}blowitz--{L}adik hierarchy, \href{http://dx.doi.org/10.1088/1751-8113/42/36/365211}{\textit{J.~Phys.~A: Math. Theor.}} \textbf{42}
  (2009), 365211, 20~pages, \href{http://arxiv.org/abs/0804.3572}{arXiv:0804.3572}.

\bibitem{CMV}
Cantero M.J., Moral L., Vel{\'a}zquez L., Five-diagonal matrices and zeros of
  orthogonal polynomials on the unit circle, \href{http://dx.doi.org/10.1016/S0024-3795(02)00457-3}{\textit{Linear Algebra Appl.}}
  \textbf{362} (2003), 29--56, \href{http://arxiv.org/abs/math.CA/0204300}{math.CA/0204300}.

\bibitem{CBGV}
Cruz-Barroso R., Gonz{\'a}lez-Vera P., A {C}hristof\/fel--{D}arboux formula and a
  {F}avard's theorem for orthogonal {L}aurent polynomials on the unit circle,
  \href{http://dx.doi.org/10.1016/j.cam.2004.09.039}{\textit{J.~Comput. Appl. Math.}} \textbf{179} (2005), 157--173.

\bibitem{D}
Damianou P.A., Master symmetries and {$R$}-matrices for the {T}oda lattice,
  \href{http://dx.doi.org/10.1007/BF00398275}{\textit{Lett. Math. Phys.}} \textbf{20} (1990), 101--112.

\bibitem{DKJM}
Date E., Kashiwara M., Jimbo M., Miwa T., Transformation groups for soliton
  equations, in Nonlinear Integrable Systems~-- Classical Theory and Quantum
  Theory ({K}yoto, 1981), World Sci. Publishing, Singapore, 1983, 39--119.

\bibitem{DVV}
Dijkgraaf R., Verlinde H., Verlinde E., Loop equations and {V}irasoro
  constraints in nonperturbative two-dimensional quantum gravity,
  \href{http://dx.doi.org/10.1016/0550-3213(91)90199-8}{\textit{Nuclear Phys.~B}} \textbf{348} (1991), 435--456.

\bibitem{FG}
Faybusovich L., Gekhtman M., Poisson brackets on rational functions and
  multi-{H}amiltonian structure for integrable lattices, \href{http://dx.doi.org/10.1016/S0375-9601(00)00445-X}{\textit{Phys. Lett.~A}}
  \textbf{272} (2000), 236--244, \href{http://arxiv.org/abs/nlin.SI/0006045}{nlin.SI/0006045}.

\bibitem{Fe}
Fernandes R.L., On the master symmetries and bi-{H}amiltonian structure of the
  {T}oda lattice, \href{http://dx.doi.org/10.1088/0305-4470/26/15/028}{\textit{J.~Phys.~A: Math. Gen.}} \textbf{26} (1993),
  3797--3803.

\bibitem{FW}
Forrester P.J., Witte N.S., Bi-orthogonal polynomials on the unit circle,
  regular semi-classical weights and integrable systems, \href{http://dx.doi.org/10.1007/s00365-005-0616-7}{\textit{Constr.
  Approx.}} \textbf{24} (2006), 201--237, \href{http://arxiv.org/abs/math.CA/0412394}{math.CA/0412394}.

\bibitem{F}
Fuchssteiner B., Mastersymmetries, higher order time-dependent symmetries and
  conserved densities of nonlinear evolution equations, \href{http://dx.doi.org/10.1143/PTP.70.1508}{\textit{Progr. Theoret.
  Phys.}} \textbf{70} (1983), 1508--1522.

\bibitem{FKN}
Fukuma M., Kawai H., Nakayama R., Continuum {S}chwinger--{D}yson equations and
  universal structures in two-dimensional quantum gravity, \href{http://dx.doi.org/10.1142/S0217751X91000733}{\textit{Internat.~J.
  Modern Phys.~A}} \textbf{6} (1991), 1385--1406.

\bibitem{GHMT}
Gesztesy F., Holden H., Michor J., Teschl G., Local conservation laws and the
  {H}amiltonian formalism for the {A}blowitz--{L}adik hierarchy, \href{http://dx.doi.org/10.1111/j.1467-9590.2008.00405.x}{\textit{Stud.
  Appl. Math.}} \textbf{120} (2008), 361--423, \href{http://arxiv.org/abs/0711.1644}{arXiv:0711.1644}.

\bibitem{GH}
Gr{\"u}nbaum F.A., Haine L., A theorem of {B}ochner, revisited, in Algebraic
  Aspects of Integrable Systems, \textit{Progr. Nonlinear Differential
  Equations Appl.}, Vol.~26, Birkh\"auser Boston, Boston, MA, 1997, 143--172.

\bibitem{HH}
Haine L., Horozov E., Toda orbits of {L}aguerre polynomials and representations
  of the {V}irasoro algebra, \textit{Bull. Sci. Math.} \textbf{117} (1993),
  485--518.

\bibitem{HS}
Haine L., Semengue J.P., The {J}acobi polynomial ensemble and the {P}ainlev\'e~{VI} equation, \href{http://dx.doi.org/10.1063/1.532855}{\textit{J.~Math. Phys.}} \textbf{40} (1999), 2117--2134.

\bibitem{HV}
Haine L., Vanderstichelen D., A centerless representation of the {V}irasoro
  algebra associated with the unitary circular ensemble, \href{http://dx.doi.org/10.1016/j.cam.2010.06.006}{\textit{J.~Comput.
  Appl. Math.}} \textbf{236} (2011), 19--27, \href{http://arxiv.org/abs/1001.4244}{arXiv:1001.4244}.

\bibitem{KS}
Kac V., Schwarz A., Geometric interpretation of the partition function of
  {$2$}{D} gravity, \href{http://dx.doi.org/10.1016/0370-2693(91)91901-7}{\textit{Phys. Lett.~B}} \textbf{257} (1991), 329--334.

\bibitem{KR}
Kac V.G., Raina A.K., Bombay lectures on highest weight representations of
  inf\/inite-dimensional {L}ie algebras, \textit{Advanced Series in Mathematical
  Physics}, Vol.~2, World Scientif\/ic Publishing Co. Inc., Teaneck, NJ, 1987.

\bibitem{KM}
Kharchev S., Mironov A., Integrable structures of unitary matrix models,
  \href{http://dx.doi.org/10.1142/S0217751X92002179}{\textit{Internat.~J. Modern Phys.~A}} \textbf{7} (1992), 4803--4824.

\bibitem{KMZ1}
Kharchev S., Mironov A., Zhedanov A., Faces of relativistic {T}oda chain,
  \href{http://dx.doi.org/10.1142/S0217751X97001493}{\textit{Internat.~J. Modern Phys.~A}} \textbf{12} (1997), 2675--2724,
  \href{http://arxiv.org/abs/hep-th/9606144}{hep-th/9606144}.

\bibitem{KMZ2}
Kharchev S., Mironov A., Zhedanov A., Dif\/ferent aspects of relativistic {T}oda
  chain, in Symmetries and Integrability of Dif\/ference Equations ({C}anterbury,
  1996), \href{http://dx.doi.org/10.1017/CBO9780511569432.004}{\textit{London Math. Soc. Lecture Note Ser.}}, Vol.~255, Edi\-tors P.A.~Clarkson, F.W.~Nijhof\/f, Cambridge Univ. Press, Cambridge, 1999, 23--40,
  \href{http://arxiv.org/abs/hep-th/9612094}{hep-th/9612094}.

\bibitem{M}
Martinec E.J., On the origin of integrability in matrix models, \href{http://dx.doi.org/10.1007/BF02102036}{\textit{Comm.
  Math. Phys.}} \textbf{138} (1991), 437--449.

\bibitem{MM}
Mironov A., Morozov A., On the origin of {V}irasoro constraints in matrix
  models: {L}agrangian approach, \href{http://dx.doi.org/10.1016/0370-2693(90)91078-P}{\textit{Phys. Lett.~B}} \textbf{252} (1990),
  47--52.

\bibitem{N}
Nenciu I., Lax pairs for the {A}blowitz--{L}adik system via orthogonal
  polynomials on the unit circle, \href{http://dx.doi.org/10.1155/IMRN.2005.647}{\textit{Int. Math. Res. Not.}} \textbf{2005}
  (2005), 647--686, \href{http://arxiv.org/abs/math-ph/0412047}{math-ph/0412047}.

\bibitem{R}
Rains E.M., Increasing subsequences and the classical groups, \textit{Electron.~J. Combin.} \textbf{5} (1998), R12, 9~pages.

\bibitem{S1}
Simon B., Orthogonal polynomials on the unit circle. {P}art~1. Classical
  theory, \textit{American Mathematical Society Colloquium Publications},
  Vol.~54, American Mathematical Society, Providence, RI, 2005.

\bibitem{S2}
Simon B., Orthogonal polynomials on the unit circle. {P}art~2. Spectral theory,
  \textit{American Mathematical Society Colloquium Publications}, Vol.~54,
  American Mathematical Society, Providence, RI, 2005.

\bibitem{TW}
Tracy C.A., Widom H., Fredholm determinants, dif\/ferential equations and matrix
  models, \href{http://dx.doi.org/10.1007/BF02101734}{\textit{Comm. Math. Phys.}} \textbf{163} (1994), 33--72,
  \href{http://arxiv.org/abs/hep-th/9306042}{hep-th/9306042}.

\bibitem{UT}
Ueno K., Takasaki K., Toda lattice hierarchy, in Group Representations and
  Systems of Dif\/ferential Equations ({T}okyo, 1982), \textit{Adv. Stud. Pure
  Math.}, Vol.~4, North-Holland, Amsterdam, 1984, 1--95.

\bibitem{V}
Vanderstichelen D., Virasoro symmetries for the Ablowitz--Ladik hierarchy and
  non-intersecting Brownian motion models, Ph.D. Thesis, Universit\'e
  Catholique de Louvain, 2011.

\bibitem{MZ}
Zubelli J.P., Magri F., Dif\/ferential equations in the spectral parameter,
  {D}arboux transformations and a~hierarchy of master symmetries for {K}d{V},
  \href{http://dx.doi.org/10.1007/BF02101509}{\textit{Comm. Math. Phys.}} \textbf{141} (1991), 329--351.

\end{thebibliography}
\end{document}